\pdfoutput=1 
\documentclass[iop,apj,tighten,revtex4]{emulateapj}
\usepackage{threeparttable}
\usepackage{multirow}
\usepackage{amsmath}
\usepackage{ulem}
\usepackage{pgffor}
\newcommand{\tabincell}[2]{\begin{tabular}{@{}#1@{}}#2\end{tabular}}
\usepackage{xcolor}
\usepackage{hyperref}
\usepackage{lscape}
\usepackage [caption=false]{subfig}
\usepackage{natbib}
\usepackage{url}
\hypersetup{
   colorlinks,
   linkcolor={blue!88!black!80},
   citecolor={blue!88!black!80},
   urlcolor={blue!88!black!80}
}

\shortauthors{Kang J. et al.}

\begin{document}

\title{NuSTAR Hard X-ray Spectra of radio galaxies}
\author{
Jialai Kang\altaffilmark{1,2}, Junxian Wang\altaffilmark{1,2}, Wenyong Kang\altaffilmark{1,2}}
\affil{
$^1$CAS Key Laboratory for Research in Galaxies and Cosmology, Department of Astronomy, University of Science and Technology of China,
Hefei, Anhui 230026, China; jxw@ustc.edu.cn\\
$^2$School of Astronomy and Space Science, University of Science and Technology of China, Hefei 230026, China\\
}

\begin{abstract}
\textit{NuSTAR} observatory, with its 3 -- 78 keV broadband spectral coverage, enables the detections of the high-energy cutoff in a number of active galaxies, including several individual radio loud ones.
In this work we present systematic and uniform analyses of 55 \textit{NuSTAR} spectra for a large sample of 28 radio galaxies, 20 of which are FR II galaxies.
We perform spectral fitting to measure the high energy cut-off $E_{cut}$, photon index $\Gamma$, reflection factor R and Fe K$\alpha$ line equivalent width. 
Measurements of $E_{cut}$ are given for 13 sources, and lower limits for the rest. We find those $E_{cut}$ non-detections could primarily be attributed to the obviously smaller net photon counts in their spectra.
This indicates that the \textit{NuSTAR} spectra of the majority of our sample are dominated by the thermal corona emission, and the $E_{cut}$ distribution of the sample is indistinguishable from that of a radio quiet one in literature. 
The flatter \textit{NuSTAR} spectra we observed, comparing with radio quiet sources,  are thus unlikely due to jet contamination. 
The radio galaxies also show weaker X-ray reflection (both in R and Fe K$\alpha$ line EW) comparing with radio quiet ones. 
Combining with the radio quiet sample we see a correlation between R and EW, but with considerably large scatter. 
Notably, the radio loud and quiet sources appear to follow a common $\Gamma$ -- R correlation trend, supporting the outflowing corona model for both populations in which higher bulk outflowing velocity yields weaker reflection and flatter X-ray slope.

\end{abstract}


\section{Introduction} \label{sec:intro}
Active Galactic Nuclei (AGNs), powered by the accretion of matter onto supermassive black holes, are the most dominant and powerful X-ray population in the sky. For radio-quiet AGNs, it is widely accepted that the primary X-ray emission originates in a hot, compact corona located above the accretion disk \citep[e.g.][]{1993ApJ...413..507H}. Besides the cutoffed power-law  \citep[e.g.][]{2013MNRAS.433.1687M,2011A&A...532A.102R}, an X-ray reflection bump at around 30 keV and an iron K$\alpha$ line emission at around 6.4 keV are also detected, produced by the surrounding material illuminated by the central X-ray source \citep[e.g.][]{1991A&A...247...25M}.

\par The circumstance is more complicated in radio loud AGNs in which relativistic jets exist \citep{Blandford_1979, 1984RvMP...56..255B} and could also produce X-ray emission. The jet contribution may vary between different classes of radio sources.  Studies have suggested that the core X-ray emission of FR II galaxies could be dominated by the corona rather than the jet, but the reverse for FR Is \citep{Evans2006,Hardcastle2009}.  Meanwhile, the jet dominance could be higher in core-dominated or young radio galaxies \citep[e.g.][]{Migliori2014}.

It is found that radio galaxies generally have  flatter X-ray spectra, weaker X-ray reflection as well as weaker Fe K$\alpha$ line \citep[e.g.][]{1999ApJ...526...60S, 10.1046/j.1365-8711.2000.03510.x,  2000ApJ...537..654E, 2001ApJ...556...35G, 2002MNRAS.332L..45B, 2005ApJ...618..139O, doi:10.1142/S0217732307024322, 2009ApJ...700.1473S, 2013MNRAS.428.2901W}. Such feathers have been broadly discussed in literature, and many possible explanations have been raised, including significantly dilution from the relativistic jet \citep{10.1046/j.1365-8711.2000.03510.x}; inner disk geometry changing \citep{2000ApJ...537..654E}; high ionization of inner disk \citep{2002MNRAS.332L..45B}; obscuration of the central accretion flow by the jet \citep{2009ApJ...700.1473S}, outflowing corona \citep{Beloborodov_outflow, Malzac_outflow}, etc.

A prominent and key feature of X-ray corona emission is the high energy cutoff in the spectrum which is an indicator of the coronal temperature $T_e$. For an optically thick thermal corona with opacity $\tau \gg 1$, $kT_e$ is approximately $E_{cut}/3$, while $kT_e \sim E_{cut}/2$ for $\tau \leq 1$ \citep[e.g.][]{Petrucci_2001}. 
The Nuclear Spectroscopic Telescope Array \citep[\textit{NuSTAR}][]{Harrison_2013}, the first focusing high-energy X-ray mission with a broad and high quality spectral coverage from 3--78 keV, 
is very powerful in probing the coronal properties of X-ray bright AGNs, as well as constraining the X-ray reflection component \citep[e.g.][]{Kamraj_2018,sampleused, refId01, 10.1093/mnras/stz156, 10.1093/mnras/stz275}. 

Particularly, \textit{NuSTAR} observations have detected the high energy cutoff $E_{cut}$ in a couple of radio loud AGNs, 
including 3C 382 \citep{Ballantyne_2014_3C382}, 3C 390.3 \citep{2015ApJ...814...24L_3C309},  4C 74.26 \citep{2017ApJ...841...80L_4C74},
3C 120 \citep{2018JApA...39...15Rb}, 3C 273 \citep{Madsen_2015_3C273}, IGR J21247+5058 \citep{2018MNRAS.481.4419B}.
This indicates that the hard X-ray emission in these individual sources are dominated by the thermal corona emission, instead of the non-thermal jet. 
Meanwhile, non-detections of the high energy cutoff were also reported with \textit{NuSTAR} spectra in several other radio galaxies, including CentaurusA\citep{F_rst_2016_CenA}, 3C 227 \citep{Kamraj_2018}, 4C +18.51 \citep{Kamraj_2018}, 
NGC 1275 \citep{Rani_2018}, 
S5 2116+81\citep{10.1093/mnras/stz156}, PKS 2331-240 and PKS 2356-61\citep{Ursini_2018}.

Systematical and statistical study of a large sample of radio galaxies is thus essential to address the following questions: 1) are the hard X-ray spectra of radio galaxies ubiquitously dominated by the thermal corona emission? 2) are their corona properties different from their radio quiet counterparts? Such a large sample study could also constrain the strength of X-ray reflection uniformly in the population utilizing the broad spectral coverage of \textit{NuSTAR}. 

In this work we present \textit{NuSTAR} spectra for a sample of 28 radio galaxies,  mostly non-blazar AGNs.
The paper is organized as follows. \S \ref{sec:OBandData} presents our sample selection along with \textit{NuSTAR} observations and data reduction. In \S \ref{sec:spec} we describe the spectral model we use and give the fitting results. In \S \ref{sec:discussion} we present the statistical analyses and  comparison with radio quiet AGNs. Finally \S \ref{sec:summary} summarizes this work. 

\section{The Sample and Data Reduction} \label{sec:OBandData}
\subsection{Radio Loud AGN Sample} 

\begin{deluxetable*}{ccccccc}
\tabletypesize{\scriptsize}
\tablecaption{\textit{NuSTAR} Observation Details for Our Radio Galaxy Sample}
\tablewidth{0.99\textwidth}
\tablehead{
\colhead{Source} & \colhead{ID} &\colhead{Radio Morpology}&\colhead{Redshift}&\colhead{Exposure} & \colhead{Total Net Counts}  & \colhead{}\\
\colhead{} & \colhead{} &  \colhead{}  &\colhead{} &\colhead{(ks)} & \colhead{($1\times10^3$)}  & \colhead{} 
}
\startdata
3C 273 & 10002020001 & core dominated & 0.158 & 244.1 &  1074.6& \\
 & 10002020003 &  &  & 49.4 & 160.6  & \\
 & 10202020002 &  &  & 35.4 & 279.0& \\
 & 10302020002 &  &  & 35.3 & 140.2 & \\
 & 10402020002 &  &  & 16.1 & 64.0& \\
 & 10402020004 &  &  & 21.1 & 69.0& \\
 & 10402020006 &  &  & 40.2 & 133.3& \\
 & 80301602002 &  &  & 60.6 & 212.8& \\
PicA & 60101047002 & FR II & 0.035 & 109.2 & 46.0 \\
CentaurusA & 60001081002 & FR I  & 0.001 & 51.4 & 1580.4 & \\
 & 60101063002 &  &  & 22.5 & 156.6& \\
 & 60466005002 &  &  & 17.3 & 258.7 & \\
3C 382 & 60001084002 & FR II & 0.058 & 82.4 & 137.2 & \\
 & 60061286002 &  &  & 16.6 & 47.1 & \\
 & 60202015002 &  &  & 23.0 & 48.6 & \\
 & 60202015004 &  &  & 24.5 & 60.3& \\
 & 60202015006 &  &  & 20.8 & 49.2& \\
 & 60202015008 &  &  & 21.7 & 49.3& \\
 & 60202015010 &  &  & 21.0 & 53.4& \\
IGR J21247+5058 & 60061305002 & FR II & 0.02 & 24.3 & 102.9 & \\
 & 60301005002 &  &  & 40.2 & 210.3 & \\
3C 390.3 & 60001082002 & FR II & 0.057 & 23.6 & 50.3& \\
 & 60001082003 &  &  & 47.5 & 99.1& \\
3C 109  & 60301011002 & FR II & 0.31 & 67.5 & 21.7 & \\
 & 60301011004 &  &  & 88.9 & 26.3 & \\
4C 74.26 & 60001080002 & FR II & 0.104 & 19.0 & 30.1& \\
 & 60001080004 &  &  & 56.5 & 92.2 & \\
 & 60001080006 &  &  & 90.8 & 128.6 & \\
 & 60001080008 &  &  & 42.8 & 62.3 & \\
3C 120 & 60001042002 & FR I  & 0.033 & 21.6 & 54.6 &\\
 & 60001042003 &  &  & 127.7 & 334.2 & \\
3C 227  & 60061329002 & FR II & 0.086 & 17.2 & 10.1& \\
 & 60061329004 &  &  & 12.1 & 6.1 & \\
3C 184.1  & 60160300002 & FR II & 0.119 & 22.2 & 4.0 & \\
3C 111  & 60202061002 & FR II & 0.049 & 21.2 & 46.0 & \\
 & 60202061004 &  &  & 49.2 & 98.1 & \\
PKS 2356-61  & 60061330002 & FR II & 0.096 & 23.0 & 4.2& \\
3C 206  & 60160332002 & FR II & 0.2 & 17.3 & 10.0& \\
3C 345  & 60160647002 & core dominated & 0.593 & 24.3 & 4.7 & \\
2MASX J03181899+6829322 & 60061342002 & FR II & 0.09 & 24.1 & 5.8 & \\
4C +18.51 & 60160672002 & FR II & 0.186 & 22.4 & 3.9&\\
S5 2116+81  & 60061303002 & FR I  & 0.084 & 18.5 & 13.7& \\
4C +21.55  & 60160740002 & FR II & 0.173 & 21.5 & 8.9& \\
PKS 2331-240 & 60160832002 & FR II & 0.048 & 21.1 & 12.1 &\\
PKS 1916-300  & 60160707002 & FR II & 0.167 & 21.7 & 6.0& \\
IGR 14488-4008  & 60463049002 & FR II & 0.123 & 20.2 & 6.8 & \\
3C 279 & 60002020002 & core dominated & 0.536 & 39.5 & 23.1 & \\
 & 60002020004 &  &  & 42.7 & 52.0& \\
Leda 100168  & 60160631002 & FR II & 0.183 & 24.1 & 4.1& \\
3C 309.1  & 60376006002 & core dominated & 0.905 & 60.9 & 5.9& \\
3C 332  & 60160634002 & FR II & 0.152 & 20.9 & 11.2& \\
PKS 0442-28  & 60160205002 & FR II & 0.147 & 22.4 & 6.0 & \\
NGC 1275 & 60061361002 & FR I  & 0.018 & 19.9 & 57.1 & \\
 & 90202046002 &  &  & 22.3 & 80.1 & \\
 & 90202046004 &  &  & 28.1 & 88.1 & \\
 \tablecomments{For sources from \citet{10.1093/mnras/stw1468}, we adopt the radio morphology given by \citet{10.1093/mnras/stw1468}. For the rest three sources from 3C Catalog, we adopt the morphology from NASA/IPAC Extragalactic Database (NED).
Total net counts are the sum of $NuSTAR$ net counts from FPMA and FPMB in the energy range of 3 -- 78 keV.
}
\label{sources}
\enddata 
\end{deluxetable*}

The main purpose of this paper is to systematically study the intrinsic hard X-ray spectra of non-blazar radio galaxies, and we are especially interested in the coronal physics in such sources. A radio galaxy sample  composed of sources without severe absorption or strong relativistic beaming is thus needed. Besides, we need \textit{NuSTAR} spectra with sufficient signal to noise ratios (S/N) to perform X-ray spectral fitting. 

Using the NVSS \citep{1998AJ....115.1693C}, FIRST \citep{1997ApJ...475..479W} and SUMSS radio surveys \citep{2003MNRAS.342.1117M}, \citet{10.1093/mnras/stw1468} searched for radio counterparts for INTEGRAL/IBIS and \textit{Swift}/BAT detected AGNs. The radio galaxies in their sample (64 radio loud AGNs, 70\% is made of high-excitation radio galaxies) are required to display a double lobe radio morphology, which guarantees a weak relativistic beaming.
We crossmatch their sample with \textit{HEASARC}\footnote{\url{https://heasarc.gsfc.nasa.gov/}} (High Energy Astrophysics Archive Research Center) for publicly released \textit{NuSTAR} observations till May 1, 2019.
In this way, we get a sample of 36  \textit{NuSTAR} observed radio galaxies.

We notice that some well known radio loud sources, such as 3C 273, are not included in this sample, because of the requirement of  displaying double lobe radio morphology by \citet{10.1093/mnras/stw1468}. 
Through cross-matching the 3C and 3CR Catalogues \citep{2002yCat.8001....0E} with \textit{Swift}/BAT 105 month hard X-ray catalog \citep[][to ensure the hard X-ray brightness]{oh_105month}  and \textit{NuSTAR} archive,
we find three more sources (3C 273, 3C 279 and 3C 345), and the \textit{NuSTAR} 
observation of 3C 345 has not been published elsewhere.
All three sources are core dominated, flat-spectrum radio quasars. The jet contributions to their \textit{NuSTAR} X-ray spectra are expected to be stronger comparing with the radio galaxies we introduced above.
However we note a high energy X-ray cutoff has been detected in 3C 273 with \textit{NuSTAR}  observation \citep{Madsen_2015_3C273}, indicating its hard X-ray emission is significantly dominated by the corona.
We note that both 3C 279 and 3C 345 show strong emission lines in UV/optical spectra \citep{2000ApJS..126..133W,2002ApJS..140..143B}, similar to 3C 273, indicating their optical emission are not dominated by the jet emission.
We speculate their hard X-ray spectra could be similarly dominated by the corona emission.
We include these three sources into this study, and stress that excluding them will not alter our main scientific results (see \S\ref{subsec:jet} for comparison and discussion).

We further exclude Compton-thick and heavily obscured sources (with $N_H$$>$$10^{23}  cm^{-2}$) which need more complicated spectral models, and are beyond the scope of this paper. Moreover, faint sources with \textit{NuSTAR}  3 -- 78 keV total net counts (FPMA $+$ FPMB, hereafter the same) $<$ 3000 are also excluded. 
Our final sample consists of 28 radio galaxies. We note that \textit{NuSTAR} spectra for 13 out of the 28 sources have not been published elsewhere. 
 We present the sample along with the corresponding 55 \textit{NuSTAR} observations in Tab. \ref{sources}.
The 11 excluded sources can be found in Tab. \ref{giveup} in Appendix \ref{app:A}.

The final sample (see Table \ref{sources}) consists of 20 FR II galaxies, 4 FR I galaxies (NGC 1275, 3C 120, Centaurus A, and S5 2116+81), and 4 core dominated sources (one compact steep spectrum quasar, 3C 309.1; and 3 flat-spectrum radio quasars, 3C 273,  3C 279 and 3C 345), thus is dominated by powerful lobe-dominated FR II galaxies and high-excitation and efficiently accreting sources (see \S\ref{subsec:jet} for further discussion). 
Further matching the whole \textit{Swift}/BAT catalog with \textit{NuSTAR} archive yields no additional sources satisfying our selection criteria. 
Note that our sample is primarily hard X-ray/soft $\gamma$-ray selected as all sources are $Swift$/BAT detected.
This is indeed a necessary condition of this study as we need sufficient $NuSTAR$ X-ray photons to perform spectral fitting, and we expect no 
significant bias due to this condition on the hard X-ray spectral properties. Meanwhile, the comparison samples of radio quiet AGNs used in this work are similarly hard X-ray/soft $\gamma$-ray selected.

\subsection{Data Reduction}\label{S:Reduction}
We download raw data from \textit{HEASARC}, derived from both \textit{NuSTAR} modules \citep[FPMA \& FPMB,][]{Harrison_2013}. The data reduction is performed using the \textit{NuSTAR} Data Analysis Software (\textit{NuSTAR}-DAS), which is part of the HEAsoft package (version 6.26). 
We use CALDB version 20190513 for calibration, and the calibrated and cleaned event files are produced using the software tool \textit{nupipeline}.

\begin{figure}
\centering

\subfloat{\includegraphics[width=1.65in]{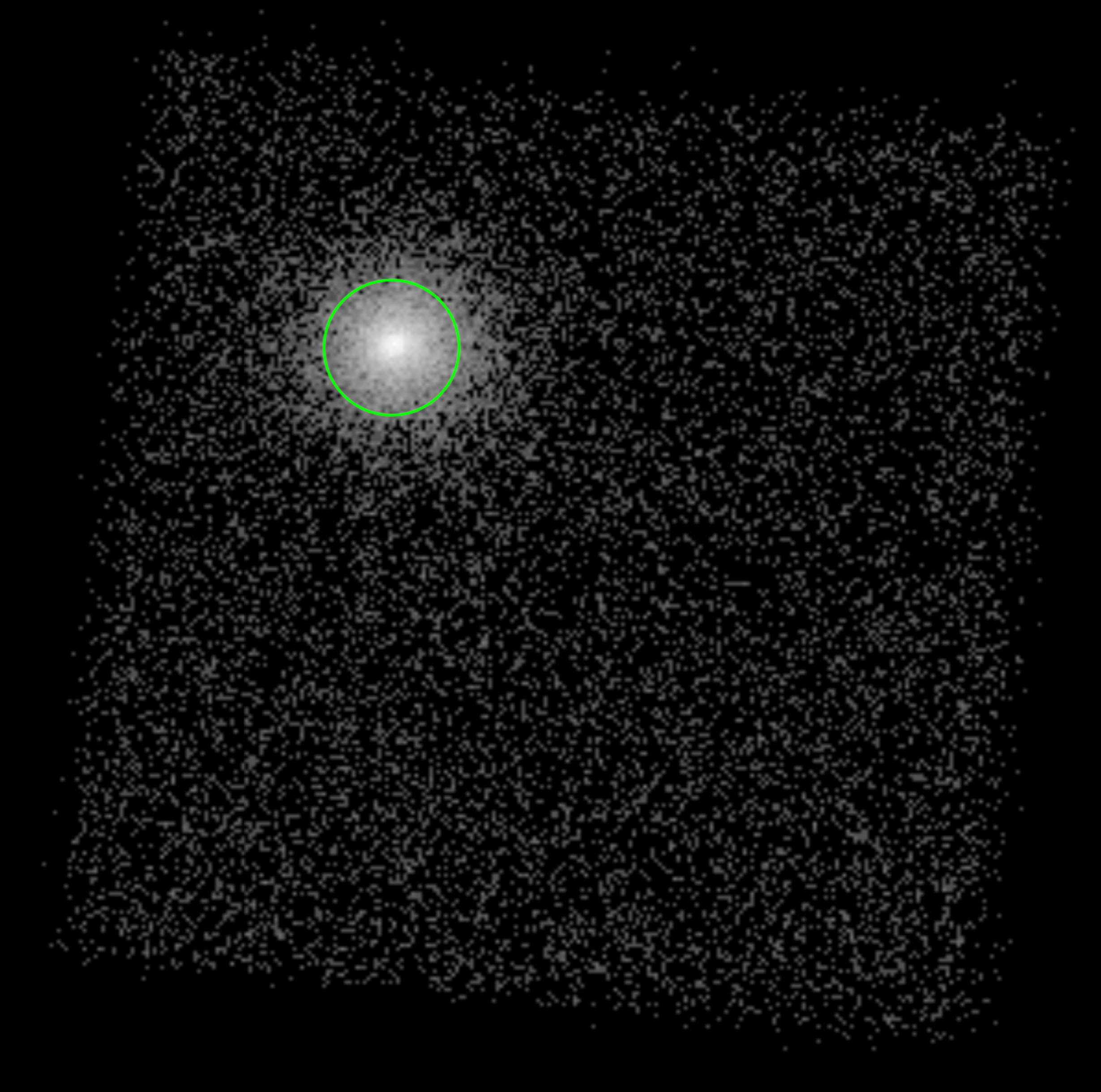}}
\hspace{1ex}
\subfloat{\includegraphics[width=1.65in]{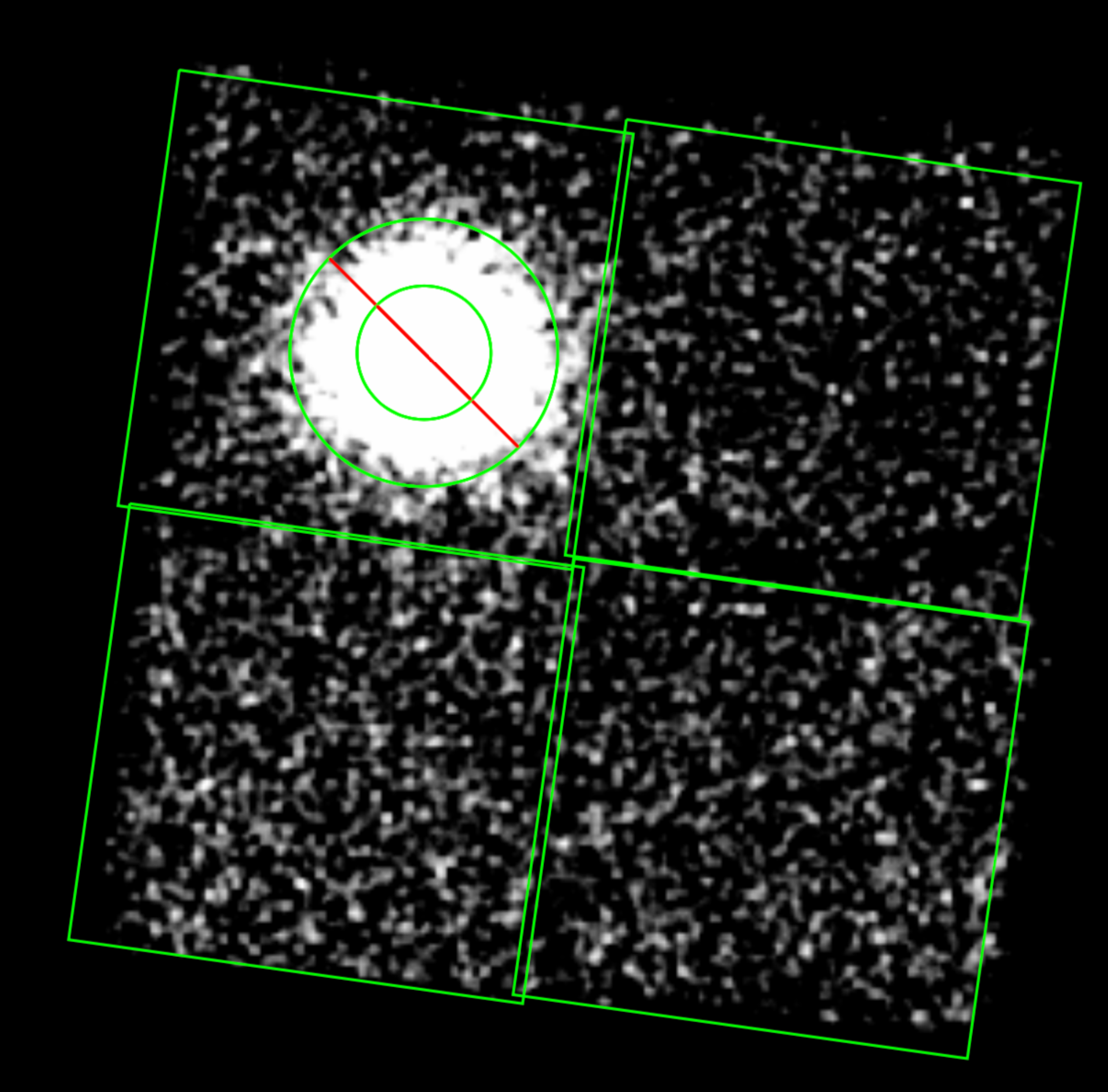}}
\caption{ 
\textit{NuSTAR} FPMA image of 3C 111 (ObsID 60202061002) as an example. The left panel shows a circle of 60\arcsec\ for source extraction.  The right panel shows source-free regions used for background simulation in \textit{NUSKYBGD}, which consist of four rectangular regions excluding a circle of 120\arcsec\ centered on the source. The image in the right panel is smoothed to demonstrate the spatially non-uniform background.
}
\label{fig:region}
\end{figure}

We use the \textit{nuproducts} module to extract source spectra within a circle centered on each source.
The S/N of the extracted net counts (subtracted given background spectrum) relies on the circle radius utilized. 
To find the optimized source extraction radius which could maximize the output S/N, we adopt a set of values of source region radii to extract the source spectra and calculate the S/N of the net counts (after subtracting the background, in the energy range of 3 -- 78 keV).
We find for our sample the median optimal radius for source extraction is $\sim$ 60\arcsec\ (slightly smaller radii for fainter sources, and vice versa), which is also the value widely used in literature.  In this work we extract source spectra within a uniform radius of 60\arcsec. 

As for background extraction, due to the spatially variable background of \textit{NuSTAR} observations \citep{Wik_2014}, the common approach adopted in literature to extract the background spectrum within a region near the source could be improved.  As shown in \citet{Wik_2014}, the \textit{NuSTAR} background consists of different components, including internal background, aperture stray light, scattered and reflected stray light, and focused cosmic X-ray background.Among them,  the aperture background (stray light) is clearly spatially variable and its spatial distribution can be fully modulated.
\citet{Wik_2014} depveloped \textit{NUSKYBGD} with built-in models for each background component,  and background spectra extracted from different source-free regions within the FOV could be fitted with the built-in models using XSPEC \citep{1996ASPC..101...17A}. With the best-fit parameters,  \textit{NUSKYBGD} could simulate the background spectra in any region within the FoV, taking care of the spatial variation of the background. In this work we use \textit{NUSKYBGD} to calculate background spectra within the source extraction region for each source.
An example of source and background spectra extraction is shown in Fig. \ref{fig:region}. As a final step, we re-bin the source spectra to achieve a minimum of 50 counts per bin using \textit{grppha}.

\section{Spectral Fitting}\label{sec:spec}

Spectral fitting is performed in 3--78 keV band using XSPEC \citep{1996ASPC..101...17A}, ${\chi}^2$ statistics, and relative element abundances given by \citet{ANDERS1989197}. Unless otherwise noted, all the errors along with upper/lower limits reported throughout the paper are calculated using ${\Delta\chi}^2$ = 2.71 criterion (90\% confidence range). For each source,  the spectra obtained by the two \textit{NuSTAR} detectors, FPMA and FPMB, are fitted simultaneously, with a cross-normalization difference typically less than 5\% \citep{2015ApJS..220....8M}.

\par We would like to measure the key hard X-ray spectral properties (including photon index $\Gamma$, high energy cutoff E$_c$, Fe K$\alpha$ line and the strength of reflection component) of non-blazar radio galaxies, and make comparison with those of radio-quiet AGNs. 
We adopt a base model \textit{pexrav} + \textit{zgauss}, in which \textit{pexrav} describes the exponentially cut-off power-law plus the reflection component \citep[a widely used model in literature,][]{1995MNRAS.273..837M}, and \textit{zgauss} the Fe K$\alpha$ line. 
Intrinsic absorption is modeled with $zphabs$, and Galactic absorption is ignored as it has little and negligible impact to \textit{NuSTAR} spectra.

We note in some X-ray models,  the continuum reflection component and the Fe K$\alpha$ line are jointly and self-consistently fitted (under certain assumptions),  
such as $pexmon$ \citep{2007MNRAS.382..194N} and $relxill$ \citep{Garc_a_2014}. In this work, we adopt the more general approach to model the continuum reflection  and the Fe K$\alpha$ line separately,  because:

\begin{itemize}
 
\item{
Most previous studies of \textit{NuSTAR} AGN samples adopted \textit{pexrav} \citep[e.g.][]{Kamraj_2018,sampleused,10.1093/mnras/stz156, 10.1093/mnras/stz275}. Fitting with identical models is essential to validate the direct comparison of our results with those studies.

}
\item{The wide energy band of \textit{NuSTAR} enables us to constrain the continuum reflection independent to the Fe K$\alpha$ line. This also enables us to potentially investigate the yet-unclear coupling between the continuum reflection and the Fe K$\alpha$ line emission. Furthermore, in case of clear decoupling between the Fe K$\alpha$ line and the continuum reflection, fitting with them simply tied may hinder precise measurement of $E_{cut}$ \citep{Zhangjx2018}.

 }

\end{itemize}

In \textit{pexrav} the parameter R, the reflection scaling factor, describes the strength of the reflection with respect to the primary emission. For simplicity, we adopt the solar element abundance for the reflector and an inclination of $\cos i = 0.45$, which is the default value of the model. 
We fix the Gaussian emission line at 6.4 keV in the rest frame, and the line width at 19 eV (the mean Fe K$\alpha$ line width in AGNs measured with Chandra HETG, \citealt{Shu_2010}), to model the neutral and narrow Fe K$\alpha$ line. 
The free parameters during the fitting include the cross-normalization constant between FPMA and FPMB, photon index $\Gamma$, $E_{cut}$, the reflection scaling factor R of \textit{pexrav}, and the normalizations of the cutoff powerlaw and the Gaussian line.

\par Our base model provides adequate fits ($P_{null} > 5\%$) to all but one source. The only exception is  NGC 1275 (3C 84). Following \citet{Rani_2018}, we add an \textit{apec} component to account for significant radiation from hot diffuse gas. 
We fix the abundance parameter of $apec$ at 0.39 and set the $kT$ parameter free to vary. With this updated model, its spectra are well fitted with $P_{null} > 40\%$ for all three \textit{NuSTAR} observations.

\begin{table*}[h!]
\caption{Spectral Fitting Results}
\label{tab:Results}
\centering
\resizebox{150mm}{105mm}{`
\begin{tabular}{l c c c c c c c}
\hline
\hline
\tabincell{l}{Source\\ \ } & \tabincell{c}{ID\\ \ } &\tabincell{c}{nH\\ ${(10^{22} cm^{-2})}$ }&\tabincell{c}{${\Gamma}$\\ \ } & \tabincell{c}{$E_{cut}$\\ (keV)} &
\tabincell{c}{    R\\ \    } & \tabincell{c}{Fe K$\alpha$ EW\\ (eV)} & \tabincell{c}{$\chi^2_\nu$\\ \ } \\
\hline

3C 273$^\gamma$ & 10002020001 &  $<  0.3$ & 1.62$_{-0.01}^{+0.02}$  & 226$_{-26}^{+42}$  & 0.05$_{-0.03}^{+0.03}$  &  $<  9.6$ & 1.02\\
 & 10002020003 &  $<  1.0$ & 1.72$_{-0.05}^{+0.02}$  &  $>  438$ &  $<  0.137$ &  $<  14.6$ & 0.96\\
 & 10202020002 &  $<  0.3$ & 1.54$_{-0.01}^{+0.02}$  & 225$_{-40}^{+69}$  &  $<  0.053$ & 10.9$_{-8.6}^{+8.6}$  & 1.02\\
 & 10302020002 & 0.7$_{-0.6}^{+0.7}$  & 1.64$_{-0.03}^{+0.05}$  &  $>  294$ &  $<  0.077$ &  $<  15.1$ & 1.00\\
 & 10402020002 &  $<  0.9$ & 1.64$_{-0.03}^{+0.07}$  & 221$_{-69}^{+265}$  & 0.21$_{-0.13}^{+0.14}$  &  $<  19.3$ & 1.05\\
 & 10402020004 &  $<  1.4$ & 1.64$_{-0.05}^{+0.08}$  &  $>  341$ &  $<  0.212$ &  $<  29.8$ & 0.94\\
 & 10402020006 &  $<  0.3$ & 1.67$_{-0.02}^{+0.02}$  & 300$_{-93}^{+246}$  &  $<  0.146$ &  $<  9$ & 0.94\\
 & 80301602002 &  $<  0.2$ & 1.58$_{-0.01}^{+0.02}$  & 271$_{-64}^{+113}$  &  $<  0.07$ &  $<  16.1$ & 0.95\\
 & total &  $<  0.2$ & 1.61$_{-0.00}^{+0.01}$  & 248$_{-20}^{+32}$  & 0.03$_{-0.02}^{+0.02}$  & 6.3$_{-3.3}^{+3.1}$  & 1.05\\
PicA$^\star$$^\gamma$ & 60101047002 &  $<  0.9$ & 1.72$_{-0.04}^{+0.04}$  & 202$_{-87}^{+527}$  &  $<  0.104$ & 49.0$_{-11.1}^{+11.1}$  & 0.98\\
CentaurusA$\odot$$^\gamma$ & 60001081002 & 8.5$_{-0.1}^{+0.1}$  & 1.74$_{-0.01}^{+0.01}$  & 335$_{-56}^{+85}$  &  $<  0.003$ & 42.3$_{-3.8}^{+3.8}$  & 1.00\\
 & 60101063002$^\bullet$ & 9.6$_{-0.5}^{+0.5}$  & 1.75$_{-0.03}^{+0.03}$  & 190$_{-55}^{+129}$  &  $<  0.017$ & 140.8$_{-12.1}^{+12.6}$  & 1.13\\
 & 60466005002 & 9.7$_{-0.5}^{+0.2}$  & 1.80$_{-0.02}^{+0.02}$  &  $>  334$ &  $<  0.024$ & 70.2$_{-9.5}^{+9.4}$  & 1.03\\
 & total & 8.8$_{-0.1}^{+0.1}$  & 1.75$_{-0.01}^{+0.01}$  & 343$_{-54}^{+78}$  &  $<  0.002$ & 52.4$_{-3.1}^{+3.1}$  & 1.11\\
3C 382 & 60001084002$^\bullet$ & 1.1$_{-0.5}^{+0.5}$  & 1.76$_{-0.04}^{+0.04}$  &  $>  297$ &  $<  0.125$ & 49.4$_{-13.5}^{+13.4}$  & 0.95\\
 & 60061286002 &  $<  1.5$ & 1.88$_{-0.08}^{+0.08}$  &  $>  184$ & 0.27$_{-0.2}^{+0.24}$  &  $<  41.9$ & 0.95\\
 & 60202015002 & 1.1$_{-0.9}^{+0.6}$  & 1.83$_{-0.08}^{+0.05}$  &  $>  225$ &  $<  0.376$ & 27.3$_{-23.0}^{+22.7}$  & 0.99\\
 & 60202015004 &  $<  0.7$ & 1.72$_{-0.03}^{+0.07}$  & 174$_{-59}^{+256}$  &  $<  0.291$ & 29.3$_{-20.3}^{+18.9}$  & 1.02\\
 & 60202015006$^\bullet$ &  $<  0.9$ & 1.78$_{-0.04}^{+0.02}$  &  $>  270$ &  $<  0.172$ & 45.8$_{-22.3}^{+21.5}$  & 0.96\\
 & 60202015008 &  $<  1.6$ & 1.74$_{-0.07}^{+0.08}$  &  $>  337$ &  $<  0.184$ & 43.0$_{-22.9}^{+21.7}$  & 0.97\\
 & 60202015010$^\bullet$ &  $<  1.0$ & 1.80$_{-0.05}^{+0.06}$  &  $>  358$ &  $<  0.246$ & 47.0$_{-21.6}^{+21.3}$  & 1.07\\
 & total & 0.7$_{-0.3}^{+0.3}$  & 1.78$_{-0.02}^{+0.02}$  &  $>  343$ & 0.07$_{-0.06}^{+0.06}$  & 39.6$_{-7.5}^{+7.5}$  & 1.01\\
IGR J21247+5058 & 60061305002$^\bullet$ & 2.9$_{-0.5}^{+0.5}$  & 1.56$_{-0.04}^{+0.04}$  & 97$_{-20}^{+33}$  &  $<  0.039$ & 36.2$_{-15.1}^{+15.0}$  & 1.01\\
 & 60301005002 & 2.6$_{-0.4}^{+0.4}$  & 1.62$_{-0.01}^{+0.03}$  & 100$_{-15}^{+22}$  &  $<  0.107$ & 31.7$_{-10.7}^{+10.7}$  & 1.07\\
 & total & 2.7$_{-0.3}^{+0.3}$  & 1.60$_{-0.02}^{+0.02}$  & 98$_{-12}^{+16}$  &  $<  0.043$ & 33.8$_{-8.6}^{+8.6}$  & 1.08\\
3C 390.3 & 60001082002 &  $<  1.3$ & 1.75$_{-0.07}^{+0.08}$  & 197$_{-82}^{+432}$  & 0.19$_{-0.17}^{+0.2}$  & 80.4$_{-28.6}^{+28.1}$  & 0.98\\
 & 60001082003 &  $<  1.2$ & 1.72$_{-0.06}^{+0.06}$  & 208$_{-73}^{+232}$  & 0.14$_{-0.12}^{+0.14}$  & 67.1$_{-20.2}^{+20.0}$  & 0.98\\
 & total & 0.5$_{-0.5}^{+0.5}$  & 1.73$_{-0.05}^{+0.05}$  & 203$_{-60}^{+145}$  & 0.16$_{-0.1}^{+0.11}$  & 71.5$_{-16.5}^{+16.3}$  & 0.98\\
3C 109$^\star$& 60301011002 &  $<  4.3$ & 1.69$_{-0.16}^{+0.20}$  &  $>  131$ &  $<  0.486$ & 41.4$_{-40.0}^{+35.7}$  & 1.05\\
 & 60301011004 &  $<  2.6$ & 1.63$_{-0.07}^{+0.16}$  & 87$_{-24}^{+86}$  & 0.32$_{-0.24}^{+0.32}$  & 45.6$_{-33.1}^{+28.7}$  & 0.95\\
 & total &  $<  2.4$ & 1.64$_{-0.06}^{+0.14}$  & 97$_{-24}^{+87}$  & 0.23$_{-0.17}^{+0.23}$  & 46.1$_{-26.5}^{+22.0}$  & 1.00\\
4C 74.26 & 60001080002 &  $<  2.4$ & 1.91$_{-0.12}^{+0.15}$  & 137$_{-57}^{+283}$  & 1.0$_{-0.39}^{+0.49}$  &  $<  126.5$ & 1.00\\
 & 60001080004 & 1.2$_{-0.9}^{+0.9}$  & 1.89$_{-0.08}^{+0.08}$  & 240$_{-97}^{+427}$  & 0.55$_{-0.18}^{+0.21}$  & 86.0$_{-42.4}^{+41.2}$  & 0.99\\
 & 60001080006$^\bullet$ &  $<  0.9$ & 1.79$_{-0.03}^{+0.07}$  & 121$_{-22}^{+48}$  & 0.66$_{-0.15}^{+0.18}$  & 145.1$_{-35.3}^{+28.3}$  & 0.99\\
 & 60001080008 &  $<  1.7$ & 1.85$_{-0.08}^{+0.10}$  &  $>  309$ & 0.5$_{-0.21}^{+0.25}$  & 92.1$_{-51.4}^{+48.4}$  & 0.94\\
 & total & 0.6$_{-0.5}^{+0.5}$  & 1.84$_{-0.04}^{+0.04}$  & 166$_{-33}^{+54}$  & 0.63$_{-0.11}^{+0.11}$  & 108.3$_{-22.6}^{+22.3}$  & 0.99\\
3C 120$^\gamma$ & 60001042002 &  $<  0.5$ & 1.86$_{-0.03}^{+0.05}$  &  $>  232$ & 0.32$_{-0.18}^{+0.2}$  & 66.5$_{-20.0}^{+20.0}$  & 0.93\\
 & 60001042003$^\bullet$ & 0.5$_{-0.3}^{+0.3}$  & 1.85$_{-0.03}^{+0.03}$  & 300$_{-85}^{+188}$  & 0.4$_{-0.08}^{+0.09}$  & 54.6$_{-8.6}^{+8.6}$  & 1.01\\
 & total & 0.4$_{-0.3}^{+0.3}$  & 1.85$_{-0.02}^{+0.02}$  & 313$_{-87}^{+189}$  & 0.39$_{-0.08}^{+0.08}$  & 56.8$_{-8.0}^{+8.0}$  & 0.99\\
3C 227$^\odot$& 60061329002 & 3.1$_{-1.9}^{+1.9}$  & 1.87$_{-0.18}^{+0.15}$  &  $>  87$ &  $<  0.83$ &  $<  19$ & 0.95\\
 & 60061329004 &  $<  3.5$ & 1.39$_{-0.14}^{+0.24}$  & 58$_{-23}^{+230}$  &  $<  0.11$ &  $<  90.4$ & 1.00\\
 & total & 1.8$_{-1.6}^{+1.6}$  & 1.65$_{-0.13}^{+0.12}$  &  $>  52$ &  $<  0.166$ &  $<  26$ & 1.05\\
3C 184.1$^\star$& 60160300002 & 7.7$_{-5.1}^{+4.9}$  & 1.34$_{-0.38}^{+0.33}$  & 38$_{-18}^{+109}$  &  $<  0.384$ &  $<  85.8$ & 1.19\\
3C 111$^\star$$^\gamma$& 60202061002 & 1.1$_{-0.8}^{+0.7}$  & 1.79$_{-0.07}^{+0.02}$  &  $>  228$ &  $<  0.109$ & 43.3$_{-22.7}^{+22.0}$  & 0.98\\
 & 60202061004$^\bullet$ & 0.7$_{-0.4}^{+0.6}$  & 1.69$_{-0.04}^{+0.06}$  & 165$_{-47}^{+202}$  &  $<  0.08$ & 48.5$_{-15.6}^{+15.5}$  & 1.07\\
 & total & 0.9$_{-0.4}^{+0.4}$  & 1.73$_{-0.04}^{+0.04}$  & 252$_{-94}^{+357}$  &  $<  0.055$ & 46.9$_{-12.2}^{+9.6}$  & 1.04\\
PKS 2356-61$^\odot$& 60061330002 & 7.9$_{-4.6}^{+4.8}$  & 1.42$_{-0.37}^{+0.37}$  & 59$_{-30}^{+531}$  &  $<  1.347$ & 149.5$_{-91.7}^{+85.9}$  & 0.97\\
3C 206$^\star$& 60160332002 &  $<  3.1$ & 1.81$_{-0.11}^{+0.24}$  &  $>  68$ & 0.51$_{-0.47}^{+0.75}$  &  $<  64.3$ & 0.81\\
3C 345$^\star$$^\gamma$& 60160647002 &  $<  4.2$ & 1.53$_{-0.12}^{+0.18}$  &  $>  290$ &  $<  0.694$ &  $<  153.0$ & 1.01\\
2MASX J03181899+6829322$^\odot$& 60061342002 & 9.0$_{-4.0}^{+3.8}$  & 1.77$_{-0.25}^{+0.21}$  &  $>  57$ &  $<  1.038$ & 78.7$_{-74.6}^{+69.3}$  & 1.18\\
4C +18.51 & 60160672002 &  $<  3.9$ & 1.55$_{-0.18}^{+0.10}$  &  $>  62$ &  $<  0.476$ &  $<  111.0$ & 0.94\\
S5 2116+81& 60061303002$^\bullet$ &  $<  2.8$ & 1.82$_{-0.14}^{+0.13}$  &  $>  174$ &  $<  0.562$ & 67.7$_{-44.3}^{+43.3}$  & 1.08\\
4C +21.55$^\star$& 60160740002 &  $<  5.6$ & 1.75$_{-0.20}^{+0.10}$  &  $>  76$ &  $<  0.274$ &  $<  86.7$ & 0.95\\
PKS 2331-240 & 60160832002 &  $<  2.5$ & 1.95$_{-0.11}^{+0.13}$  &  $>  260$ &  $<  0.616$ &  $<  85.5$ & 1.06\\
PKS 1916-300$^\star$& 60160707002 &  $<  2.8$ & 1.76$_{-0.14}^{+0.22}$  &  $>  69$ &  $<  1.125$ &  $<  95.9$ & 0.88\\
IGR 14488-4008$^\star$& 60463049002 & 2.6$_{-2.1}^{+2.2}$  & 2.06$_{-0.20}^{+0.20}$  &  $>  238$ & 2.22$_{-1.19}^{+2.02}$  & 78.8$_{-64.3}^{+62.5}$  & 1.15\\
3C 279$^\odot$$^\gamma$& 60002020002 &  $<  4.2$ & 1.68$_{-0.05}^{+0.08}$  &  $>  188$ &  $<  0.161$ & 33.1$_{-31.1}^{+31.1}$  & 1.01\\
 & 60002020004 & 2.2$_{-1.8}^{+1.7}$  & 1.76$_{-0.04}^{+0.03}$  &  $>  670$ &  $<  0.086$ &  $<  29.0$ & 0.89\\
 & total & 2.0$_{-1.7}^{+1.5}$  & 1.76$_{-0.05}^{+0.02}$  &  $>  620$ &  $<  0.08$ &  $<  30.8$ & 0.93\\
Leda 100168$^\star$& 60160631002 & 3.6$_{-3.6}^{+4.9}$  & 1.86$_{-0.28}^{+0.34}$  &  $>  76$ &  $<  1.375$ & 135.0$_{-90.3}^{+83.2}$  & 0.93\\
3C 309.1$^\star$$^\gamma$& 60376006002 &  $<  8.9$ & 1.71$_{-0.32}^{+0.10}$  &  $>  101$ &  $<  0.962$ &  $<  153.3$ & 0.86\\
3C 332$^\star$& 60160634002 &  $<  3.9$ & 1.83$_{-0.18}^{+0.19}$  &  $>  78$ & 0.5$_{-0.48}^{+0.7}$  &  $<  35$ & 0.90\\
PKS 0442-28$^\star$& 60160205002 &  $<  5.2$ & 1.78$_{-0.24}^{+0.10}$  &  $>  66$ &  $<  0.798$ &  $<  103.4$ & 1.09\\
NGC 1275$^\gamma$& 60061361002 &  $<  0.5$ & 1.82$_{-0.35}^{+0.17}$  &  $>  76$ &  $<  1.382$ & 41.9$_{-19.4}^{+19.3}$  & 0.94\\
 & 90202046002 &  $<  0.3$ & 1.79$_{-0.07}^{+0.06}$  &  $>  438$ &  $<  0.152$ & 70.3$_{-13.8}^{+13.9}$  & 0.95\\
 & 90202046004 &  $<  0.9$ & 1.99$_{-0.21}^{+0.12}$  &  $>  156$ &  $<  0.624$ & 62.6$_{-16.7}^{+16.6}$  & 1.01\\
 & total &  $<  0.3$ & 1.82$_{-0.05}^{+0.07}$  &  $>  298$ &  $<  0.227$ & 61.2$_{-9.8}^{+9.3}$  & 1.10\\
\hline
\end{tabular}
}
\begin{tablenotes}
\item{The Fe K$\alpha$ line EW is measured with a fixed rest frame central energy (6.4 keV) and a fixed line width (19 eV).}
\item{$\star$}{: Sources for which  \textit{NuSTAR} observations are reported for the first time.\\}
\item{$\odot$}{: Sources with \textit{NuSTAR} observations reported in literature, but $E_{cut}$ detection or lower limits reported for the first time in this work.\\}
\item{$\bullet$}{: Observations for which a broad Fe K$\alpha$ line is required. The updated fitting results are  given in Tab. \ref{tab:freesigmaResults}. \\}
\item{$^\gamma$}{: Sources in Fermi-LAT catalog.\\}

\end{tablenotes}
\end{table*}

\begin{table*}
\caption{Updated Spectral Fit Results for Observations with Broad Fe K$\alpha$ Lines}
\label{tab:freesigmaResults}
\centering
\begin{tabular}{l c c c c c c c c}
\hline
\hline
\tabincell{l}{Source\\ \ } & \tabincell{c}{ID\\ \ } &\tabincell{c}{nH\\ ${(10^{22} cm^{-2})}$ }&\tabincell{c}{${\Gamma}$\\ \ } & \tabincell{c}{$E_{cut}$\\ (keV)} &
\tabincell{c}{    R\\ \    } &\tabincell{c}{Fe K$\alpha$ $\sigma$\\ (keV)} & \tabincell{c}{Fe K$\alpha$ EW\\ (eV)} & \tabincell{c}{$\chi^2_\nu$\\ \ } \\
\hline
CentaurusA & 60101063002 & 9.1$_{-0.5}^{+0.5}$  & 1.71$_{-0.04}^{+0.04}$  & 144$_{-36}^{+70}$  &  $<  0.01$ & 0.19$_{-0.06}^{+0.06}$  & 187.7$_{-24.2}^{+25.6}$  & 1.11\\
3C382 & 60001084002 & 0.7$_{-0.5}^{+0.6}$  & 1.72$_{-0.04}^{+0.05}$  &  $>  320$ &  $<  0.08$ & 0.28$_{-0.14}^{+0.15}$  & 84.1$_{-26.8}^{+29.4}$  & 0.94\\
3C382 & 60202015006 &  $<  0.5$ & 1.76$_{-0.03}^{+0.02}$  &  $>  280$ &  $<  0.14$ & 0.32$_{-0.18}^{+0.25}$  & 85.2$_{-36.4}^{+42.1}$  & 0.95\\
3C382 & 60202015010 &  $<  0.3$ & 1.77$_{-0.03}^{+0.03}$  &  $>  380$ &$< 0.15$  & 0.35$_{-0.16}^{+0.2}$  & 96.6$_{-35.8}^{+40.6}$  & 1.06\\
IGRJ21247p5058 & 60301005002 & 2.2$_{-0.4}^{+0.3}$  & 1.58$_{-0.04}^{+0.03}$  & 90$_{-13}^{+18}$  &  $<  0.06$ & 0.4$_{-0.16}^{+0.18}$  & 74.2$_{-24.8}^{+28.5}$  & 1.06\\
4C74d26 & 60001080006 & $< 0.8$  & 1.79$_{-0.04}^{+0.07}$  & 124$_{-24}^{+50}$  & 0.62$_{-0.15}^{+0.18}$  & 0.4$_{-0.11}^{+0.11}$  & 122.7$_{-33.9}^{+31.0}$  & 0.99\\
3C120 & 60001042003 &  $<  0.3$ & 1.80$_{-0.01}^{+0.03}$  & 228$_{-46}^{+98}$  & 0.33$_{-0.07}^{+0.08}$  & 0.33$_{-0.08}^{+0.08}$  & 103.7$_{-15.9}^{+16.9}$  & 0.99\\
3C111 & 60202061004 &$< 0.6$  & 1.67$_{-0.05}^{+0.05}$  & 147$_{-45}^{+111}$  &  $<  0.06$ & 0.26$_{-0.1}^{+0.12}$  & 85.0$_{-26.2}^{+29.9}$  & 1.06\\
S52116p81 & 60061303002 &  $<  1.3$ & 1.73$_{-0.07}^{+0.06}$  &  $>  140$ & 0.06$< 0.35$  & 0.44$_{-0.2}^{+0.25}$  & 192.3$_{-81.6}^{+92.8}$  & 1.06\\
\hline
\end{tabular}

\end{table*}

We present our best-fit results  in Tab. \ref{tab:Results}, the spectra in Appendix \ref{app:B}, and the $\Gamma$ vs $E_{cut}$ contours in Appendix \ref{app:C}.
For spectra with no intrinsic obscuration, X-ray reflection, narrow Fe K$\alpha$ line or $E_{cut}$ detected (significance $<$ 90\%), we present their high or low limits (90\% confidence level) respectively. 
In addition, as we do  not see dramatic variations in obscuration, photon index $\Gamma$, line EW and high energy cut-off $E_{cut}$ between \textit{NuSTAR} exposures for sources with multiple observations, we also fit those multiple exposures simultaneously with $N_H$, $\Gamma$, $E_{cut}$, R and EW tied. The results (labelled as `total') are also given in  Tab. \ref{tab:Results}.

Since the Fe K$\alpha$ line could be relativistically broadened in some sources, we also perform spectral fitting allowing the line width to vary freely. 
We find in 9 observations, allowing the line width to vary can significantly improve the fit ($\Delta {\chi}^2 > 5$). The updated spectral fitting results are presented in Tab. \ref{tab:freesigmaResults}.
Such an approach could elevate the best-fit line equivalent width (EW) by a factor of $\sim$ 2--3, but barely affects the best-fit continuum properties.

We finally note that in \S\ref{subsec:YReq} we compare our Fe K$\alpha$ line EW with a radio quiet sample of \citet{sampleused}, who fit $NuSTAR$ spectra with the same model as ours except for that they fixed the Fe K$\alpha$ line width at 50 eV. We find that using 50 eV instead of 19 eV (adopted in this work) negligibly alters the fitting results as  both values are far smaller than the spectral resolution of $NuSTAR$, $\sim$ 0.4 keV at around 6.4 keV \citep{Harrison_2013}.
Also, for consistency, we only adopt our measurements with line width fixed for comparison (those reported in Table \ref{tab:Results}).

\section{Results and Discussion}\label{sec:discussion}

In this work we present the first systematic study of \textit{NuSTAR} hard X-ray spectra of radio loud AGNs. In this section we present the distribution of their measured spectral parameters, together with discussion and comparison with radio quiet ones. 
As we did not reveal strong variations in most spectral parameters concerned in individual sources, the comparison presented below is primarily between sources. 
For those with multiple exposures, we adopt the following parameters for further analyses, unless explicitly stated otherwise:  
the best-fit $\Gamma$, reflection factor $R$ and Fe K line EW from the `total' fit in table \ref{tab:Results} which generally yield better constraints to these parameters; 
the lower limit to $E_{cut}$ from the `total' fit for sources with $E_{cut}$ non-detected, and the best-fit $E_{cut}$ (and other spectral parameters if involved) from the corresponding individual exposure for sources with $E_{cut}$ detections. 
Following \citealt{10.1093/mnras/stz275} we adopt the lowest best-fit $E_{cut}$ for sources with $E_{cut}$ detected in multiple exposures.

\subsection{High Energy Cutoff $E_{cut}$}
\label{sec:cutoff}

Through performing uniform spectral fitting to \textit{NuSTAR} spectra, we detect high energy cut-off $E_{cut}$ (in at least one exposure for sources with multiple exposures, with a confidence level $>$ 90\%, ) in 13 out of 28 radio galaxies, and present lower limits for the rest 15 sources. 
The `total' fit in Tab. \ref{tab:Results} does not yield new detections of $E_{cut}$.
Note for a couple of exposures, the derived error bars of the best-fit $E_{cut}$ are rather large. For instance Pic A and PKS 2356-61 (see Tab. \ref{tab:Results}) show $E_{cut}$ reaching $\sim$ 600 keV within 90\% confidence ranges, which is out of the NuSTAR energy range (because the cutoff is not a very sharp feature). For such exposures the detections of $E_{cut}$ are marginal (with confidence level slightly above 90\%, see the contour plots in Appendix \ref{app:C}). Note they also have smaller number of  \textit{NuSTAR}  photons comparing with the majority sources with $E_{cut}$ detections (see \S\ref{sec:4.2}). We keep the marginal $E_{cut}$ detections for these two sources for statistical consistency. Excluding them would not alter the key results presented in this work.

\subsubsection{Comparison of $E_{cut}$ measurements with literature studies}
\label{sec:compare}

\par
\begin{figure}
\includegraphics[width=3.5in]{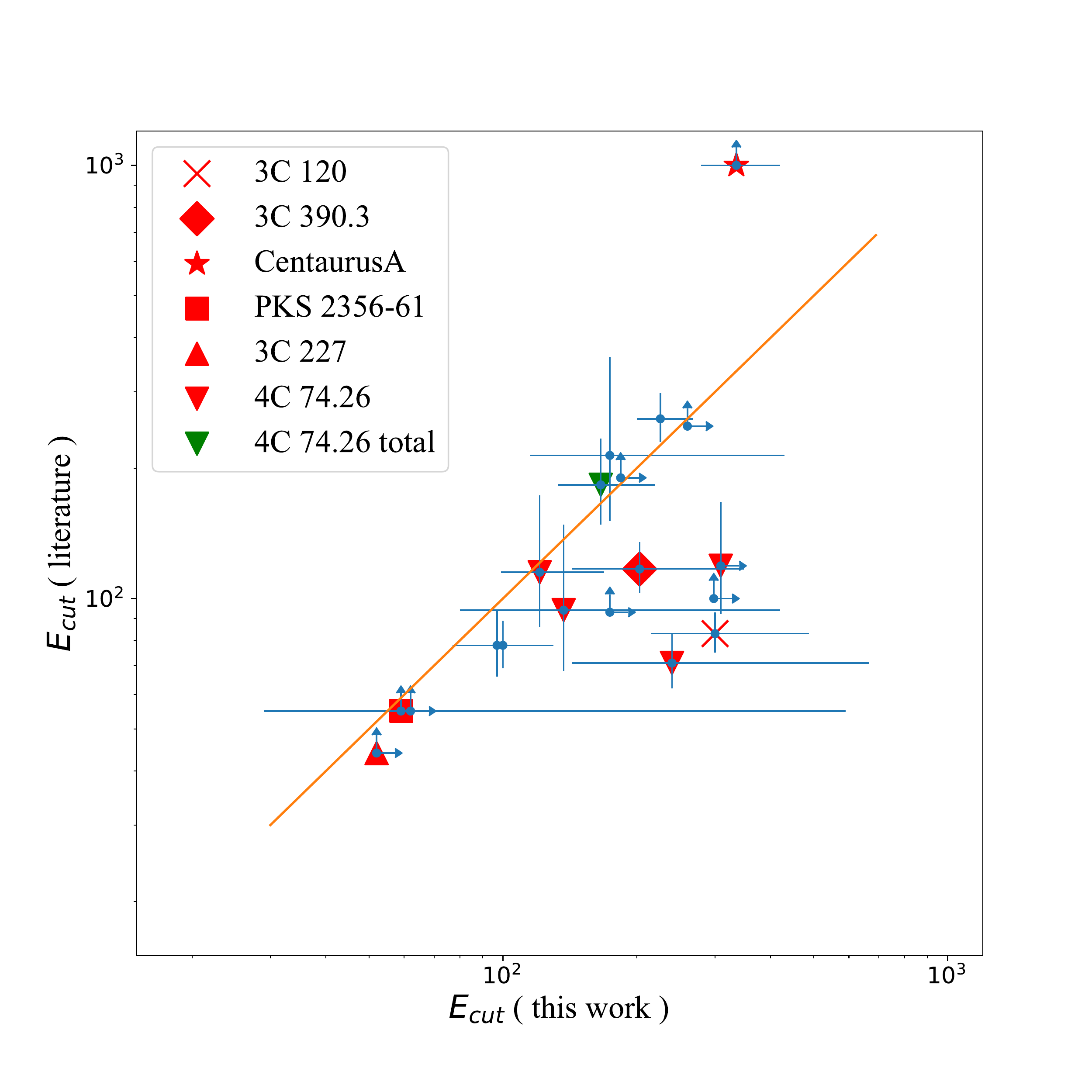}
\caption{A comparison of our $E_{cut}$ measurement with those reported in literature (also see Tab. \ref{tab:compare}). 
 The 1:1 line is shown for reference. The sources briefly noted in \S\ref{sec:compare} are marked.
 }

\label{fig:cutoffcompare}
\end{figure}

  21 out of the 55 \textit{NuSTAR} exposures presented in this work have been reported with $E_{cut}$ measurements in literature. In Fig. \ref{fig:cutoffcompare} and Tab. \ref{tab:compare} 
we present comparison between our $E_{cut}$ with those published for the 21 exposures of 13 sources. Among them, we yield $E_{cut}$ detections for 3C 227, CentaurusA and PKS 2356-61 for which literature studies of 
\textit{NuSTAR}  spectra only provided lower limits. Below we provide brief notes on the three sources. 

\begin{itemize}
 \item{
For 3C 227, \cite{Kamraj_2018} co-added two \textit{NuSTAR} spectra and reported a lower limit to $E_{cut}$ ($\geq$ 44 keV), while we presented an $E_{cut}$ measurement (58$^{+230}_{-23}$ keV) for one exposure and a lower limit ($>$ 87 keV) for another. }
\item{
For Centaurus A, \cite{F_rst_2016_CenA} claimed non-detection of $E_{cut}$ ($>$ 1 MeV) using one \textit{NuSTAR} exposure (ID: 60001081002) together with a partially simultaneous XMM exposure, though a coronal temperature $kT_e = 216^{+19}_{-22}$ keV was derived using a Comptonization model.
However, the \textit{NuSTAR} exposure only slightly overlap with XMM exposure (MJD range of  56510.54-56511.67 versus 56511.53-56511.66) thus the fitting results could have been biased by potential variation of the spectral slope\footnote{For the same \textit{NuSTAR} exposure, \cite{F_rst_2016_CenA} yielded a powerlaw spectral slope of $\Gamma$=1.815$\pm{0.005}$ through fitting together with the XMM spectra, considerably steeper than $\Gamma$ = 1.74$\pm0.01$ we obtained using \textit{NuSTAR} spectra alone}, and the inter-instrument calibration may also have played a role \citep[e.g.][]{F_rst_2016_CenA,gamma_xmm_1,gamma_xmm_2}. 
 Meanwhile we derived $E_{cut}$ measurements from three \textit{NuSTAR} exposures of Centaurus A (335$^{+85}_{-56}$ keV, 190$^{+129}_{-59}$ keV, $>$ 334 keV). 
 }
 \item{
For PKS 2356-61, our marginal detection of $E_{cut}$ (59$^{+531}_{-30}$ keV) is statistically consistent with the lower limit ($>55$ keV) given by \cite{Ursini_2018}.
}
\end{itemize}

While our measurements are generally consistent with literature studies (Fig. \ref{fig:cutoffcompare}), we report considerably higher $E_{cut}$ and larger error bars for 3C 120, 3C 390.3 and individual exposures of 4C 74.26. 
\begin{itemize}
\item{
For 3C 120, this is likely because \citet{2018JApA...39...15Rb} did not re-bin the spectra of 3C 120 (presumably, based on the degree of freedom reported) but adopted $\chi ^2$ statistics. 
Utilizing C-statistic on un-rebinned spectra would yield results well consistent with what we present in this work.
}
\item{
For 3C 390.3, this could be due to the fact that the Fe K$\alpha$ line and continuum reflection component were fitted jointly in literature \citep{2015ApJ...814...24L_3C309}, while we adopt a conservative approach with two components decoupled. 
}
\item{

For 4C 74.26 our measurements (or lower limit) from the 3 individual exposures are higher than those reported by \citet{10.1093/mnras/stz156} which fitted simultaneous $Swift$/XRT and NuSTAR spectra,
however our `total' fit result is well consistent with that reported by \citet{2017ApJ...841...80L_4C74}.
}
\end{itemize}

Meanwhile, in our sample, \textit{NuSTAR} observation(s) are reported for the first time\footnote{Here the ``first time'' claim is obtained through crossmatching with \textit{NuSTAR} bibliographies: \url{https://heasarc.gsfc.nasa.gov/docs/heasarc/biblio/pubs/nustar.html}} for 13 sources in our sample, yielding 4 detections of $E_{cut}$ and 9 lower limits, which
are marked with asterisks in Tab. \ref{tab:Results}.
Additionally, for 3C 279 and 2MASX J03181899+6829322, for which the literature studies, though fitted the spectra with cutoff powerlaw, did not provide constraints to $E_{cut}$.
\citep{Hayashida_2015_3C279,sampleused}.
Over all, this work presents for the first time $E_{cut}$ detections in 7 sources and lower limits in 11 sources.

\begin{table*}[h]
\caption{ $E_{cut}$ comparison with previous work}
\label{tab:compare}
\centering
\setlength{\tabcolsep}{3mm}{
\begin{tabular}{c c c c c c}
\hline
\hline
\tabincell{c}{Source\\ \ } & \tabincell{c}{ID\\ \ } &\tabincell{c}{$E_{cut}$ (this work)\\ (keV) }&\tabincell{c}{$E_{cut}$ (Literature)\\ (keV) } & \tabincell{c}{Ref.\\ \ } & \tabincell{c}{Notes on\\previous work \ }\\
\hline
3C 273 & 10002020001 & 226$_{-26}^{+42}$   & 202$_{-34}^{+51}$  & \citet{Madsen_2015_3C273} &  \\
CentaurusA & 60001081002 & 335$_{-56}^{+85}$ & $> 1000$ & \citet{F_rst_2016_CenA} & with \textit{XMM}\\
3C 382 & 60061286002 &  $>  184$ & $> 190$ & \citet{Ballantyne_2014_3C382} & \\
 & 60001084002 & $>  297$ & $214^{+147}_{-63}$ & \citet{Ballantyne_2014_3C382} & \\
IGR J21247+5058 & 60061305002 & 97$_{-20}^{+33}$ & $78^{+16}_{-12}$ & \citet{2018MNRAS.481.4419B} &\\
& 60301005002 &100$_{-15}^{+22}$  & $80^{+11}_{-9}$ & \citet{2018MNRAS.481.4419B} & \\
3C 390.3$^*$ & 60001082002 & 197$_{-82}^{+432}$ \\
 & 60001082003 & 208$_{-73}^{+232}$ \\
 & total & 203$^{+145}_{-60}$  &$117^{+18}_{-14}$ & \citet{2015ApJ...814...24L_3C309} & addspec, with Suzaku\\
4C 74.26 & 60001080002 & 137$^{+283}_{-57}$ & $94^{+54}_{-26}$ & \citet{10.1093/mnras/stz156} & with XRT\\
 & 60001080004 & 240$^{+427}_{-97}$ & $71^{+12}_{-9}$ & \citet{10.1093/mnras/stz156} & with XRT \\
  & 60001080006 & 121$^{+48}_{-22}$ & $115^{+58}_{-29}$ & \citet{10.1093/mnras/stz156} & with XRT \\
   & 60001080008 & $>309$ & $119^{+48}_{-27}$ & \citet{10.1093/mnras/stz156} & with XRT \\
     & total & $166^{+54}_{-33}$ & $183^{+51}_{-35}$      & \citet{2017ApJ...841...80L_4C74} & with XRT \\
3C 120 & 60001042003 & 300$_{-85}^{+188}$ & $83^{+10}_{-8}$ & \citet{2018JApA...39...15Rb} & un-binned with $\chi^2$ statistics\\
3C 227$^*$ & 60061329002 & $> 87$ \\
 & 60061329004 & 58$_{-23}^{+230}$ \\
  & total & $>  52$ & $> 44$ & \citet{Kamraj_2018} & addspec \\
PKS 2356-61 & 60061330002 & 59$_{-30}^{+531}$ & $> 55 $ & \citet{Ursini_2018} & with BAT \\
4C +18.51   & 60160672002 & $ >62 $ & $> 55 $ & \citet{Kamraj_2018} & \\
S5 2116+81 & 60061303002 & $ >174 $ & $> 93  $ & \citet{10.1093/mnras/stz156} & with XRT   \\
PKS 2331-240 & 60160832002 & $ >260 $ & $> 250  $ & \citet{Ursini_2018} & with BAT and  \textit{XMM} \\
NGC 1275$^*$ & 90202046002 & $> 438$ \\
 & 90202046004 & $> 156$ \\
 & total & $ >298^*$ & $> 100  $ & \citet{Rani_2018} & with BAT \\

\hline
\end{tabular}
}
\begin{tablenotes}
\item {\textit *}{
 For 3C 390.3, 3C 227 and NGC 1275, the corresponding literature studies did not provide $E_{cut}$ measurements for individual NuSTAR exposures.
 For NGC 1275, our `total' fitting result is derived using three NuSTAR exposures (see Tab. \ref{tab:Results}), while \citet{Rani_2018} used the two listed in this table.
 } \\

\end{tablenotes}
\end{table*}

\begin{figure}
\includegraphics[width=3.5in]{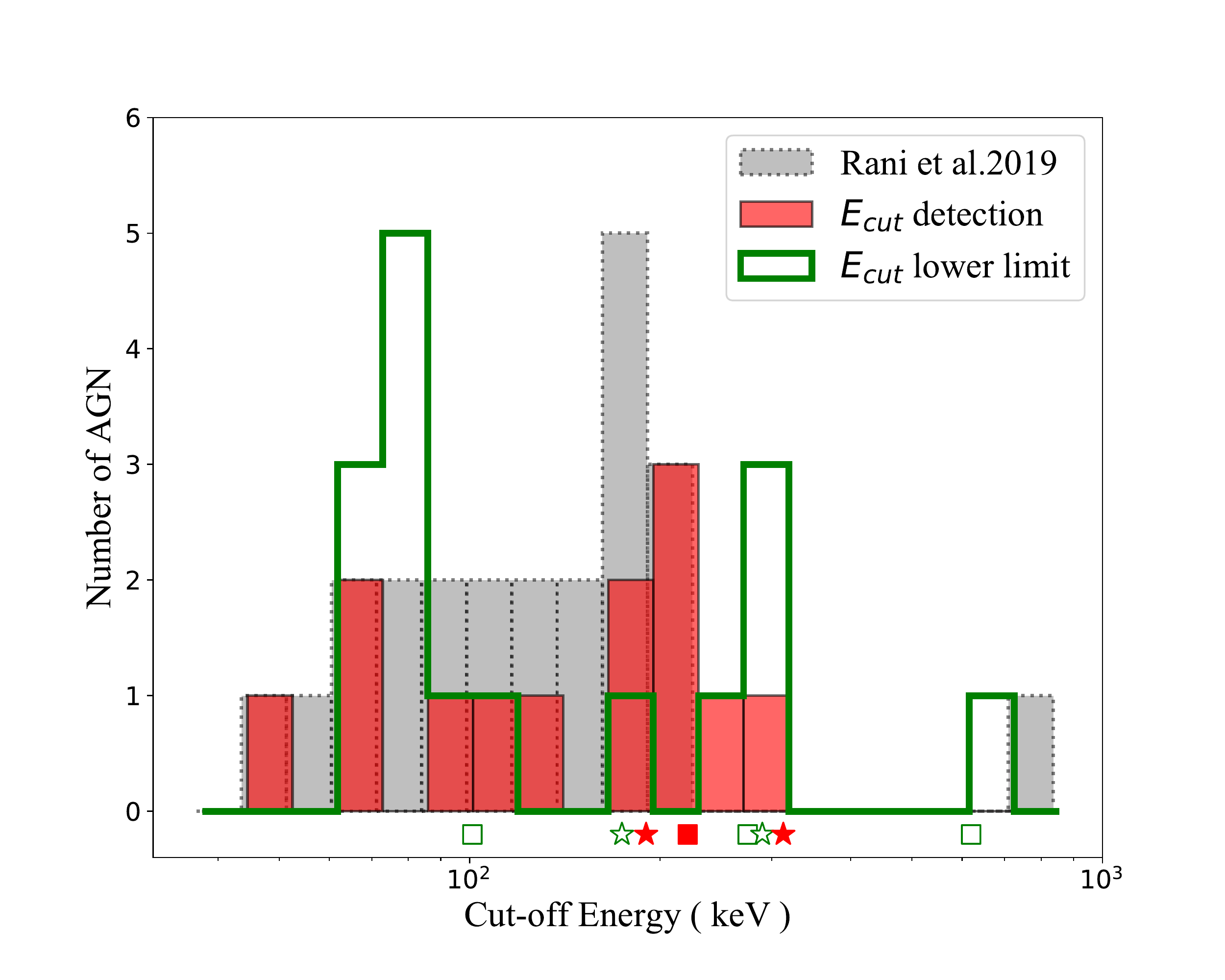}\
\caption{The distribution of the best-fit cut-off energy $E_{cut}$ (red) and lower limits (green line) we measured for our sample. We find no significant difference of $E_{cut}$ distribution between our radio galaxy sample and the radio quiet sample (grey) of \citet{10.1093/mnras/stz275}. 
 The four FR I galaxies and four core-dominated sources are marked with stars and boxes (open for lower limits, and solid for E$_{cut}$ detections) respectively.
}
\label{fig:cutoff}
\end{figure}

\subsubsection{Comparison with radio quiet AGNs}

We plot in Fig. \ref{fig:cutoff} the distribution of our $E_{cut}$ measurements. Following \citet{10.1093/mnras/stz275}, we plot the lowest $E_{cut}$ value for sources with more than one $E_{cut}$ measurements.

It is useful to compare the $E_{cut}$ of our radio galaxies with those of radio quiet ones.
Up to now, there are a number of sample studies of \textit{NuSTAR} spectra of (mostly radio quiet) AGNs. 
For example, \citet{Kamraj_2018} studies \textit{NuSTAR} spectra of 46 Seyfert 1 AGNs with \textit{NuSTAR} and reports $E_{cut}$ detection in 2 sources and lower limits in 44 sources.
\citet{refId01}, \citet{10.1093/mnras/stz156} and \citet{10.1093/mnras/stz275} respectively measure and detect $E_{cut}$ in \textit{NuSTAR} samples, with all three added up to about 25 sources with $E_{cut}$ detections. Besides, there are many works focus on the corona property of individual sources \citep[e.g.][]{ 2018MNRAS.481.4419B, NGC7469,PG1247,Ark120}.

\citet{10.1093/mnras/stz275} collected from literature a list of \textit{NuSTAR} sources with $E_{cut}$ measurements \citep[e.g.][]{ 2018JApA...39...15Rb, 2018JApA...39...15Ra, refId01, 10.1093/mnras/stx2457, 2018MNRAS.481.4419B, NGC7469,PG1247,Ark120}. Their sample consists of 28 sources, of which 5 overlaps with our sample. We use the rest 23 sources (all are radio quiet) for comparison with our work (Fig. \ref{fig:cutoff}).
Our $E_{cut}$ detections result in an arithmetically averaged value and scatter of $\langle E \rangle = 147 \pm 72$ keV, showing no difference from that of  \citet[][$\langle E \rangle = 143 \pm 131$ keV]{10.1093/mnras/stz275}. Kolmogorov-Smirnov test shows no significant difference either between two samples. 

\subsection{$E_{cut}$ detections versus non-detections}\label{sec:4.2}

\begin{figure}
\includegraphics[width=3.5in]{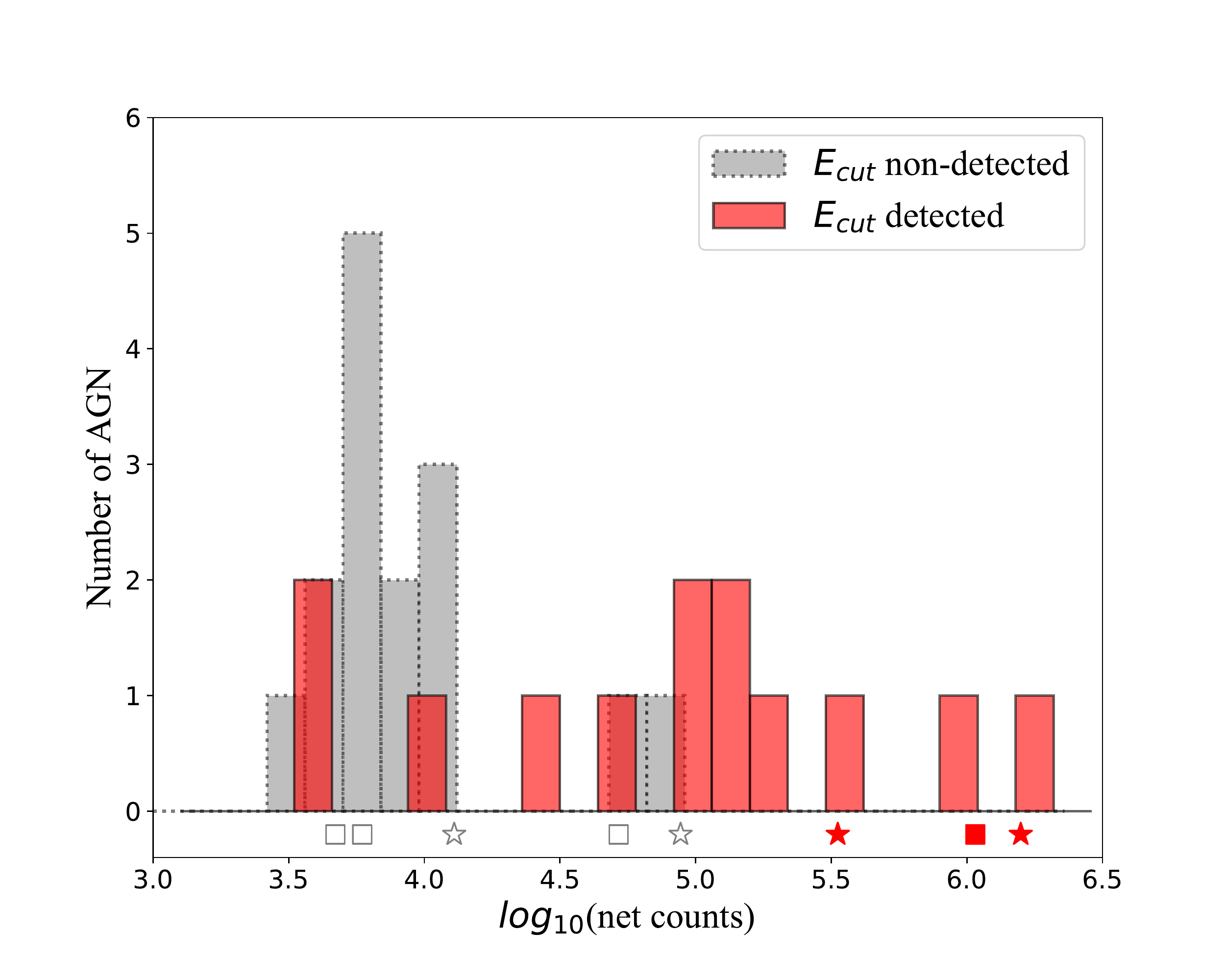}\
\includegraphics[width=3.5in]{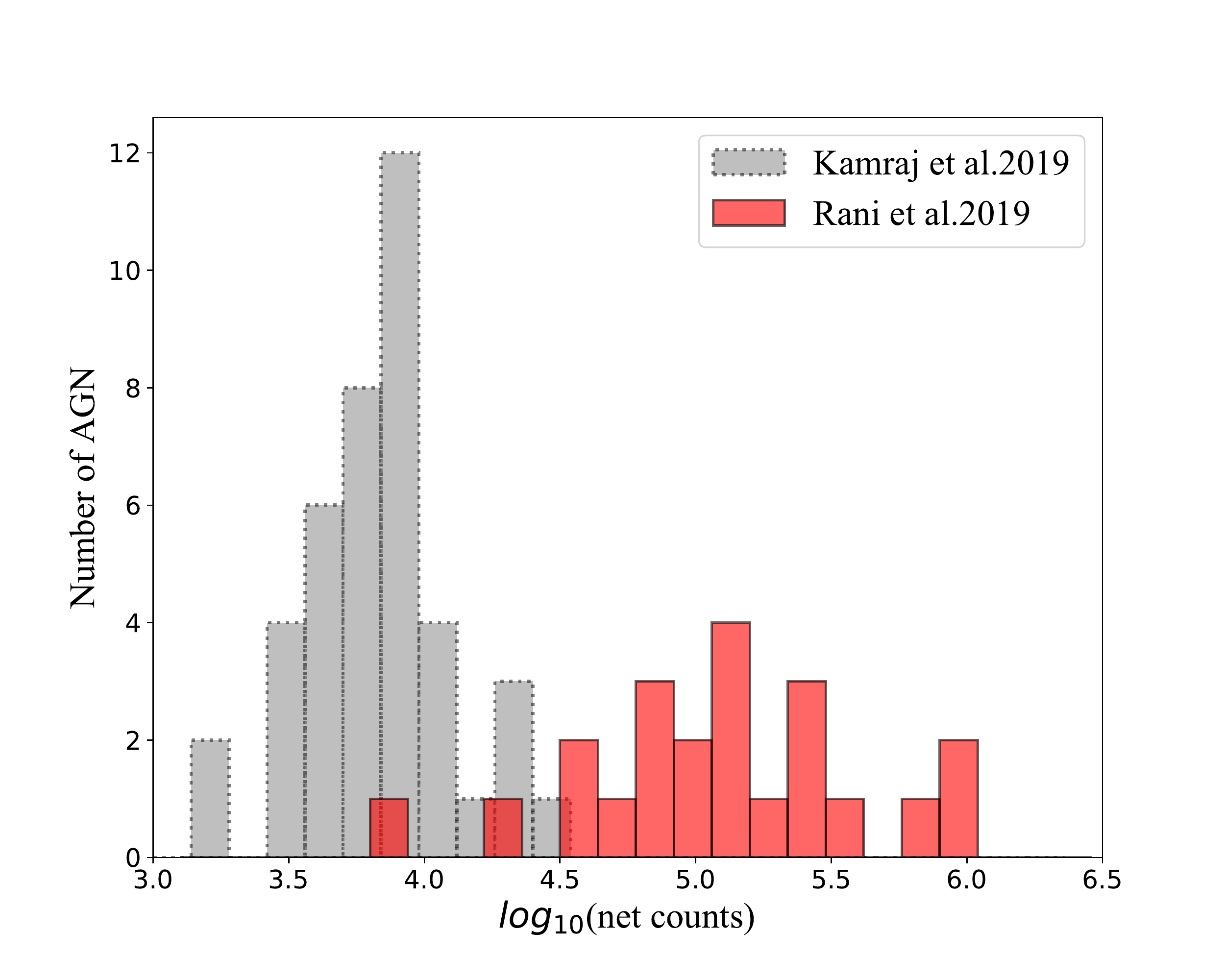}
\caption{The distribution of (FPMA$+$FPMB) net counts for $E_{cut}$ detections and non-detections.
Upper panel: our radio galaxy sample; The 4 FR I and 4 core-dominated galaxies are also marked, with stars and boxes respectively. Lower panel: the $E_{cut}$ non-detected sample in \citet{Kamraj_2018} and the $E_{cut}$-detected sample in \citet{10.1093/mnras/stz275}. 
}
\label{fig:count}
\end{figure}

\begin{figure*}[]
\centering
\subfloat{\includegraphics[width=0.33\textwidth]{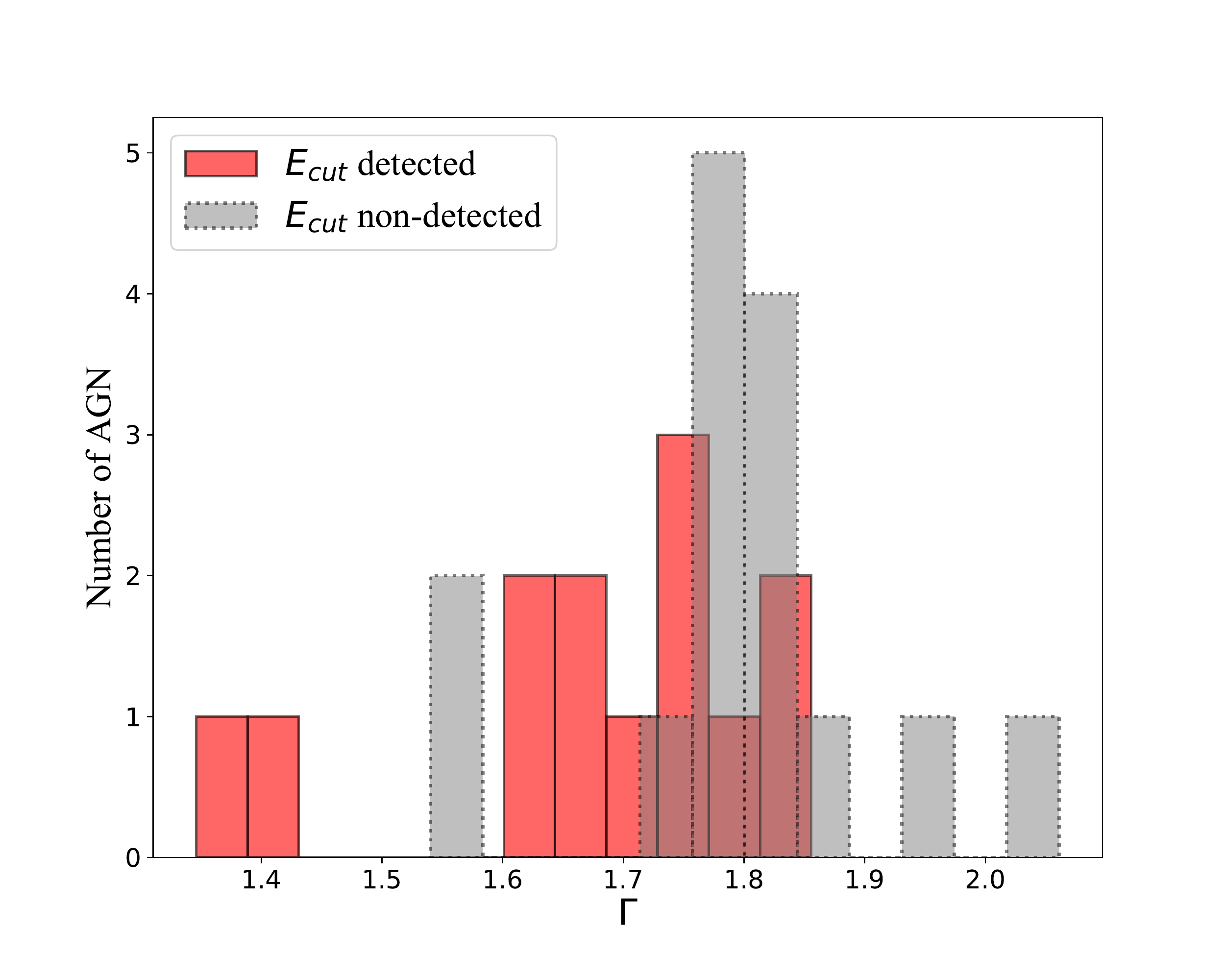}}
\subfloat{\includegraphics[width=0.33\textwidth]{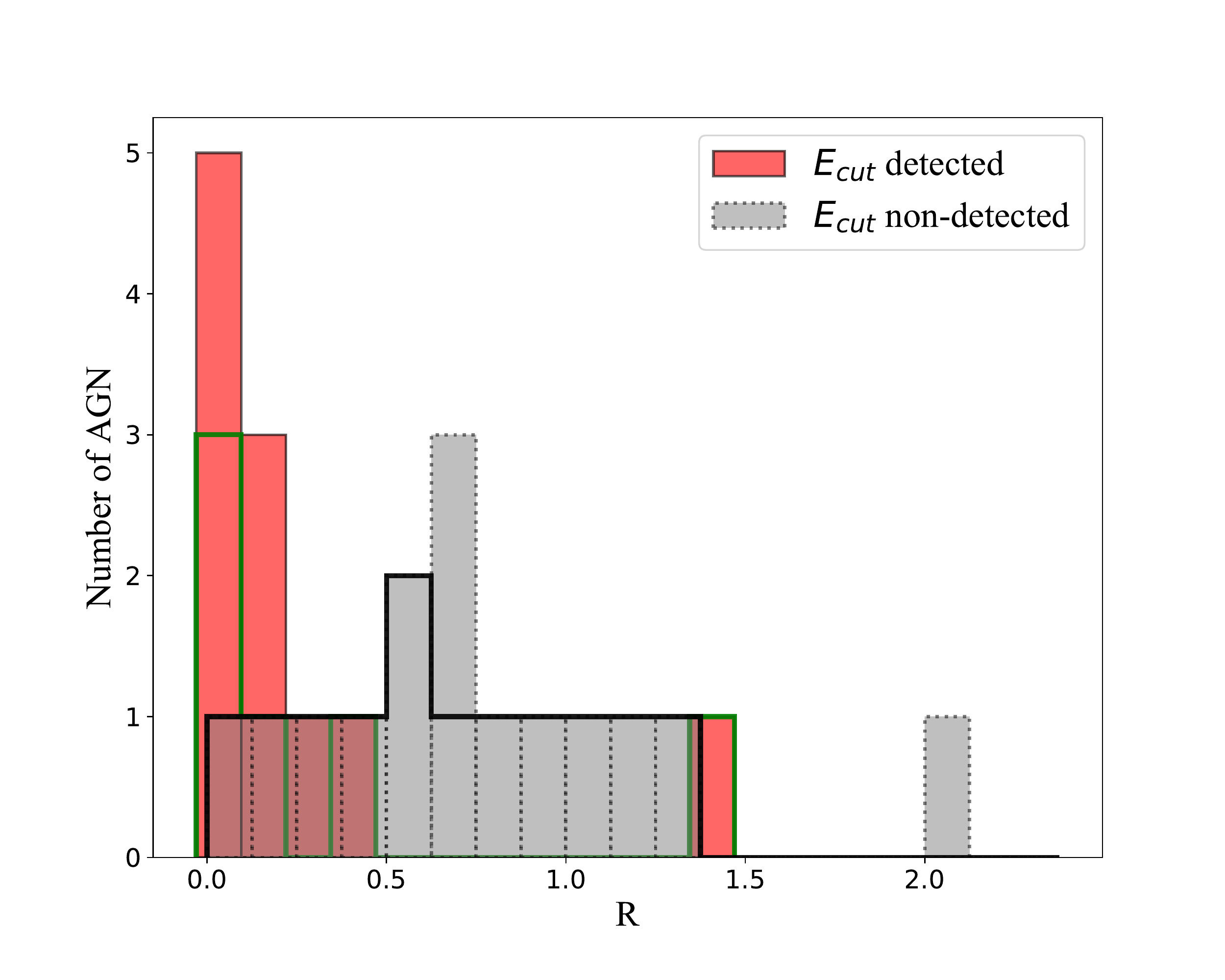}}
\subfloat{\includegraphics[width=0.33\textwidth]{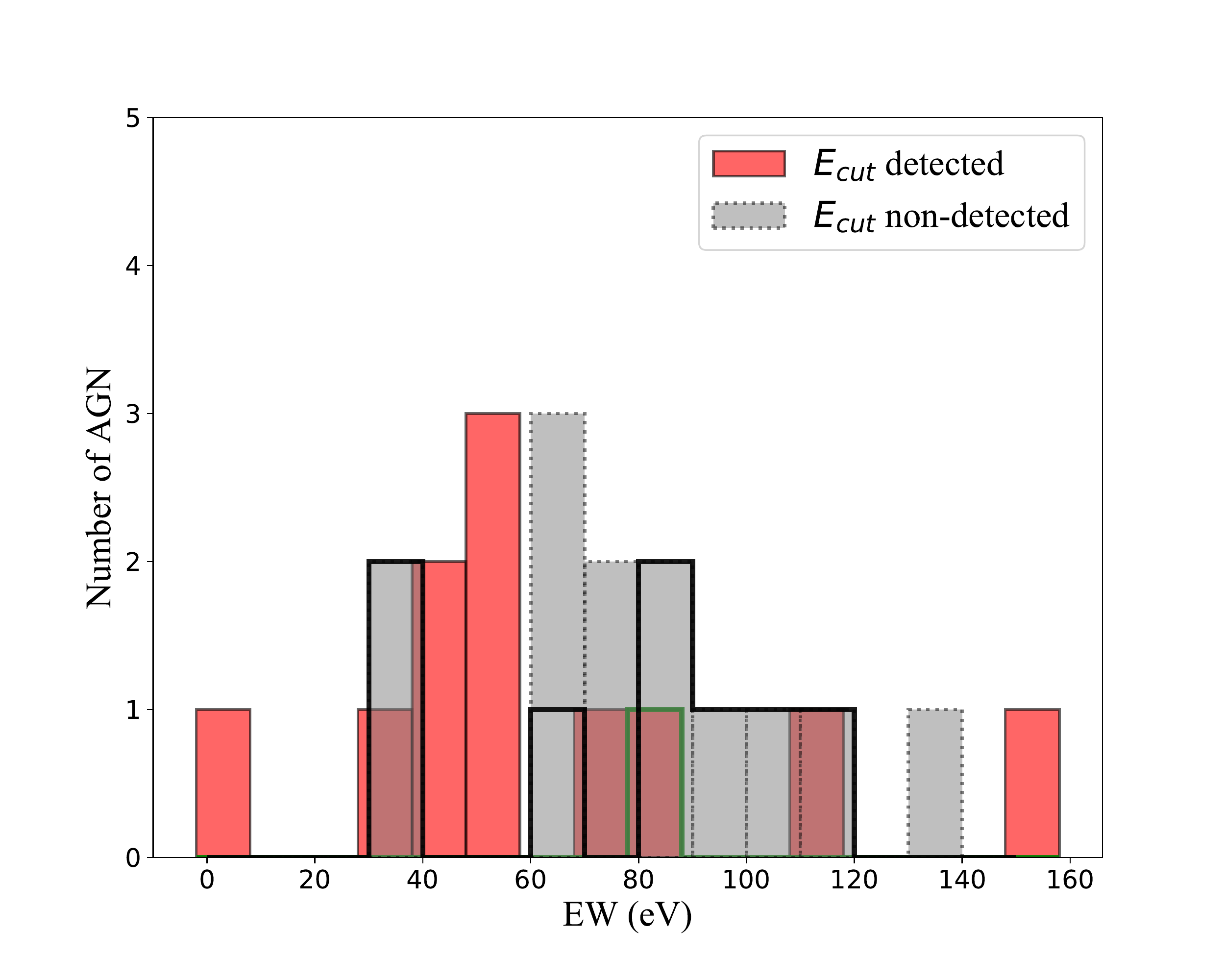}}\\
\caption{\label{fig:parawithE}The distribution of spectral parameters of our sources with and without $E_{cut}$ detections. 
In the middle and right panel, there are a number of  sources with upper limits to R or EW. These sources are marked with green (for $E_{cut}$ detections) and black lines (for $E_{cut}$ non-detections) respectively.
}
\end{figure*}

It is valuable to check whether those $E_{cut}$ non-detections are due to intrinsically larger $E_{cut}$ (or no cutoff at all) or insufficient S/N in the spectra. 
In Fig. \ref{fig:cutoff} we over-plot the distribution of the lower limits to our $E_{cut}$ non-detections,  which are indistinguishable from the distribution of the 
$E_{cut}$ detections. This indicates that such non-detections do not necessarily correspond to intrinsically higher $E_{cut}$.

We show the distribution of \textit{NuSTAR} net counts (FPMA $+$ FPMB) in Fig. \ref{fig:count}. For sources with multiple observations, we use the exposure with maximum net counts. We clearly see less net counts in sources without $E_{cut}$ detections, and at net counts $>$ 10$^{5.0}$, the $E_{cut}$ detection fraction is 100\%, showing the non-detections are dominantly due to their lower S/N. KS test shows that the distributions of the two subsamples (with and w/o $E_{cut}$ detections) are significantly different with a confidence level of 99.8\%. 
We obtain a mean net count and scatter of $10^ {4.9 \pm 0.8}$ for the sources with $E_{cut}$ detections, and $10^ {4.0 \pm 0.4}$ for the rest. Excluding the two brightest sources (3C273 and  CentaurusA) does not alter the results. We conclude that the $E_{cut}$ non-detections in our sample are primarily due to the smaller S/N of the corresponding spectra, demonstrating that the hard X-ray spectra of non-blazar radio galaxies are ubiquitously dominated by the thermal corona. 

\par We also plot in Fig. \ref{fig:count} the net counts distribution of the large sample in \citet[][mostly $E_{cut}$ non-detected]{Kamraj_2018} and the $E_{cut}$-detected sample in \citet{10.1093/mnras/stz275}. 
To estimate their \textit{NuSTAR} net counts, we download their \textit{NuSTAR} data from \textit{HEASARC} and perform the same data reduction process as 
in \S \ref{S:Reduction}\footnote{For simplicity, here we do not use \textit{NUSKYBGD} to calculate the background.  Instead we use the common approach, i.e., an annular region around the source to extract the background. This may be inaccurate for spectral analysis but sufficient for the purpose here.}. Finally, after excluding radio loud samples, we derive the total net counts (FPMA $+$ FPMB) of the 41 $E_{cut}$ non-detections in  \citet{Kamraj_2018}  and 22 detections in \citet{10.1093/mnras/stz275}. PG 1247+268 is excluded because its $E_{cut}$ detection was obtained through joint fitting with \textit{XMM-Newton} data \citep{PG1247}. Again for those with multiple observations, the maximum net counts value is adopted. 
Similarly, we conclude that those $E_{cut}$  non-detections in radio quiet AGNs are also to a great extent due to lower S/N in the spectra.

We note in our sample there are 4 $E_{cut}$ detections with net counts $<$ 10$^{4.5}$, while all but one $E_{cut}$ detections from \citet{10.1093/mnras/stz275} have net counts $>$ 10$^{4.5}$.
This is likely because $E_{cut}$ detections are more sensitive to net counts at higher energy instead of 3 -- 78 keV, thus for radio galaxies with flatter X-rays spectra,
relatively less 3 -- 78 keV net counts are required to enable $E_{cut}$ detections. In deed, all the 4 $E_{cut}$ detections with net counts $<$ 10$^{4.5}$
have $\Gamma$ $<$ 1.63. 

In Fig. \ref{fig:parawithE} we plot the distribution of $\Gamma$, reflection factor $R$ and Fe K$\alpha$ EW for our sources with and without $E_{cut}$ detections.
Note only upper limits to $R$ or $EW$ could be derived for the majority of $E_{cut}$ non-detections because of their small photon counts (see the upper panel of Fig. \ref{fig:count}).
Log--rank test within ASURV \citep{1985ApJ...293..192F}, which handles the censored data points, shows the two subsamples are indistinguishable in $R$ and Fe K$\alpha$ EW (with p-value = 0.45 and 0.80 respectively). 
 But sources with $E_{cut}$ detections tend to have flatter
X-ray spectra (with a statistical confidence level of 98\%).
This could also be attributed to the fact that $E_{cut}$ detection is easier (harder) in flatter (steeper) spectra. 
 
\subsection{$\Gamma$, R and Fe K$\alpha$ EW} 
\label{subsec:YReq}

\citet[][hereafter PW19]{sampleused} studied the X-ray reflection in both absorbed and unabsorbed AGNs, consisting of 87 sources of different types (e.g., narrow-line Seyfert 1, Seyfert 1, Seyfert 1.5, Seyfert 2, Compton-thick, etc). 
The sample of PW19 contains 46 unabsorbed sources ($N_H <10^{23}  cm^{-2}$, the same definition as in this paper), among which the only radio-loud one is 2MASX J03181899+6829322 (which is also in our sample). In this section we choose their 45 unabsorbed and radio quiet sources for comparison (hereafter PW19 sample).  

\begin{figure*}
\centering
\subfloat{\includegraphics[width=0.33\textwidth]{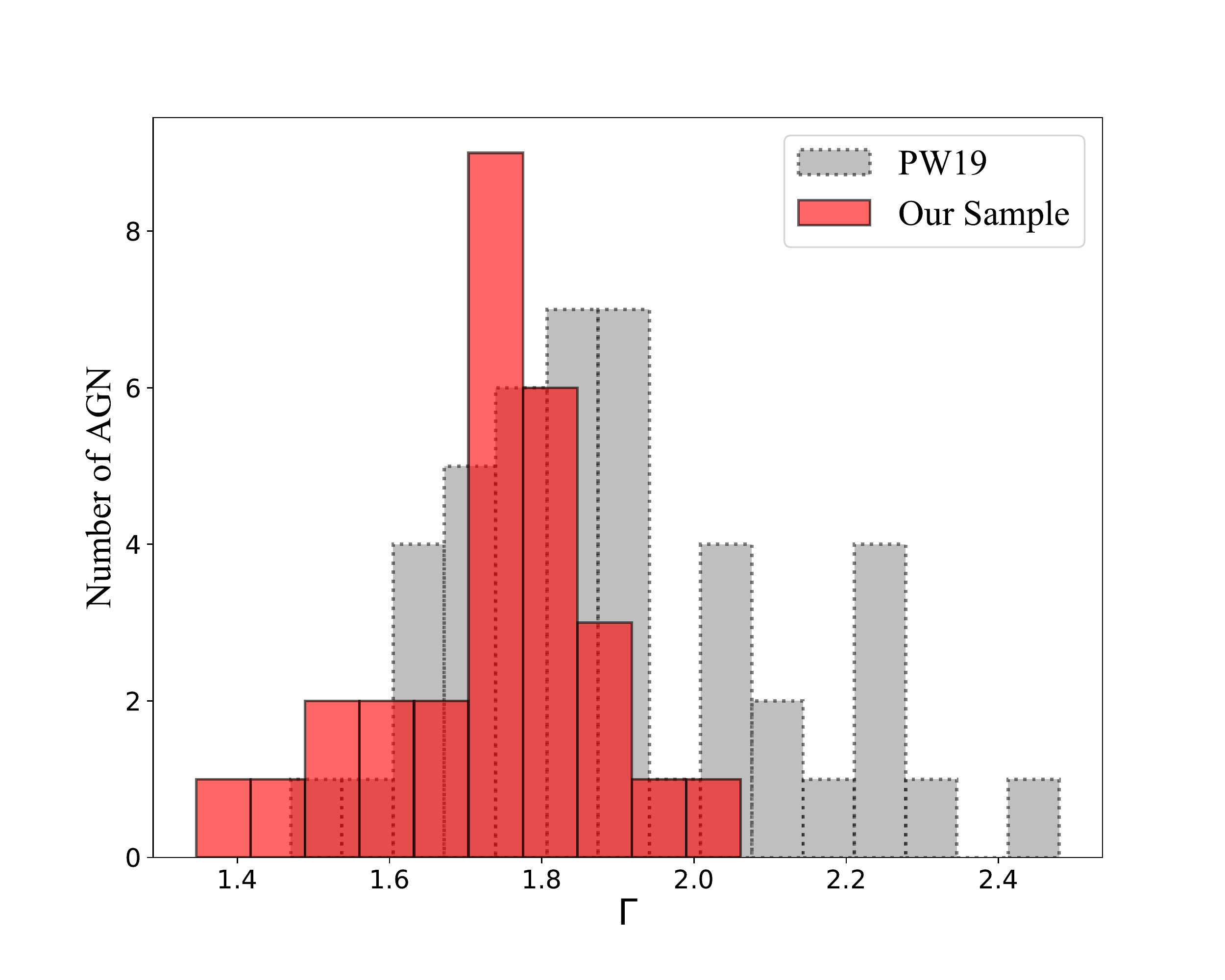}}
\subfloat{\includegraphics[width=0.33\textwidth]{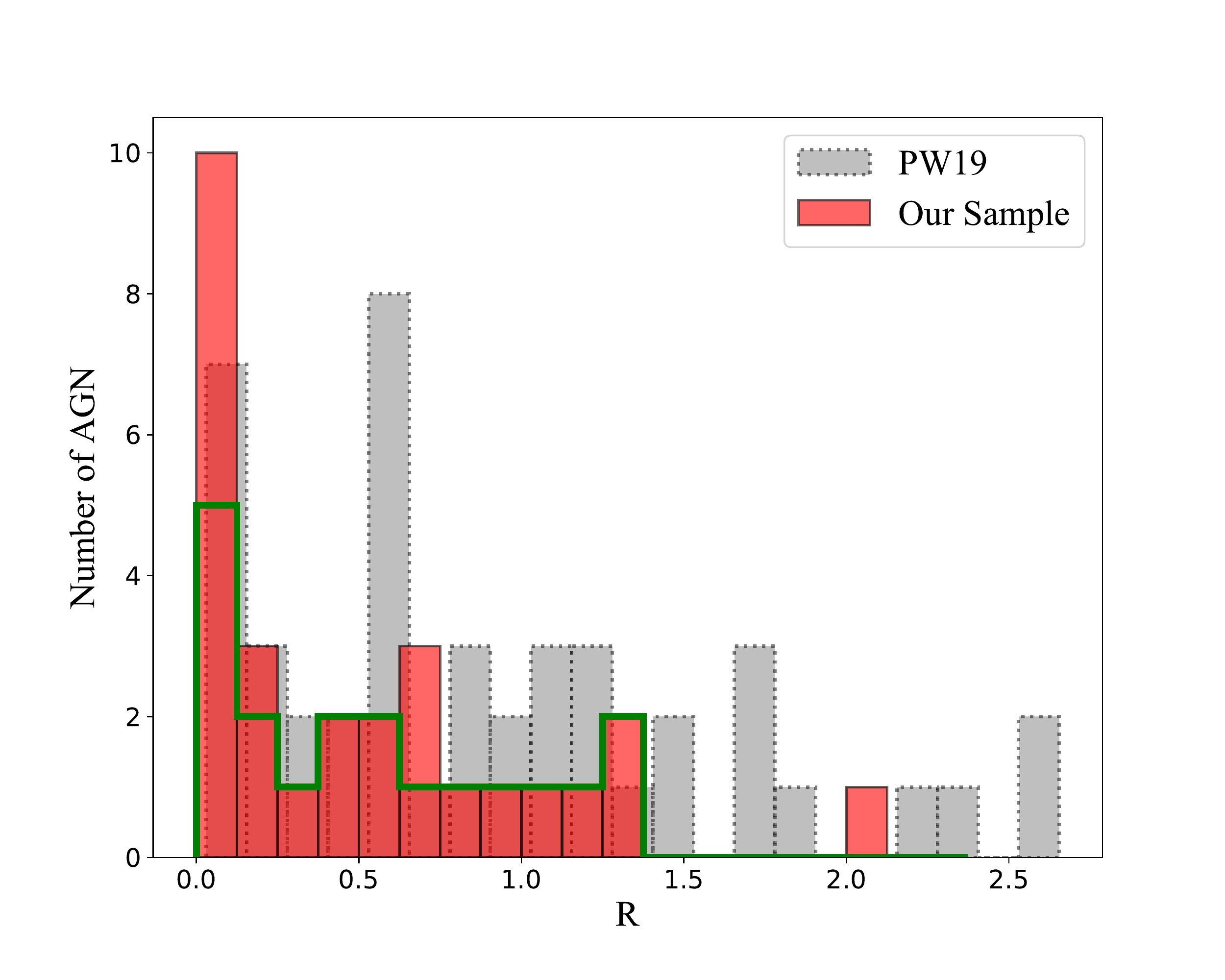}}
\subfloat{\includegraphics[width=0.33\textwidth]{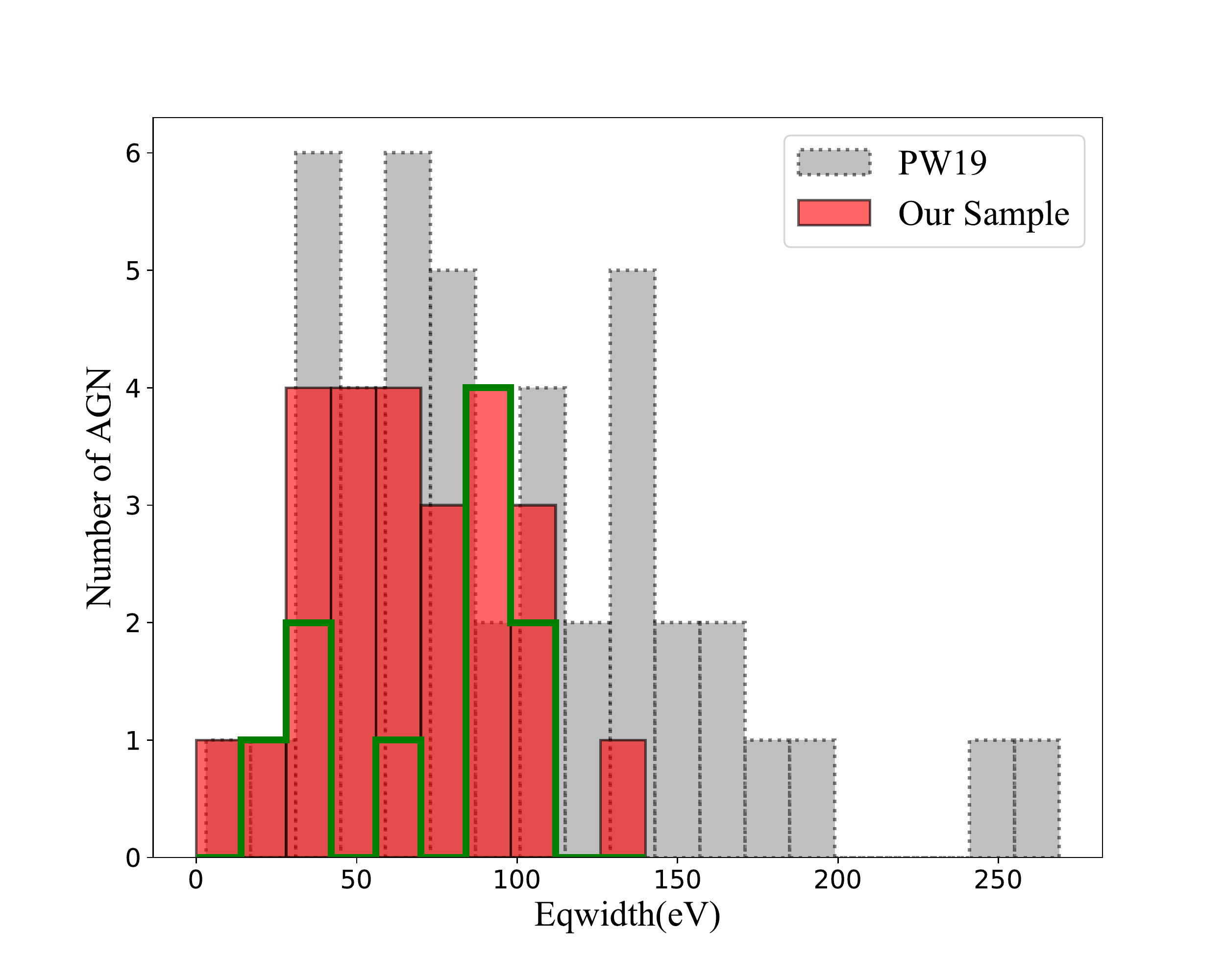}}\\
\caption{The distribution of photon index $\Gamma$, $pexrav$ reflection factor R  and Fe K$\alpha$ line EW (from left to right) for our sample and radio quiet sources in PW19.  The green lines mark sources with only upper limits to R or EW in our sample. 
Apparently, all three parameters are smaller in our sample than in PW19. The difference is significant at $99.7$\% confidence for $\Gamma$ (KS test), and $>99.9$\% confidence for both R and EW (using log--rank test within ASURV to handle the censored data points).}
\label{fig:YREW}
\end{figure*}

The distribution of photon index $\Gamma$, R and Fe K$\alpha$ EW from our sample and PW19 are given in Fig. \ref{fig:YREW}. It is obviously seen that radio galaxies in our sample show flatter \textit{NuSTAR} spectra comparing with radio quiet AGNs in PW19. More precisely, the mean value and scatter of $\Gamma$ is $1.73 \pm 0.15$ for our sample, while that of PW19 is $1.90 \pm 0.21$. With KS test, we find that the difference in the spectral slope between our sample and PW19 is significant at $\sim$ 99.7\% confidence.
This is in good agreement with previous studies showing radio loud AGNs have harder X-ray spectra \citep[e.g.][]{10.1046/j.1365-8711.2000.03510.x,10.1111/j.1365-2966.2005.09550.x,hardgamma2,hardgamma1}, but for the first time with $NuSTAR$ observations which extend the spectral range to 78 keV.

\begin{figure}
\includegraphics[width=3.5in]{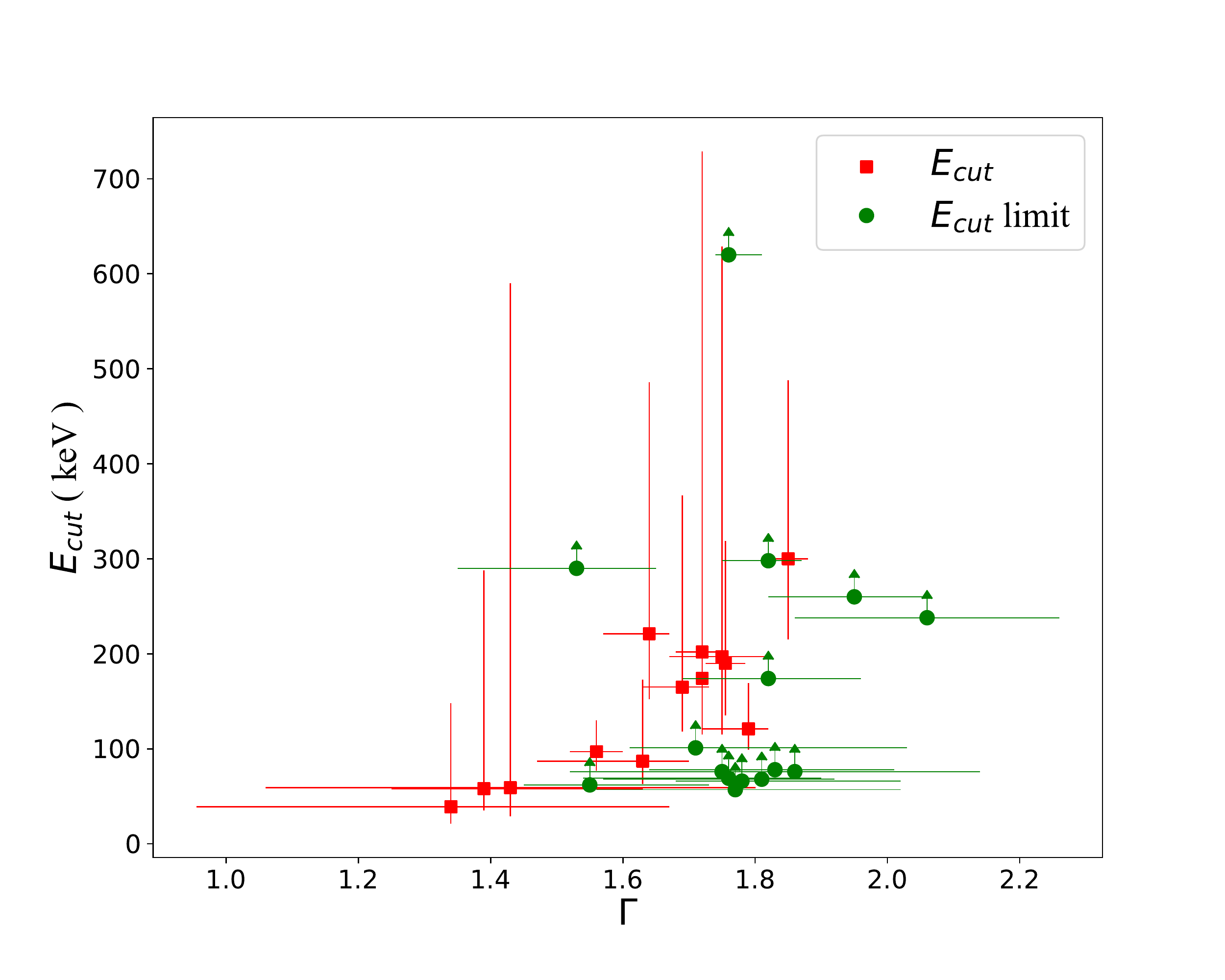}
\caption{Best-fit $E_{cut}$ vs. $\Gamma$ for the radio galaxies in our sample. A correlation between $E_{cut}$ and $\Gamma$ is seen for the 13 sources with $E_{cut}$ detections. The correlation however disappears when those sources withou $E_{cut}$ detections are included.
}
\label{fig:YEcor}
\end{figure}

We examine for potential correlations between each two of the spectral parameters ($E_{cut}$, $\Gamma$, $R$, and EW), but only find statistically significant correlation 
between $\Gamma$ and $E_{cut}$ for the 13 sources with $E_{cut}$ detections, for which Spearman test gives $\rho$ = 0.764 and P-value = 0.006 (but note the large statistical uncertainties in both $E_{cut}$  and $\Gamma$, Fig. \ref{fig:YEcor}).
Such a correlation, i.e., smaller $E_{cut}$ in flatter spectra,  was similarly reported in a recent study (mostly radio quiet AGNs) by \citet{10.1093/mnras/stz156}, but not seen in earlier works 
\citep[e.g.][]{10.1111/j.1365-2966.2009.15257.x, refId01}. As \citet{10.1093/mnras/stz156} pointed out, such a correlation however could be artificial since there is clear degeneracy between 
$E_{cut}$ and $\Gamma$ during the spectral fitting.

Our radio galaxies also show weaker X-ray reflection (both in R and Fe K$\alpha$ line EW) while comparing with radio quiet ones in PW19 (both with confidence level $>$ 99.9\%, see Fig. \ref{fig:YREW}).
To properly deal with sources with only upper limits to line EW and/or R, we use the Kaplan-Meier estimator within the ASURV package to calculate the mean value and scatter of both parameters in both samples. We get $\langle R \rangle = 0.23 \pm 0.40$ and $\langle EW \rangle = 52.9 \pm 39.0$ eV for our sample, while for radio quiet ones in PW19, $\langle R \rangle = 0.91 \pm 0.70$ and $\langle EW \rangle = 102 \pm 62$ eV. This also confirms previous findings that radio galaxies tend to have weak X-ray reflection components\citep[e.g.][]{2013ApJ...772...83L, 1999ApJ...526...60S, 10.1046/j.1365-8711.2000.03510.x}, including both the continuum reflection and the Fe K$\alpha$ line emission. 

\begin{figure}[]
\centering
\subfloat{\includegraphics[width=3.5in]{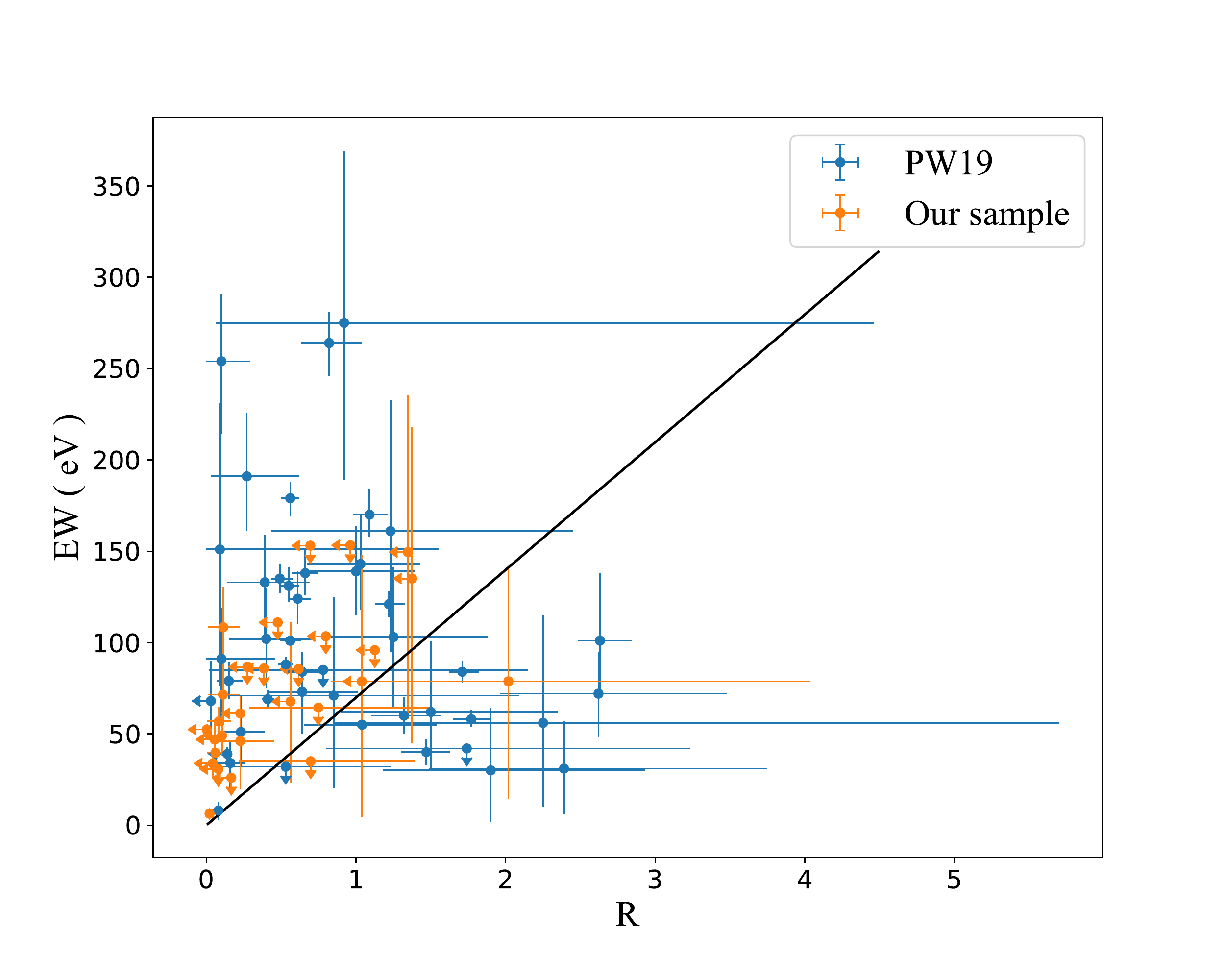}}
\caption{\label{fig:eqRcor}Relection factor R vs. EW of our sample and PW19 sample. The black line shows the best linear fit for all sources. 
}
\end{figure}

Positive correlation between the reflection factor R and  Fe K$\alpha$ line strength EW is expected, as both are expected to correlate with the covering factor of the surrounding cold medium,
e.g., the accretion disc and the torus,  which is illuminated by the central X-ray source and reradiates the reflection continuum and Fe K line 
\citep[e.g.][]{1991A&A...247...25M}. 
However, we are unable to detect significant correlation using our sample alone (Fig. \ref{fig:eqRcor}), likely because the large statistical uncertainties. 
We find no correlation either between R and EW using the radio quiet PW19 sample, though PW19 reported clear correlation between the Fe K line flux and the flux of the reflected emission at 30 keV (likely because both fluxes positively correlate with source brightness). Combining our sample with PW19 sample, we find a marginal correlation between R and EW (generalized Spearman test within \textit{ASURV} finding $\rho$ = 0.260 and P-value = 0.014). However the scatter is also remarkable (see Fig. \ref{fig:eqRcor}),
with only 65.7\% of sources agree with the best-fit linear correlation within 90\% statistical uncertainties. 
This suggests that the coupling between continuum reflection and Fe K$\alpha$ line is more complicated than a simple linear relation. For instance an anti-correlation between R and EW had been revealed in literature \citep[e.g. in NGC 5548,][]{Chiang_2000}. Fitting the continuum reflection and Fe K$\alpha$ line strictly tied, which thus should only be used with caution, may bias the measurements of continuum reflection and then $E_{cut}$ \citep[e.g.][]{Mantovani_Feline, Zhangjx2018}.

\subsection{Jet contamination}
\label{subsec:jet}
\par In this work we have shown that a significant fraction of of our radio galaxies (13 out of 28) show high energy cutoff in their $NuSTAR$ X-ray spectra, and those non-detections of E$_{cut}$ could be dominantly attributed to their low $NuSTAR$ counts. This suggests that the X-ray spectra of our sample is dominated by the thermal corona emission instead of jet. 

\par However, strong jet contamination are still possible in a small number of individual sources. The jet contamination could be particularly severe in some sources, from which the known extended jet-driven X-ray emission  (Pic A, for an example, see \citealt{Hardcastle2005, Hardcastle2016, Migliori2007}) is unresolvable for $NuSTAR$. In fact the E$_{cut}$ detection in Pic A presented in this work is marginal, and we are unable to rule out severe jet contamination in this source.

\par A more proper interpretation of our results is, for the majority of our sample, the $NuSTAR$ spectra are corona dominated. 
Our sample also contains 4 FR I galaxies, and 4 core-dominated sources. We mark those sources in Fig. \ref{fig:cutoff} and Fig. \ref{fig:count}.
Likely due to the small numbers, we are unable to reveal statistical differences between their detection rates of E$_{cut}$ (2 out of 4 FR Is, 1 out of 4 core-dominated, and 10 out of 20 FR II galaxies), or between their distributions of E$_{cut}$, $\Gamma$, R or Fe K$\alpha$ line EW. Interestingly, however, from Fig. \ref{fig:count} we do see hints that the jet contribution to $NuSTAR$ spectra in our FR I and core-dominated galaxies are likely higher than in FR II galaxies:
the 3 FR I or core-dominated galaxies with  E$_{cut}$ detections have the most $NuSTAR$ counts among all sources, and the only two E$_{cut}$ non-detections at $NuSTAR$ counts $>$ 10$^{4.5}$ are
FR I or core-dominated. This suggests that while detecting E$_{cut}$ with $NuSTAR$ spectra is feasible in FR I or core-dominated galaxies, it could be more challenging (comparing to FR IIs) due to jet contamination.

\par 
 Radio galaxies with strong jet contributions could be gamma-ray bright sources.
We further match our sample with Fermi-LAT Fourth Source Catalog \citep{Fermi}  
 which includes 5,064 sources above 4$\sigma$ significance based on the first eight years of science data from the Fermi Gamma-ray Space Telescope mission in the energy range from 50 MeV to 1 TeV.
We find 9 Fermi gamma-ray sources in our sample, including all the 4 core-dominated sources, 3 out of 4 FR Is, and 2 out of 20 FR IIs. 
These 9 sources, which are marked in Tab. \ref{tab:Results}, may have stronger jet contribution to the $NuSTAR$ spectra. Among them we report E$_{cut}$ detections for 5 sources, including 1 core-dominated, 2 FR Is, and 2 FR IIs. 
Likely due to the small number of sources, we are unable to reveal significant difference between this subsample (Fermi detected) and the rest of the sample,  in the distribution of their E$_{cut}$, $\Gamma$, R and Fe K$\alpha$ line EW.
We calculate the ratio of Fermi/LAT 100 MeV -- 100 GeV flux to $NuSTAR$ 3--78 keV flux for the 9 sources (see Tab. \ref{tab:Fermi}). We find that the 3 sources with highest LAT/NuSTAR flux ratios (3C 279, 3C 345, NGC 1275) have E$_{cut}$ non-detected, meanwhile 5 out the rest 6 sources show E$_{cut}$ detections, including 3C 273 which has the 4th highest LAT/NuSTAR flux ratio ($\sim$ 0.53).  
This also hints that while the jet contamination does not affect the E$_{cut}$ detection in the majority of our sample, it may have hindered the detection of E$_{cut}$ in a few sources.

\begin{table}[h]
\caption{Fluxes of Fermi sources}
\label{tab:Fermi}
\centering
\begin{tabular}{c c c c}
\hline
\hline
\tabincell{c}{Source\\ \ } & \tabincell{c}{$flux_{Fermi}$\\ ($erg/s/cm^{2}$)} &\tabincell{c}{$flux_{NuSTAR}$\\ ($erg/s/cm^{2}$)} & \tabincell{c}{flux ratio\\ \ } \\
\hline
3C 279 & 2.8e-10 & 3.3e-11 & 8.37 \\
NGC 1275 & 2.8e-10 & 1.1e-10 & 2.56 \\
3C 345 & 2.8e-11 & 1.5e-11 & 1.92 \\
3C 273 & 1.3e-10 & 2.4e-10 & 0.53 \\
3C 309.1 & 3.0e-12 & 5.8e-12 & 0.51 \\
PicA & 4.7e-12 & 2.2e-11 & 0.22 \\
3C 111 & 1.6e-11 & 1.1e-10 & 0.14 \\
3C 120 & 1.3e-11 & 1.3e-10 & 0.10 \\
CentaurusA & 6.4e-11 & 9.1e-10 & 0.07 \\
\hline
\end{tabular}
\begin{tablenotes}
\item {} : {The Fermi/LAT flux is calculated in 100 MeV -- 100 GeV, while the $NuSTAR$ flux is in 3 -- 78 keV}. 
\end{tablenotes}

\end{table}

\par Detailed X-ray spectral fitting including both the jet and corona components \citep{Lohfink2013,Madsen_2015_3C273} or broadband SED modeling \citep[e.g.][]{Kataoka2011} might be able to quantifying the jet contamination. For instance \citet{Madsen_2015_3C273} examined possible jet contribution to $NuSTAR$ spectra of 3C 273 with simultaneous $NuSTAR$ and INTEGRAL observations. They found jet emission starts to dominate
only above 30 -- 40 keV, and fitting $NuSTAR$ spectra alone can not reveal the jet component. While measuring the jet contribution is beyond the scope of this work, we expect the jet contribution to $NuSTAR$  spectra in most of our sources be weaker than that in the blazar 3C 273.

\subsection{The physical nature of the X-ray coronae in radio galaxies?}

The facts that radio AGNs show flatter X-ray spectra as well as weaker reflection and Fe K$\alpha$ line, have been extensively discussed. 
One possibility is strong contamination to hard X-ray spectra from the jet emission \citep[e.g.][]{1997MNRAS.292..468R,2002NewAR..46..221G}, as the Doppler boosted X-ray emission from the jet is flatter and yields little illumination on the disk and equatorial material. 
However, in this work we have shown that the $NuSTAR$ spectra of the majority of our radio galaxy sample are dominated by corona emission, thus 
jet contamination is unable to explain the observed flatter $NuSTAR$ spectra and smaller R and EW.

Since the $E_{cut}$ distribution of our radio galaxies is similar to that of radio quiet AGNs (see \S\ref{sec:cutoff}),  we expect the corona temperature in
radio galaxies is similar to that in radio quiet AGNs. The spectral hardening in our radio galaxies thus can not be attributed to higher coronal temperature either.
Consequently, the X-ray coronae in radio galaxies should have geometry different from those in radio quiet AGNs. 

A further note is that in 3C 273, though the jet contamination is weak, including a jet component yields even lower $E_{cut}$ \citep{Madsen_2015_3C273}, which means we may have overestimated $E_{cut}$ (then corona temperature) in radio galaxies if the jet contamination in which is comparable to or stronger than that in 3C 273. 

Possible mechanisms to explain the weak X-ray reflection and Fe K$\alpha$ line in radio galaxies also include
highly ionized inner disk \citep{2002MNRAS.332L..45B}, different inner disc geometry \citep[in the form of ion-supported torus or advection-dominated accretion flow,][]{2000ApJ...537..654E},
obscuration of the central accretion flow (where reflection is produced) by the jet or the corona itself \citep{2009ApJ...700.1473S}, 
self-obscuration by the geometrically thick accretion disc \citep{Paltani1998}, and outflowing corona \citep{Beloborodov_outflow, Malzac_outflow, Ballantyne_2014_3C382, 2015ApJ...814...24L_3C309, 2017ApJ...841...80L_4C74, King_outflow}.

Below we show that the findings in this work favor the outflowing corona model, in which the flatter X-ray spectra and weaker X-ray reflection in radio galaxies could be consistently be attributed to higher outflowing velocity, though we are unable to robustly rule out other possibilities.    
PW19 reported a clear correlation between $\Gamma$ and the reflection factor R for their (mostly radio quiet) sample, and the correlation is genuine, e.g., 
can not be attributed to the coupling between the two parameters $\Gamma$ and R in the $pexrav$ model. 
In Fig. \ref{fig:YRcor} we plot photon index $\Gamma$ and the reflection factor R for our sample. Due to the usually small R and large uncertainties, we are unable to detect significant correlation using our sample alone. 
However, combining with the PW19 sample,   we find a strong correlation between $\Gamma$ and R (generalized Spearman test finding $\rho$ = 0.712 and P-value $< 0.0001$, Fig. \ref{fig:YRcor}). 
More importantly, while radio loud AGNs have clearly flatter spectra and smaller R, they appear well consistent with the $\Gamma$ -- R correlation defined by the radio quiet sources, suggesting 
a common underlying mechanism for both populations. 
Such a $\Gamma$ -- R correlation, firstly observed by \citet{Zdziarski_Rgamma},
appears consistent with the outflowing corona model\footnote{
Alternatively,  the observed $\Gamma$ -- R correlation may be attributed to variable overlap between hot and cold accretion flows \citep[e.g.][]{Zdziarski_Rgamma}, or a combination of distant reflection and a pivoting primary powerlaw spectrum with a nearly constant Comptonized luminosity \citep[e.g.][]{Malzac_2002}. } of \citet{Beloborodov_outflow} in which the corona is outflowing with a relativistic bulk velocity perpendicular to the disc. 
In the model, the apparent strength of the disk reflection component R is given by:
\begin{equation}
R = \frac{(1 + \beta / 2)(1 - \beta \mu_s )^3 }{(1 + \beta)^2}
\end{equation}
where $\beta = v/c$ is the bulk velocity of the outflowing away from the reflector, and $\mu_s$ determined by the coronal geometry (for example, $\mu_s = 0$ describes a slab geometry, while $\mu_s \sim 0.5 $ roughly corresponds to a blob with radius of order its height\citep[for details, please refer to][]{Beloborodov_outflow}. Assuming a typical $\mu_s = 0.5$, the photon index $\Gamma$ can be approximated as:
\begin{equation}
\Gamma \approx 2[{\frac{(1 + \beta)}{\sqrt{(1- \beta ^2 )}}}]^{-0.3}
\end{equation}
The expected $\Gamma$-R relation is shown in Fig. \ref{fig:YRcor}, generally consistent with the observed trend.

Note in the model of \citet{Beloborodov_outflow}, $\Gamma$ = 2.0 is assumed for spectra from static corona.
Based on the model, outflowing (inflowing) corona yields flatter (steeper) spectra. For instance, $\Gamma$ = 1.44 is expected  for an outflowing velocity of 0.8c, and  2.36 for an inflowing velocity of 0.6c.
However, considering possible large scatter in the corona physical properties (geometry, temperature, opacity) of various AGNs, and there are additional factors which may affect the reflection strength (such as the light bending effect, and  reflection from distant material other than the disk), we stress that spectra steeper than $\Gamma$ = 2.0 in Fig. \ref{fig:YRcor}  do not necessarily imply inflowing corona,
thus do not necessarily object the studies which reported other observational evidences supporting outflowing corona also in radio quiet AGNs \citep{Liu_2014, Wilkins2015}. 
 Nevertheless, based on the outflowing corona model, Fig. \ref{fig:YRcor} implies that the corona in radio galaxies could have higher outflow velocity 
compared with radio quiet AGNs.

\begin{figure}[]
\centering
\subfloat{\includegraphics[width=3.5in]{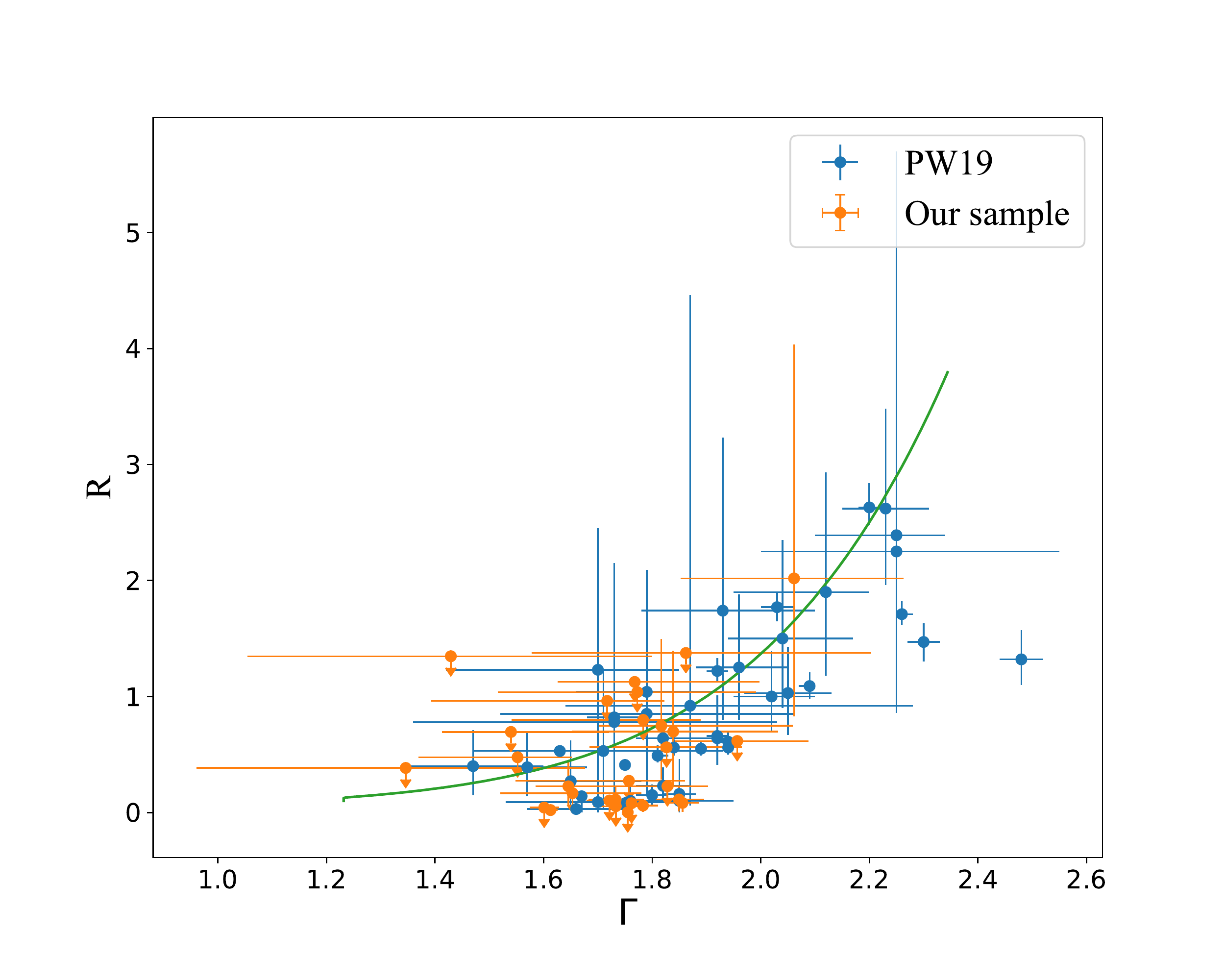}}
\caption{\label{fig:YRcor} $\Gamma$ vs. reflection factor R of our sample and PW19 sample. The green line shows the expected R vs. $\Gamma$ curve based on the model in \citet{Beloborodov_outflow}, which is a result of the relative motion of the outflowing corona with respect to the disc.
}
\end{figure}

\section{Summary}\label{sec:summary}
In this work we present 55 \textit{NuSTAR} observations of a sample of 28 radio galaxies (mostly FR II galaxies) with weak beaming effect of the relativistic jet and X-ray absorption column density N$_H$ $<$ 10$^{23}$ cm$^{-2}$.
We fit the spectra uniformly to measure the spectral photon index $\Gamma$, high energy cutoff $E_{cut}$, the X-ray reflection factor R, and the Fe K$\alpha$ line EW.  Our main results are as follows:

\begin{itemize}

\item{}We report $E_{cut}$ detections in 13 sources and lower limits to the rest 15 sources.  Over half of them (7 $E_{cut}$ detections and 11 lower limits) are reported for the first time. 

\item{}The $E_{cut}$ non-detections is primarily due to the their lower \textit{NuSTAR} net photon counts. This implies that the \textit{NuSTAR}  X-ray emission of the majority of our sample, are dominated by the thermal corona emission. 

\item{}Meanwhile, we do see hints that jet contribution may have hindered the detection of $E_{cut}$ in a couple of core-dominated or FR I galaxies.

\item{}The distribution of the detected $E_{cut}$ in our radio sample is indistinguishable from that of radio quiet AGNs reported in literature.

\item{} The radio galaxies in our sample have flatter \textit{NuSTAR} spectra, as well as smaller reflection factor R and Fe K$\alpha$ line EW, comparing with radio quiet ones. The flatter spectra however can not be attributed to jet contamination. 

\item{}Combining with a large radio quiet sample, we see a correlation between reflection factor R and Fe K$\alpha$ line EW, but with considerably large scatter, showing the coupling
between continuum reflection and Fe K$\alpha$ line is rather complicated. 

\item{}In the plot of R versus $\Gamma$, the radio galaxies follow the same correlation trend with the radio quiet sample. This supports the outflowing corona model which predicts weaker reflection and flatter X-ray spectra in case of higher corona outflowing velocity.

\end{itemize}

\acknowledgments
{We thank the anonymous referee for constructive comments which have significantly improved the manuscript. This research has made use of the NuSTAR Data Analysis Software (NuSTARDAS) jointly developed by the ASI Science Data Center (ASDC, Italy) and the California Institute of Technology (USA). This research has made use of the NASA/IPAC Extragalactic Database (NED),which is operated by the Jet Propulsion Laboratory, California Institute of Technology, under contract with the National Aeronautics and Space Administration. The work is supported by National Science Foundation of China (grants No. 11421303 $\&$ 11890693) and CAS Frontier Science Key Research Program (QYZDJ-SSW-SLH006).
}

\bibliography{sample}{}

 \begin{appendix}
 {

\section{A: Sources Excluded from Analyses}
\label{app:A}

We exclude 11 NuSTAR observed radio galaxies from analyses in this work, either because they are heavily obscured ($N_H$ $>$ $10^{23}  cm^{-2}$),
or NuSTAR spectra have too few net counts ($<$3000). The sources are listed in Table \ref{giveup}.  
\begin{table*}[!t]
\caption{\textit{NuSTAR} Observation Details for the Excluded Sources }
\label{giveup}
\centering

\setlength{\tabcolsep}{6mm}{
\begin{tabular}{c c c c c c}
\hline
\hline
\tabincell{c}{Source\\ \ } & \tabincell{c}{ID\\ \ } &\tabincell{c}{Redshift\\ \ }&\tabincell{c}{Exposure\\ (ks)} & \tabincell{c}{Total Counts\\ ($1\times10^3$)} & \tabincell{c}{Remark\\ \ }\\
\hline
Cygnus A & 60001083002 & 0.056 & 43.6 & 106.9 & \textit0 \\
 & 60001083004 &  & 20.7 & 51.2 &  \\
VII Zw 292 & 60160374002 & 0.058 & 13.0 & 2.0 & \textit1 \\
3C 452 & 60261004002 & 0.081 & 51.7 & 15.4 & \textit2 \\
3C 445 & 60160788002 & 0.056 & 19.9 & 8.3 & \textit3 \\
NGC612 & 60061014002 & 0.03 & 16.7 & 2.9 & \textit4 \\
3C 098 & 60061042002 & 0.03 & 27.3 & 4.4 & \textit5 \\
3C 234.0 & 60160382002 & 0.185 & 14.9 & 2.1 & \textit6 \\
PKS 0707-35 & 60160285002 & 0.111 & 18.9 & 1.6 & \textit7 \\
4C 29.30 & 60061083002 & 0.065 & 21.0 & 2.1 & \textit8 \\
Mrk 1498 & 60160640002 & 0.055 & 23.6 & 14.1 & \textit9 \\
3C 403   & 60061293002 & 0.059 & 20.0 & 4.6 & \textit{10}\\
\hline
\end{tabular}
}
\begin{tablenotes}
\item {\textit0} : {too strong absorption with $N_H$$\sim$$2\times10^{23}  cm^{-2}$ \citep{Reynolds_2015}. } \\
\item {\textit1} : {too faint with total net counts$\sim$1000 and compton-thick \citep{10.1093/mnras/stx3159}.} \\
\item {\textit2} : {too strong absorption with $N_H$$\sim$$5.7\times10^{23}  cm^{-2}$ \citep{10.1093/mnras/stx3159}.} \\
\item {\textit3} : {too strong absorption with $N_H$$\sim$$2\times10^{23}  cm^{-2}$ \citep{Sambruna_2007}.} \\
\item {\textit4} : {too faint with total net counts$\sim$1500 and compton-thick \citep{10.1093/mnras/stx3159}.} \\
\item {\textit5} : {too strong absorption with $N_H$$\sim$$1.5\times10^{23}  cm^{-2}$ \citep{10.1093/mnras/stw1438}.} \\
\item {\textit6} : {too faint with total net counts$\sim$1100.} \\
\item {\textit7} : {too faint with total net counts$\sim$800.} \\
\item {\textit8} : {too faint with total net counts$\sim$1100 and too strong absorption with $N_H$$\sim$$4\times10^{23}  cm^{-2}$ \citep{Siemiginowska_2012}.} \\
\item {\textit9} : {too strong absorption with $N_H$$\sim$$1.7\times10^{23}  cm^{-2}$ \citep{10.1093/mnras/stz2265}.} \\
\item {\textit{10}} : {too strong absorption with $N_H$$\sim$$2.4\times10^{23}  cm^{-2}$. The \textit{NuSTAR} spectra for this source is reported the first time in this work. The best-fit spectral parameters with our base model are $\Gamma = 1.60_{-0.40}^{+0.27}$}, $E_{cut} > 60$ keV, $R < 1.5$ and $EW< 169$ eV ($\chi^2_\nu =  1.11$).\\

\end{tablenotes}
\end{table*}

\section{B: NuSTAR Spectra}
\label{app:B}
In this section, we present NuSTAR unfolded spectra analyzed in this work, along with the best-fit models and the residual data-to-model ratios (Fig. \ref{fig:spectra}).
\begin{figure*}[!t]
\centering
\ContinuedFloat
\subfloat{\includegraphics[width=0.33\textwidth]{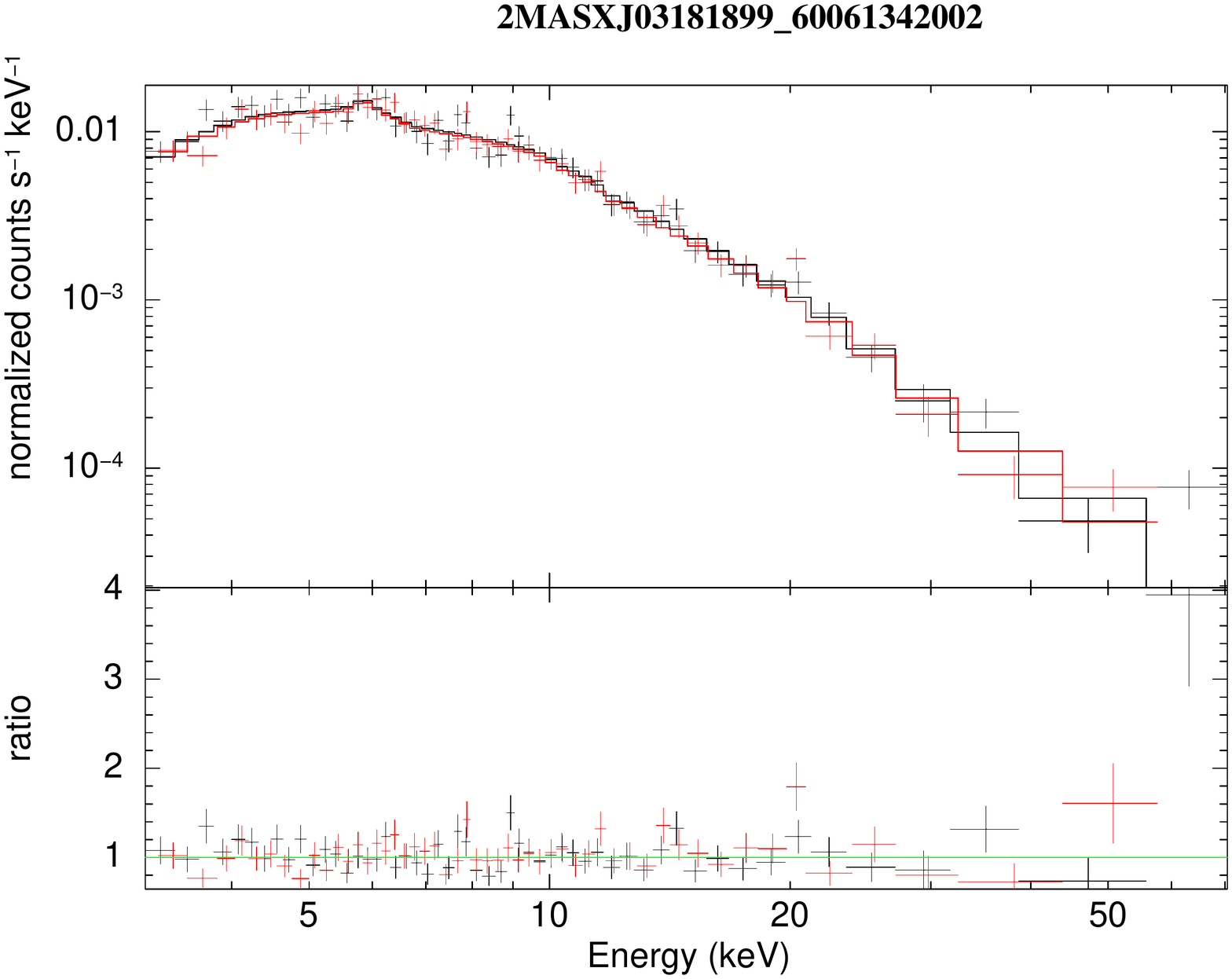}}
\subfloat{\includegraphics[width=0.33\textwidth]{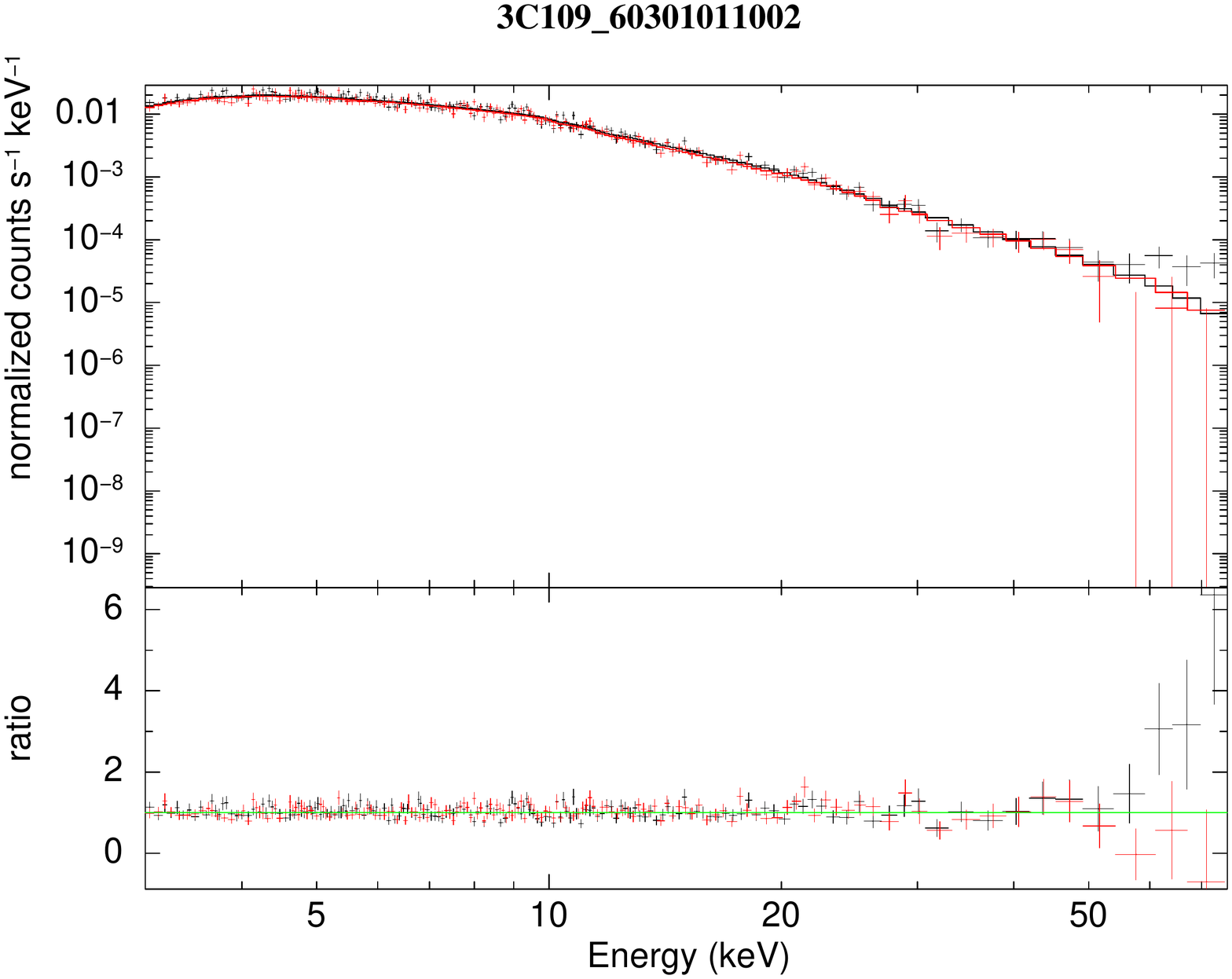}}
\subfloat{\includegraphics[width=0.33\textwidth]{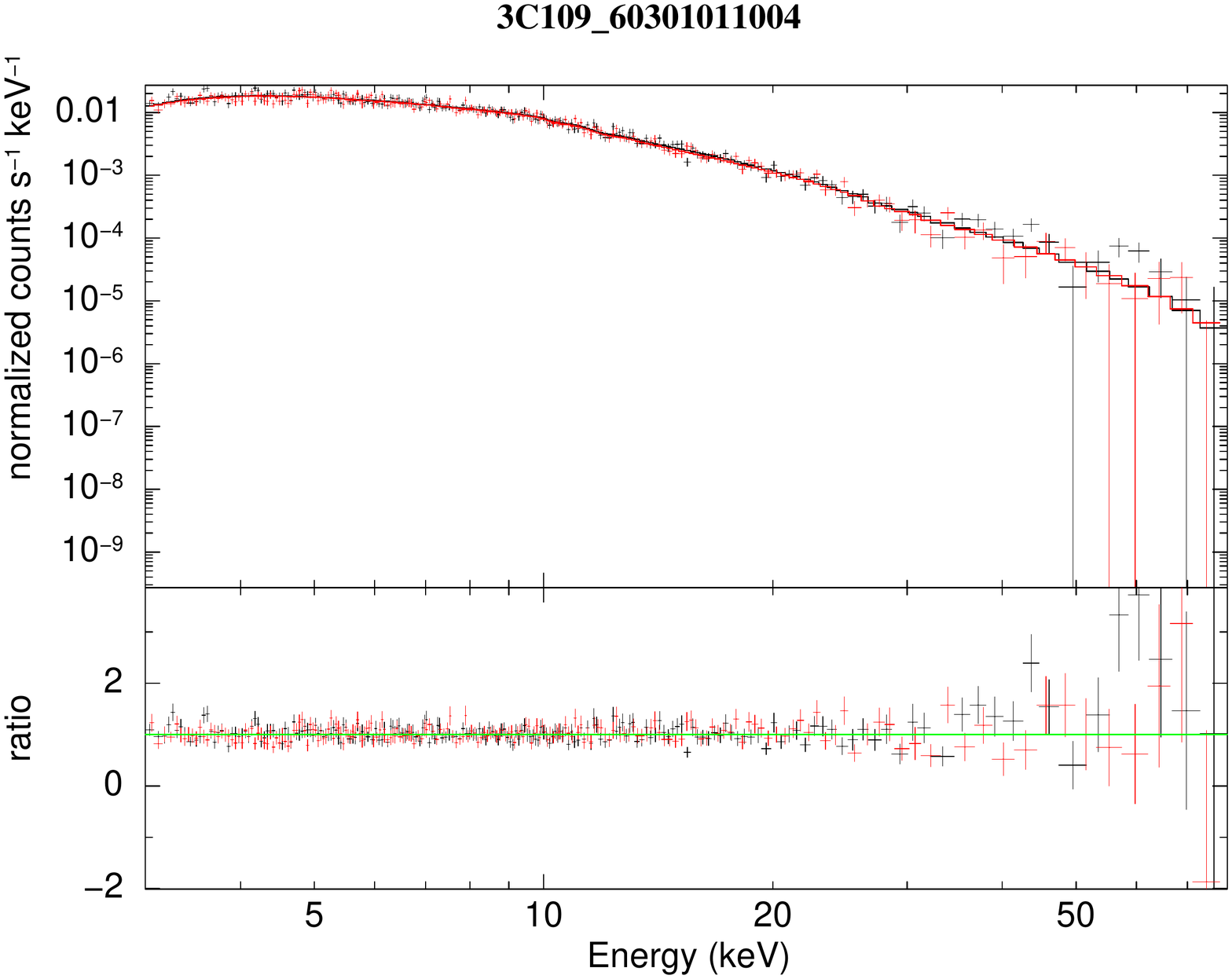}}\\
\subfloat{\includegraphics[width=0.33\textwidth]{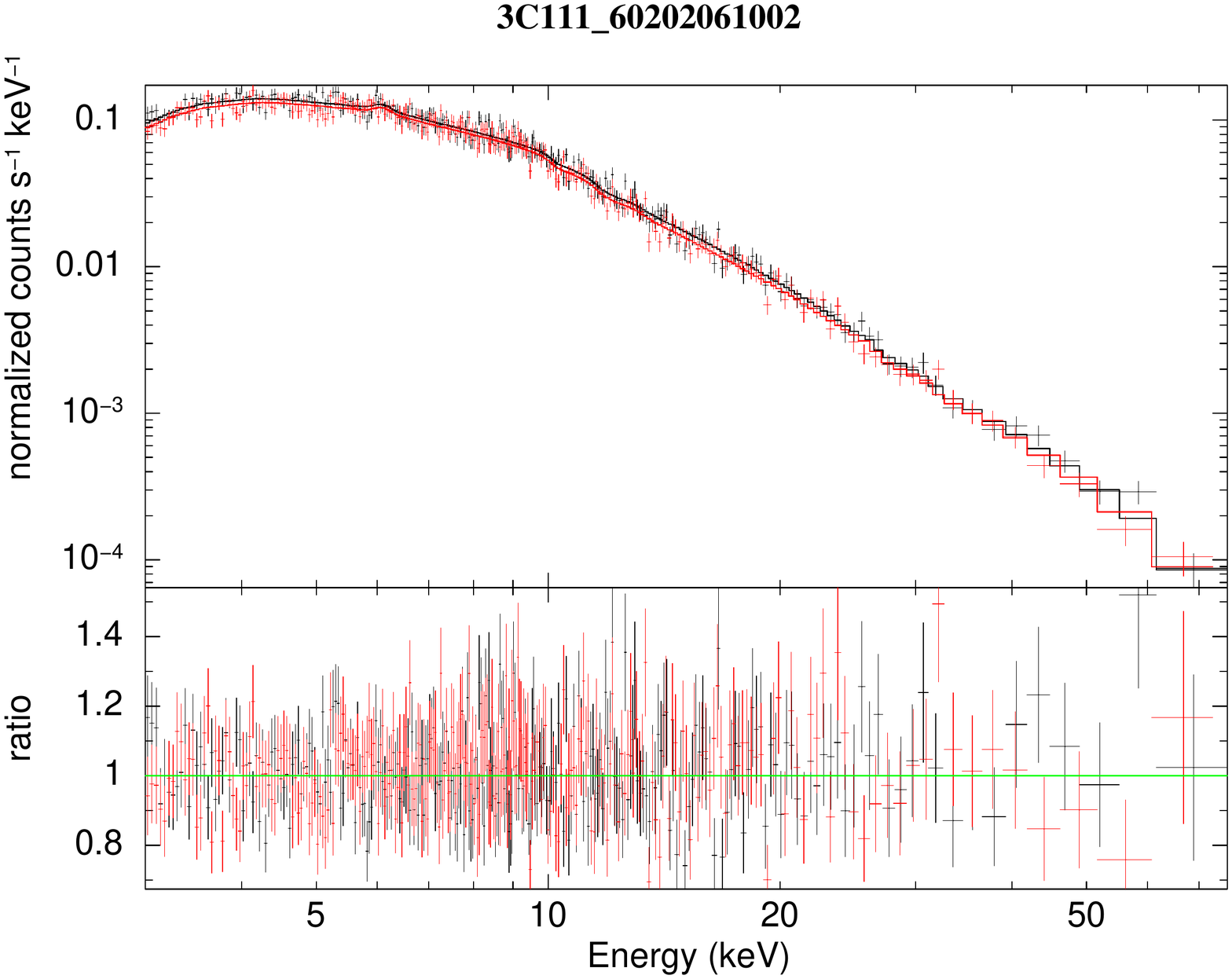}}
\subfloat{\includegraphics[width=0.33\textwidth]{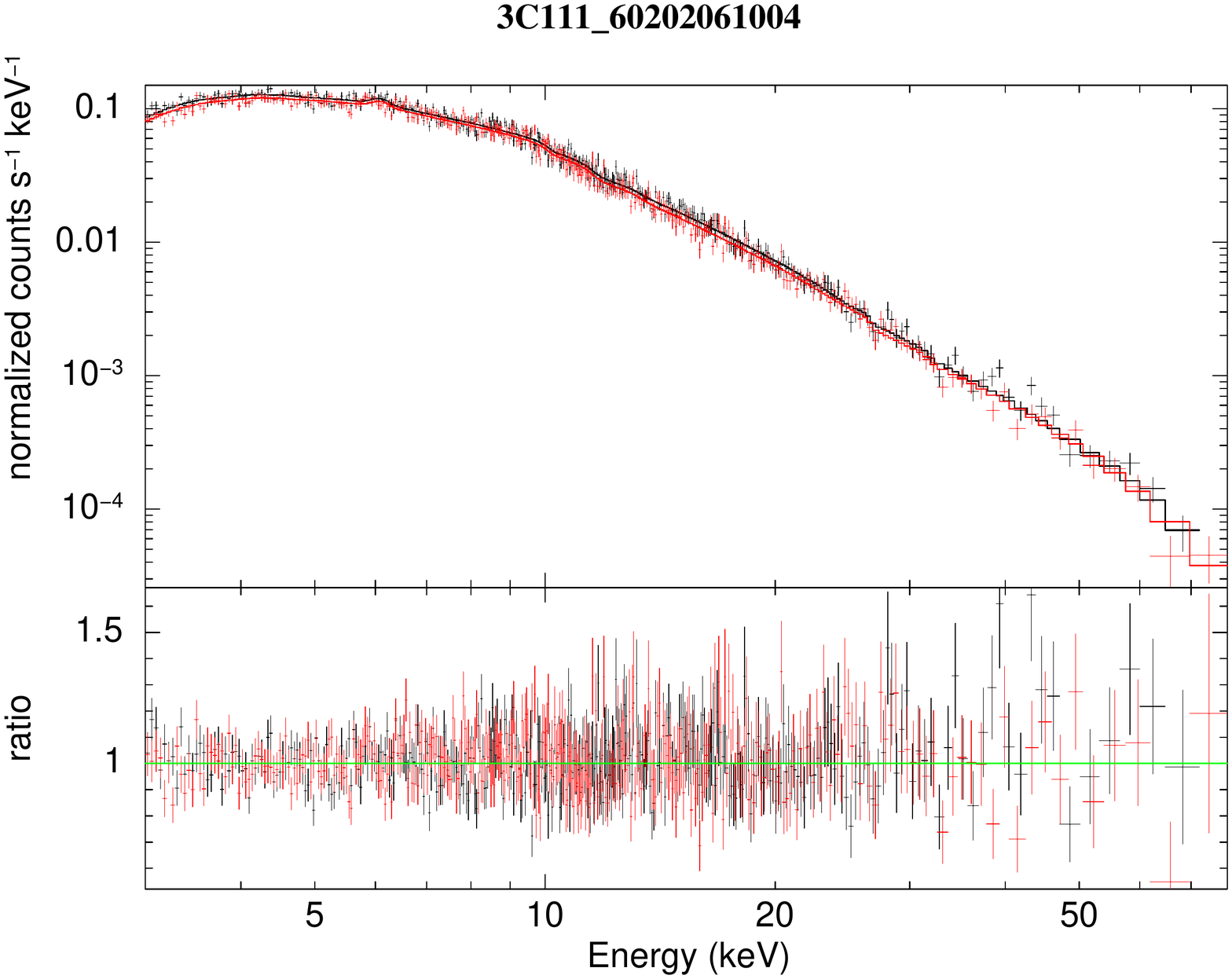}}
\subfloat{\includegraphics[width=0.33\textwidth]{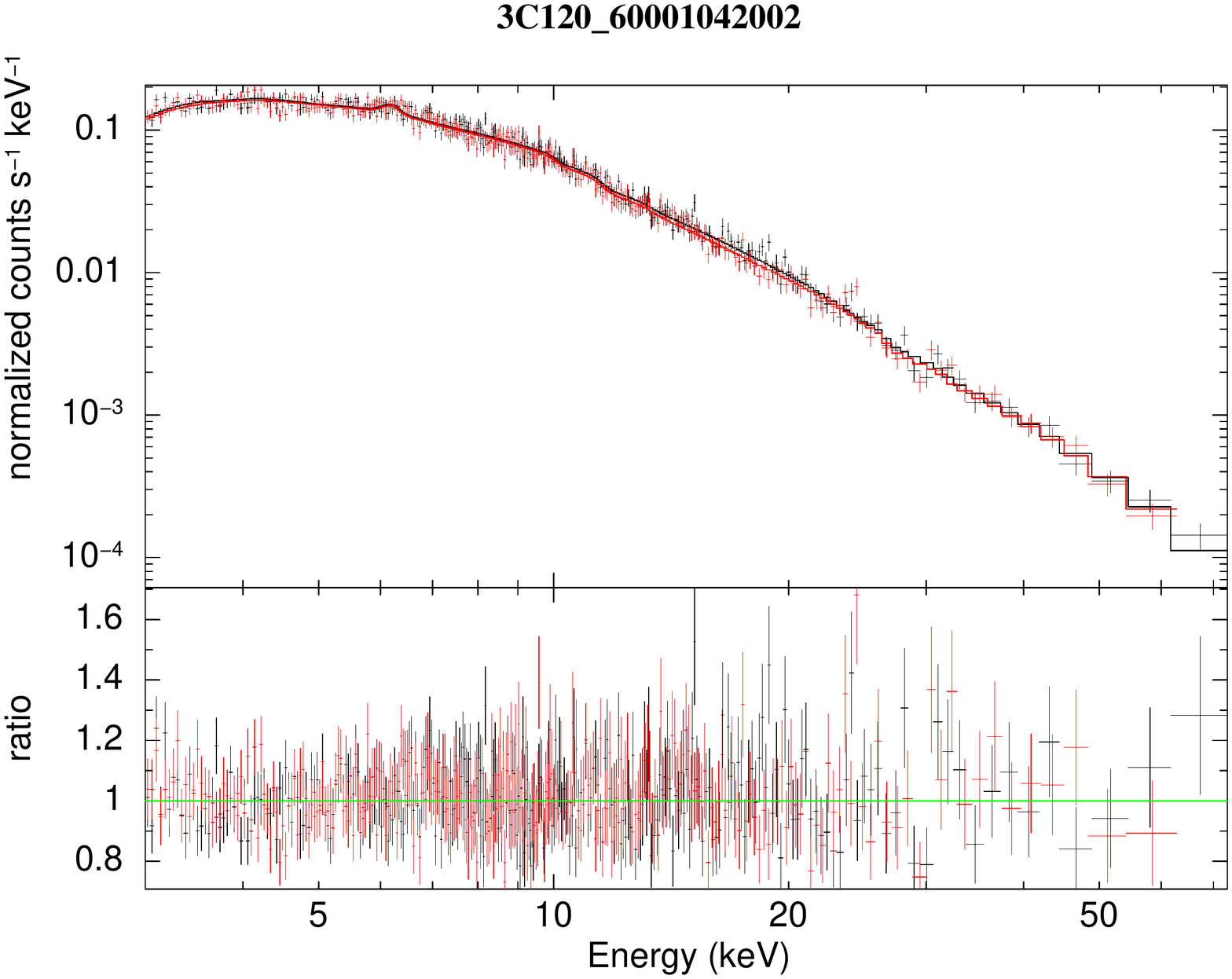}}\\
\subfloat{\includegraphics[width=0.33\textwidth]{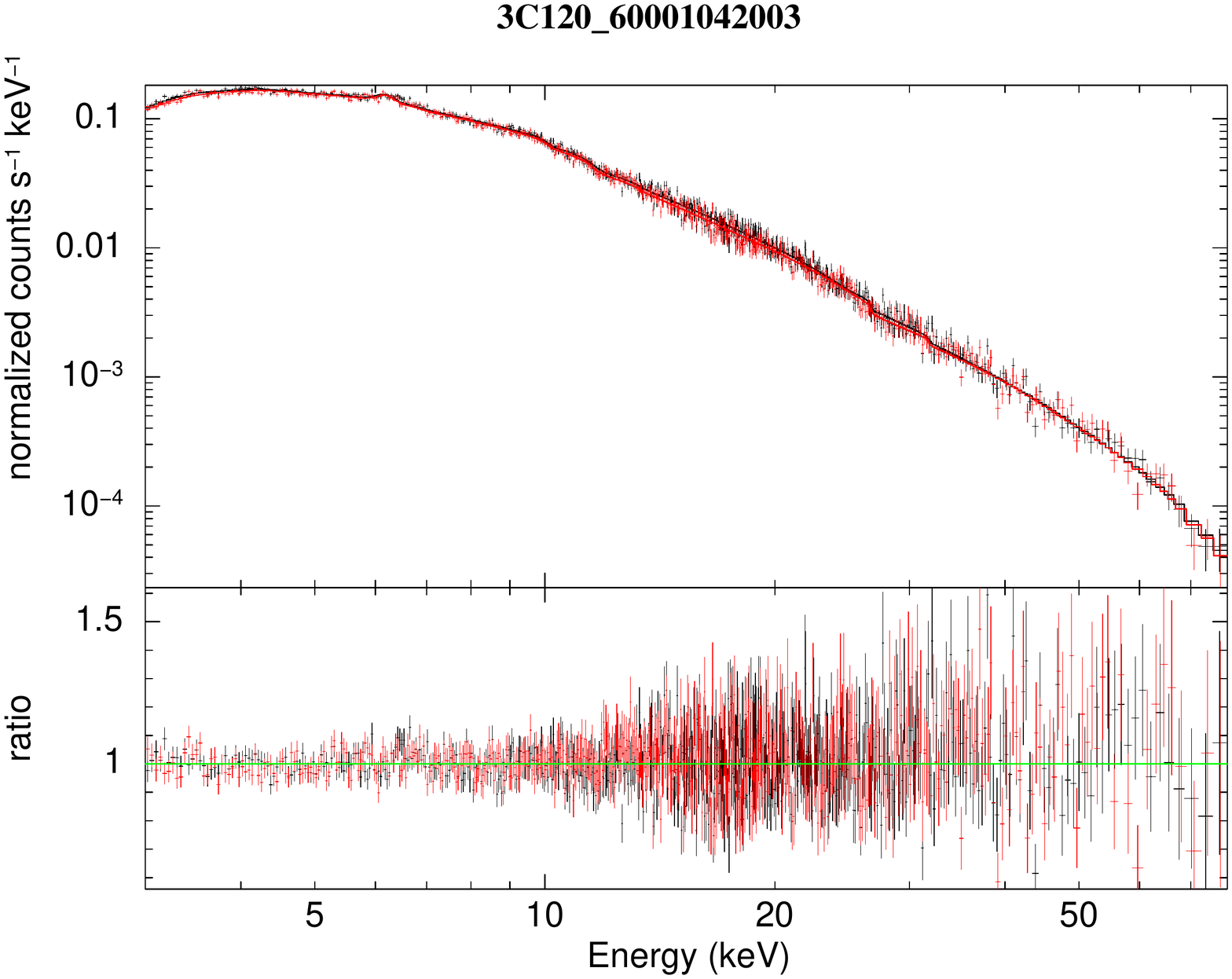}}
\subfloat{\includegraphics[width=0.33\textwidth]{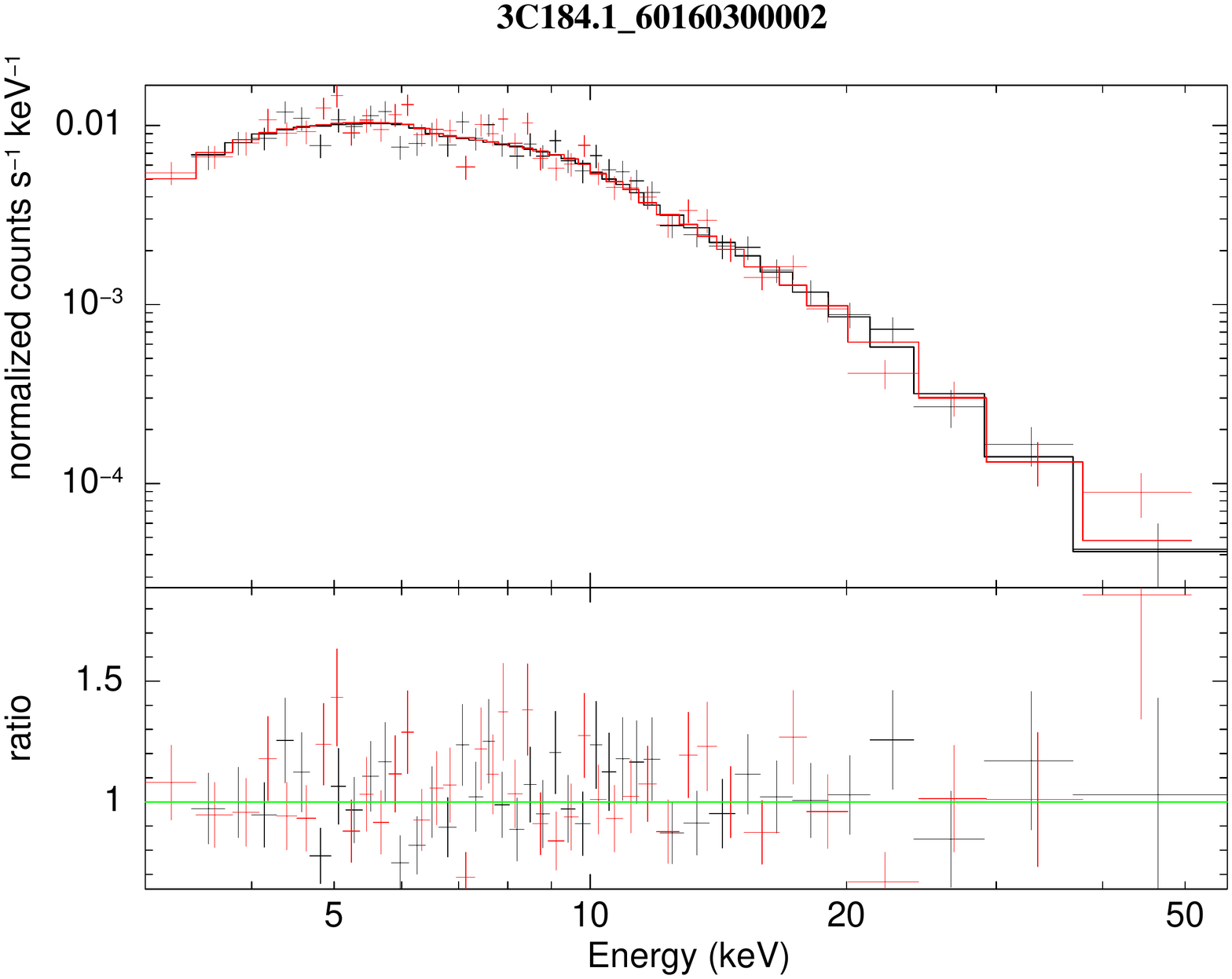}}
\subfloat{\includegraphics[width=0.33\textwidth]{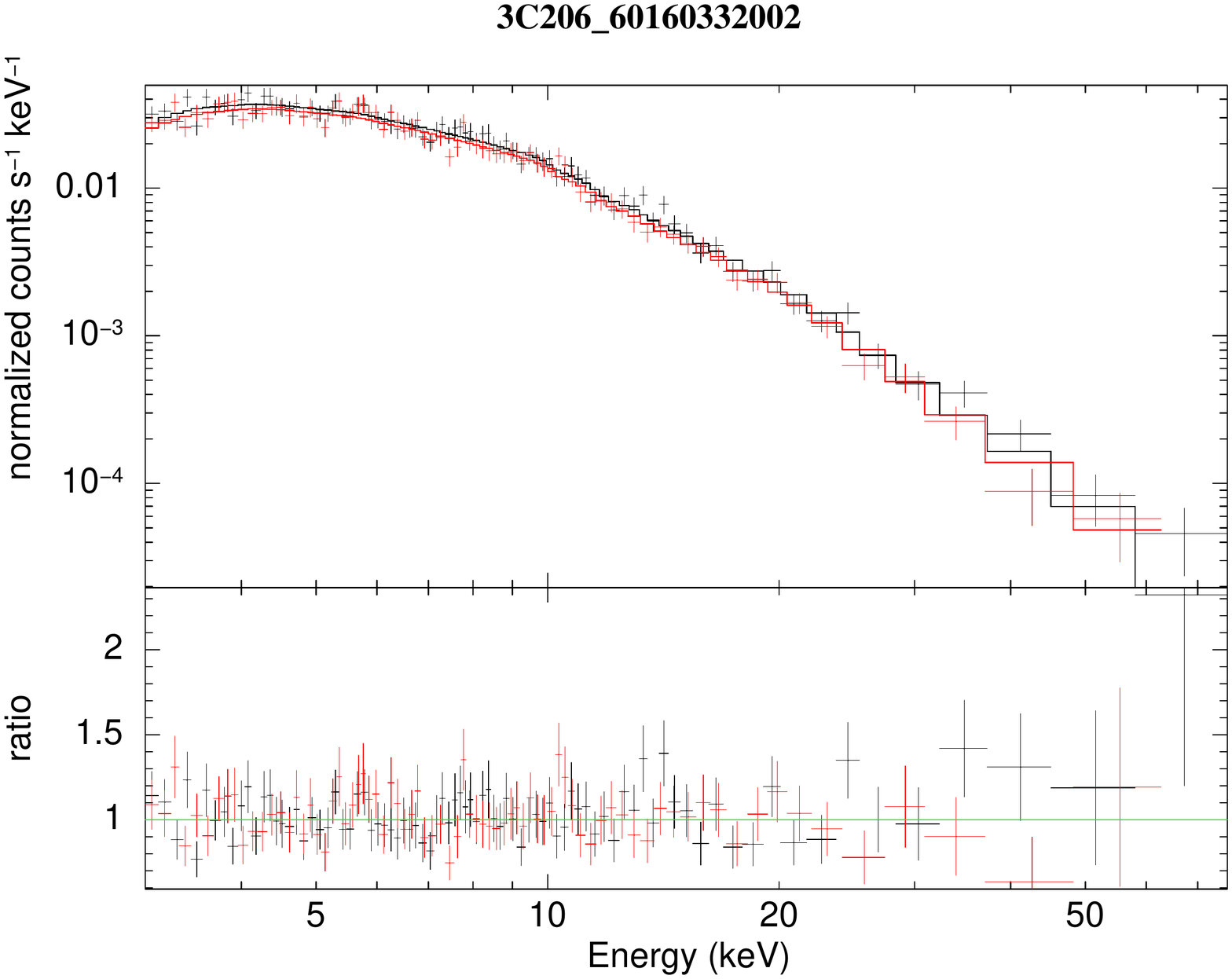}}\\
\subfloat{\includegraphics[width=0.33\textwidth]{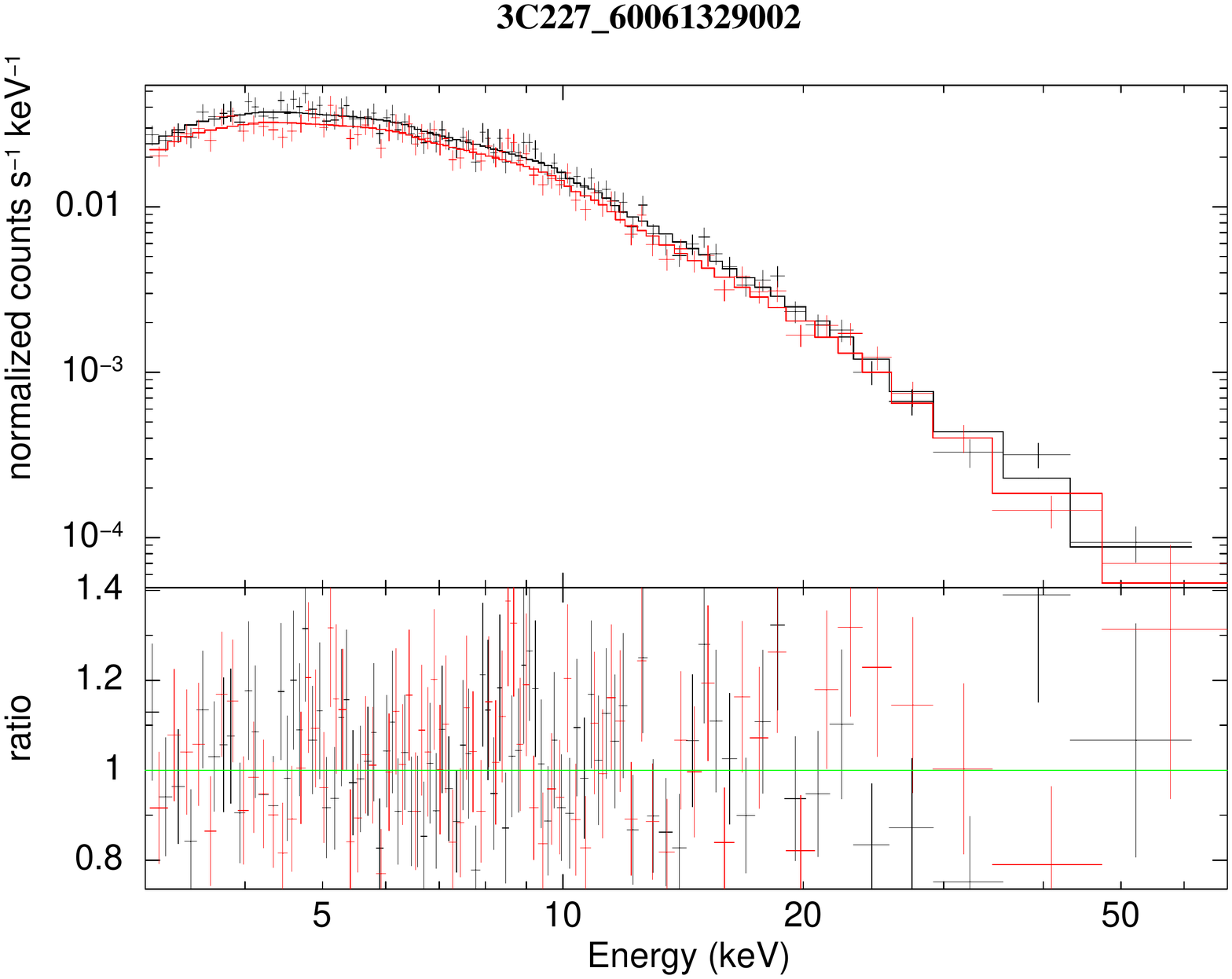}}
\subfloat{\includegraphics[width=0.33\textwidth]{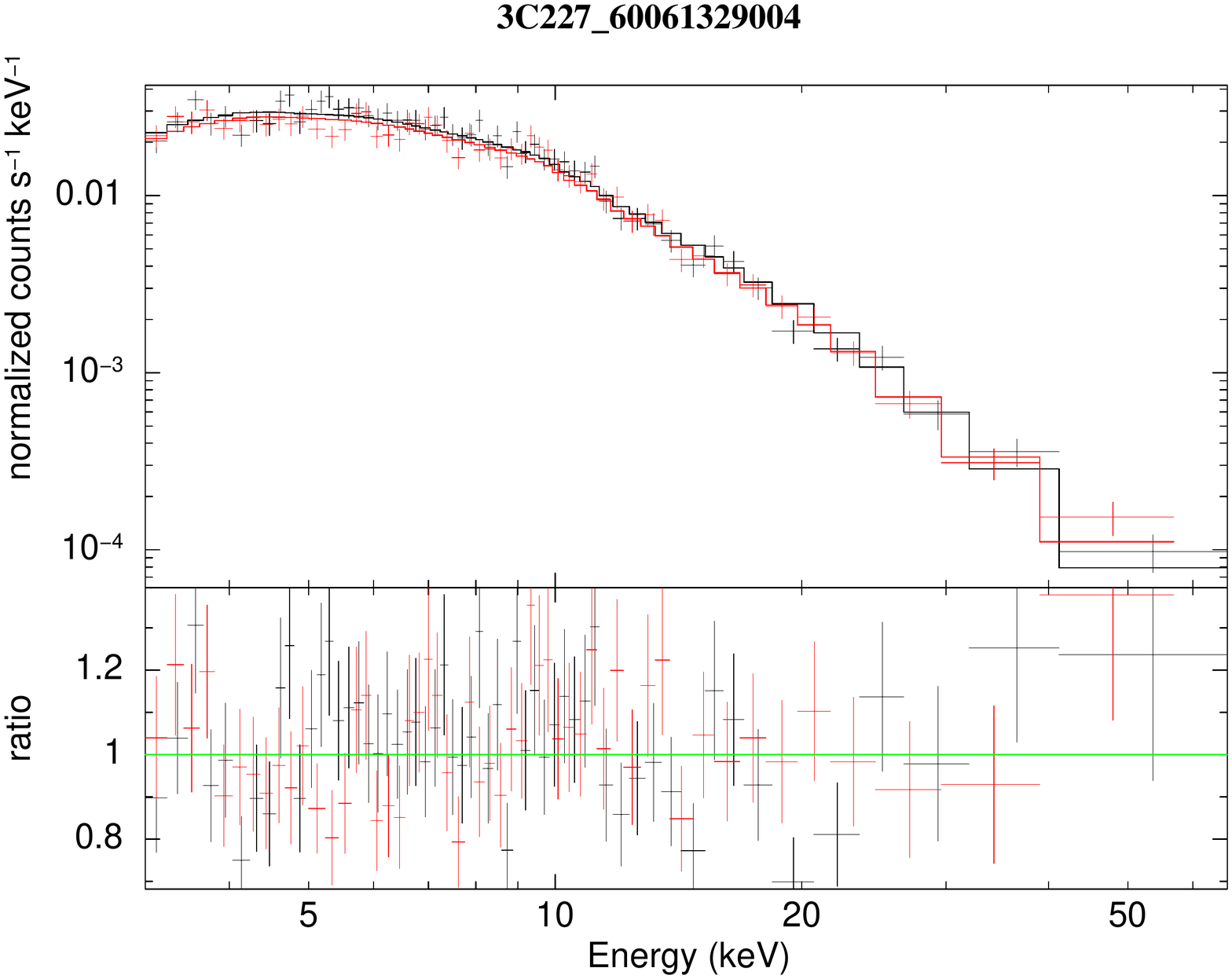}}
\subfloat{\includegraphics[width=0.33\textwidth]{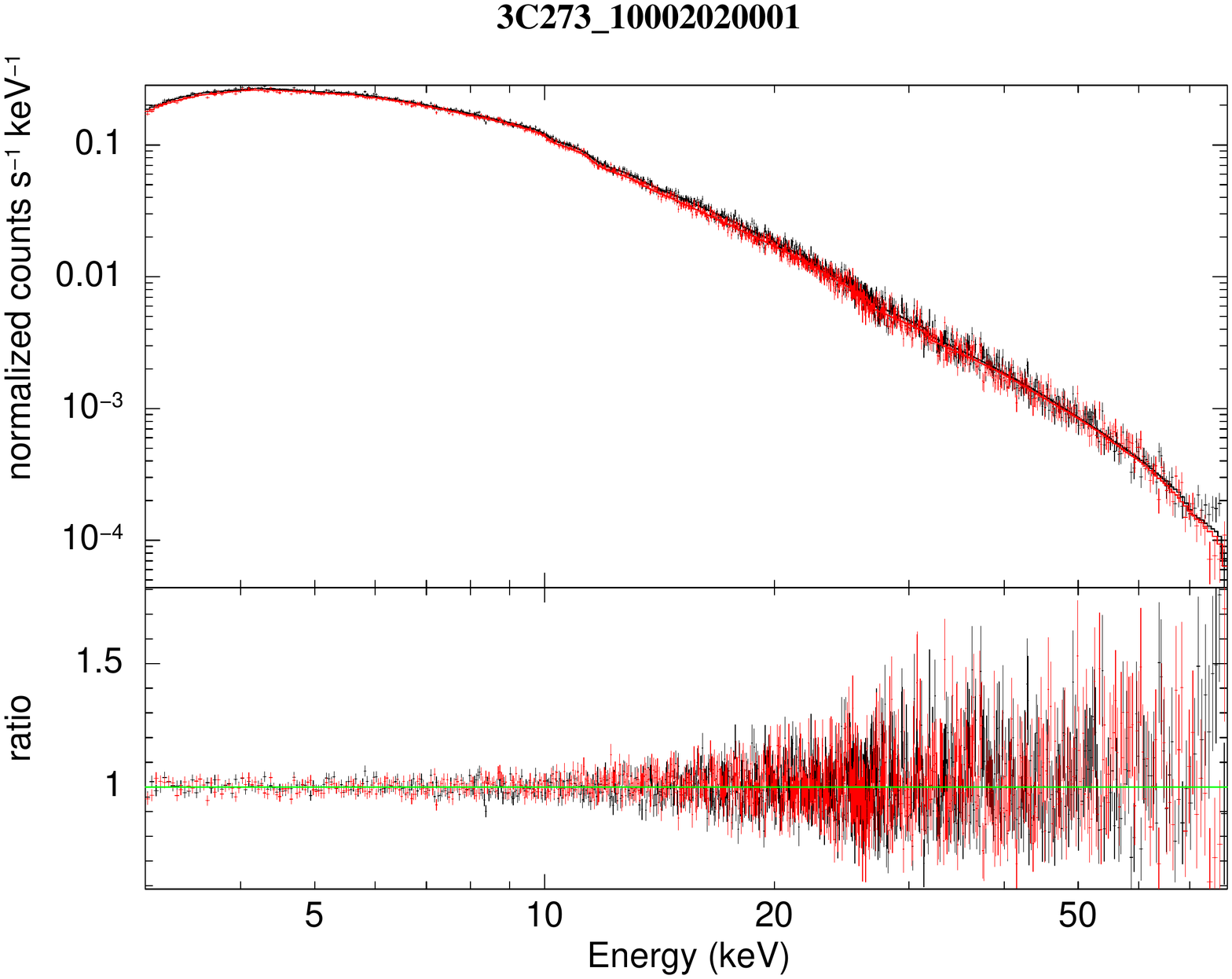}}\\
\subfloat{\includegraphics[width=0.33\textwidth]{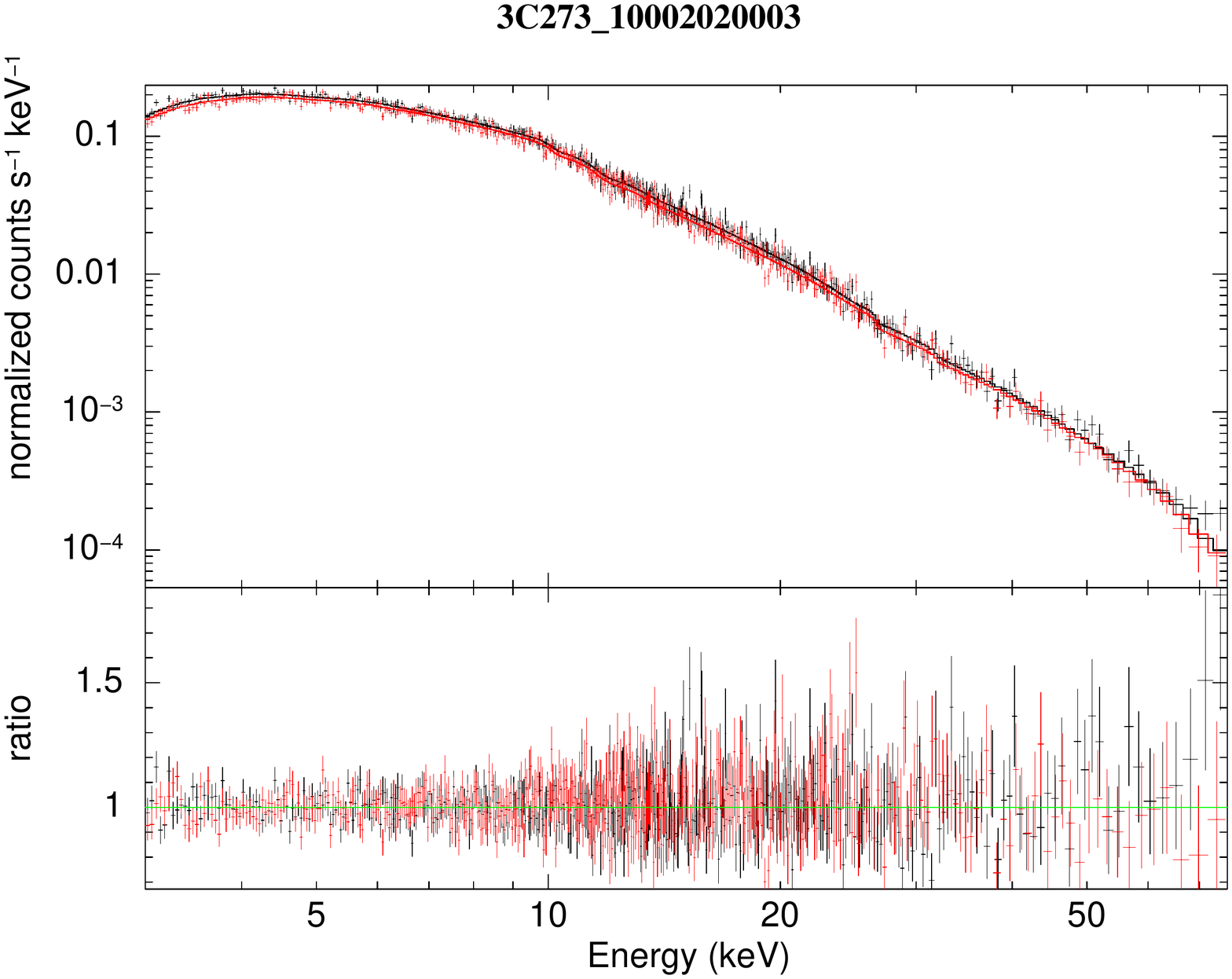}}
\subfloat{\includegraphics[width=0.33\textwidth]{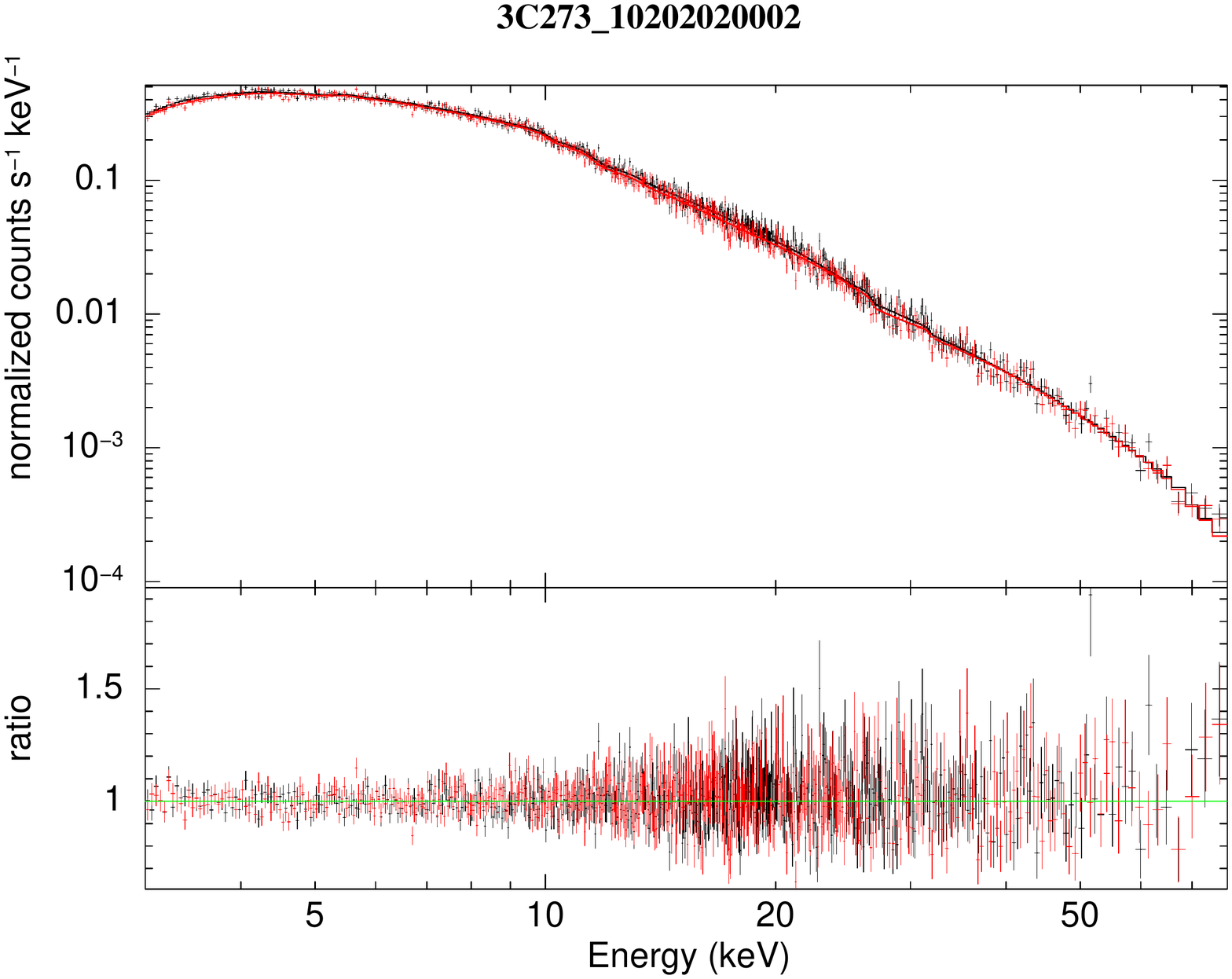}}
\subfloat{\includegraphics[width=0.33\textwidth]{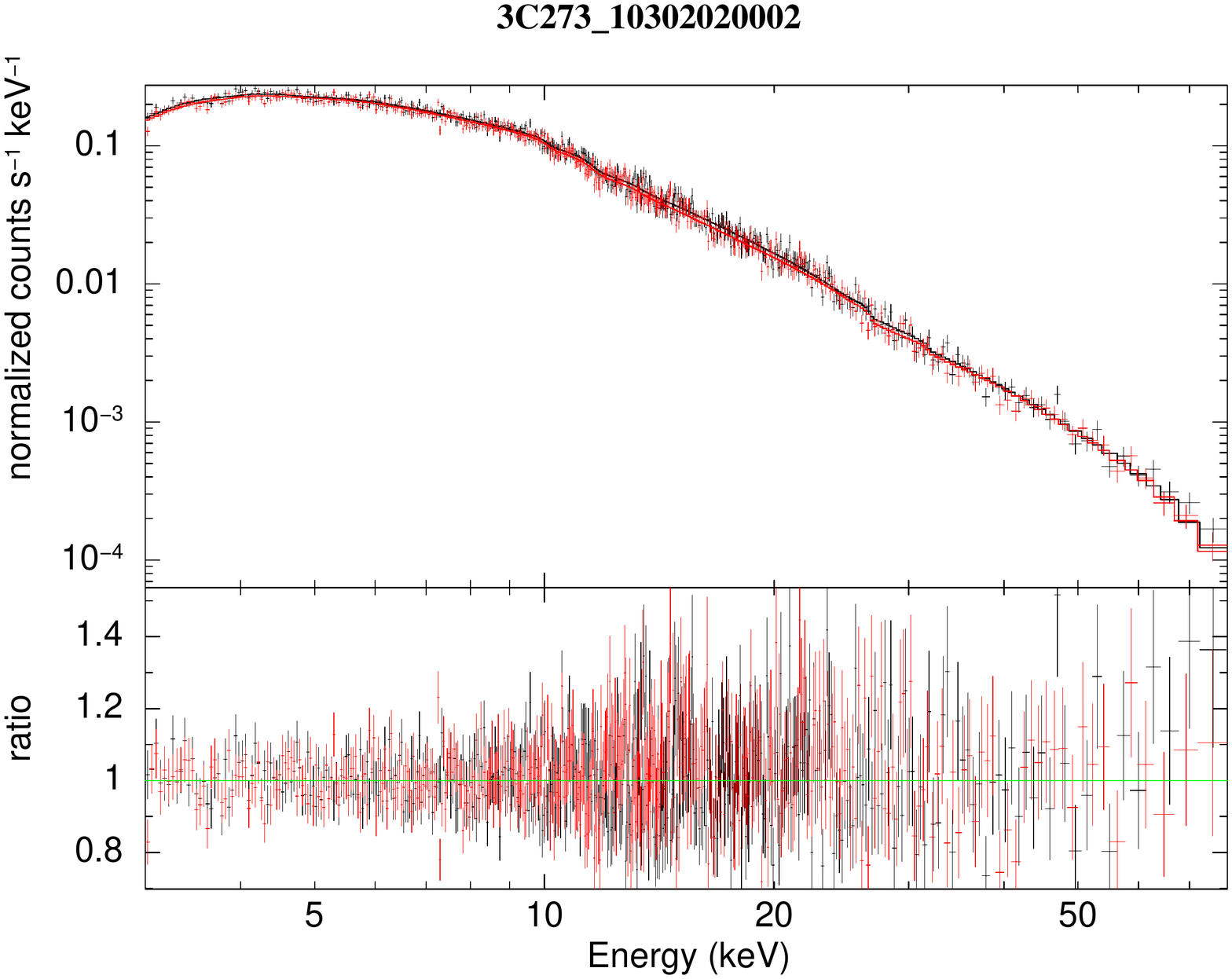}}\\

\end{figure*}
\addtocounter{figure}{-1} 
\begin{figure*}[!t]
\addtocounter{figure}{1} 
\centering
\ContinuedFloat
\subfloat{\includegraphics[width=0.33\textwidth]{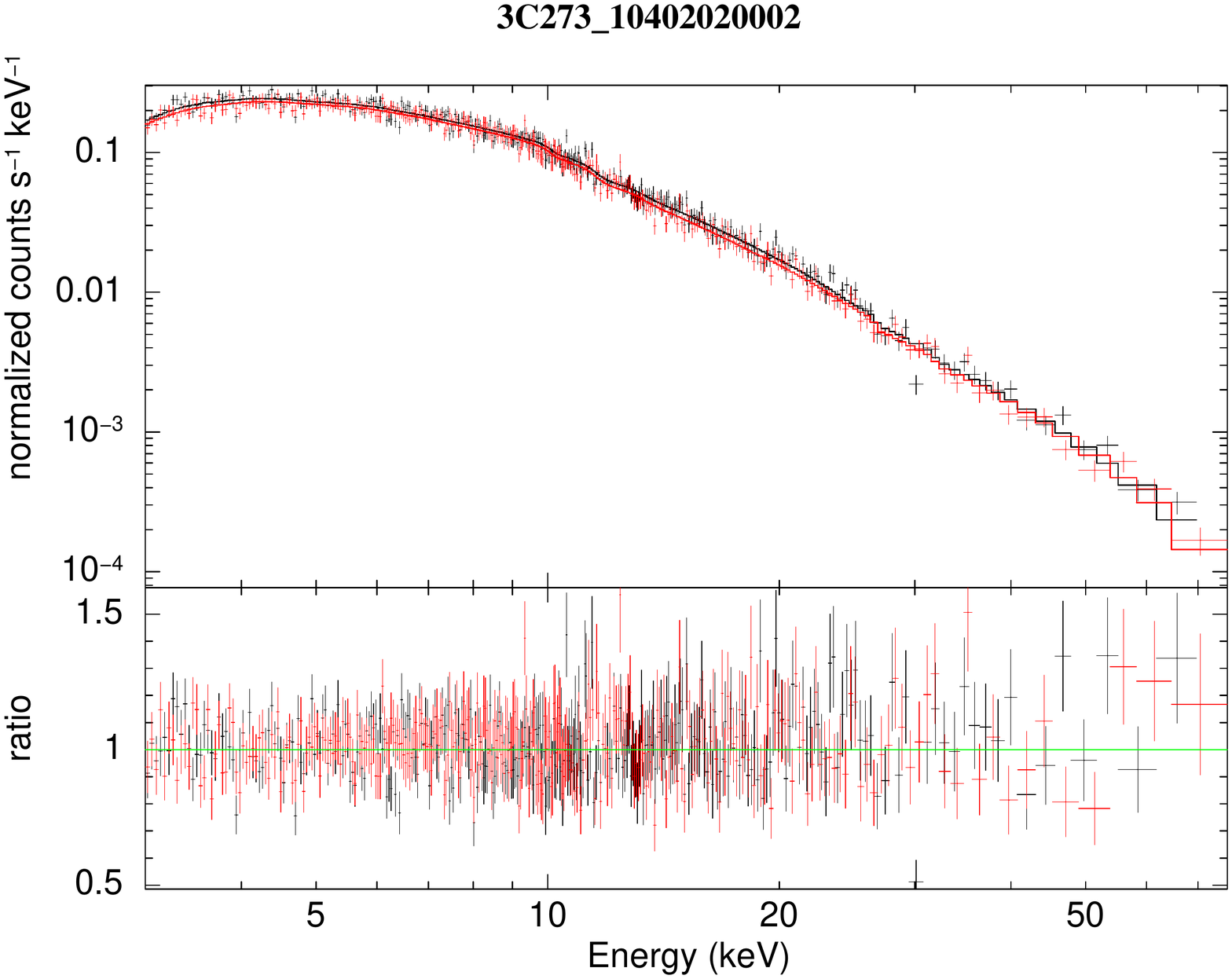}}
\subfloat{\includegraphics[width=0.33\textwidth]{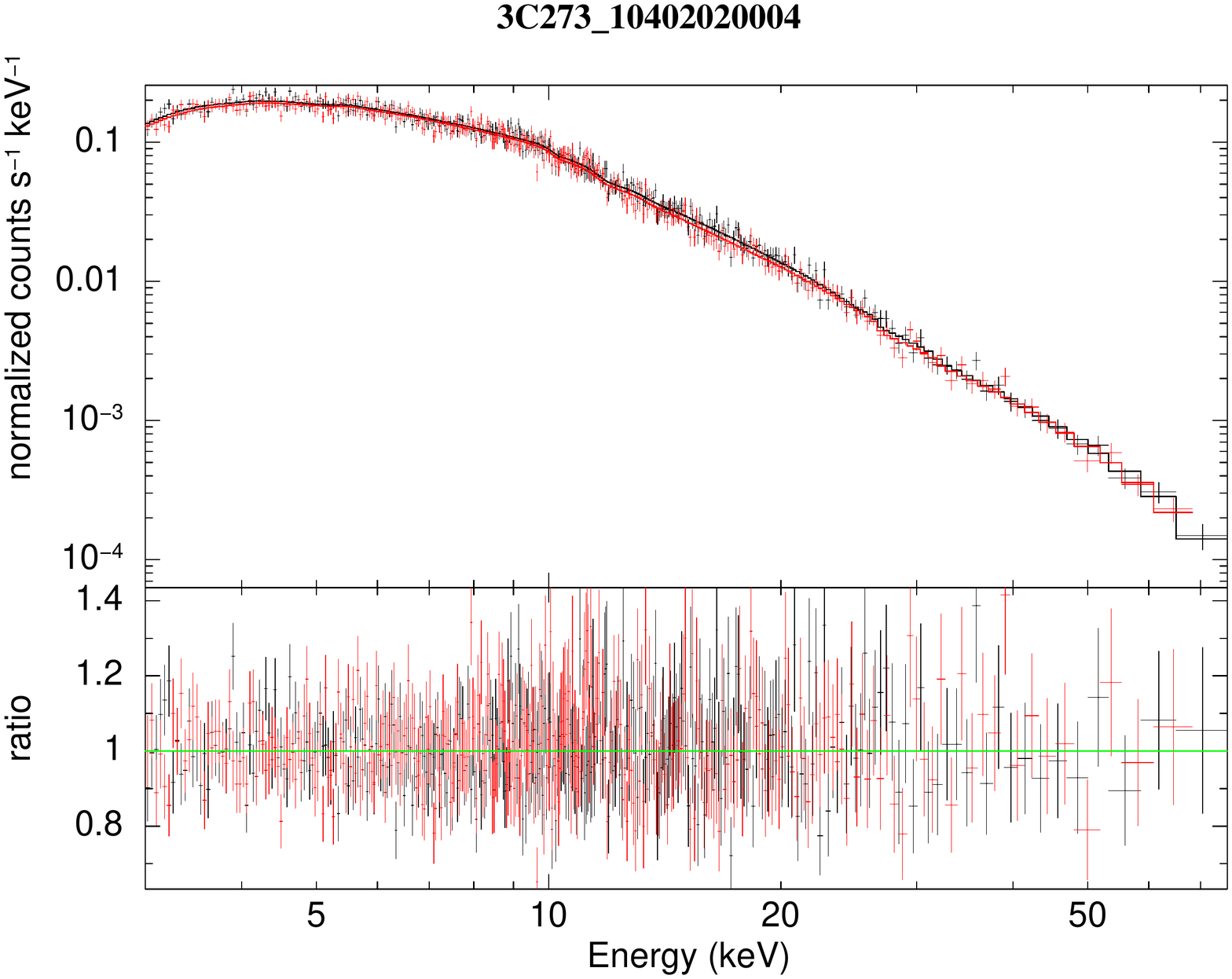}}
\subfloat{\includegraphics[width=0.33\textwidth]{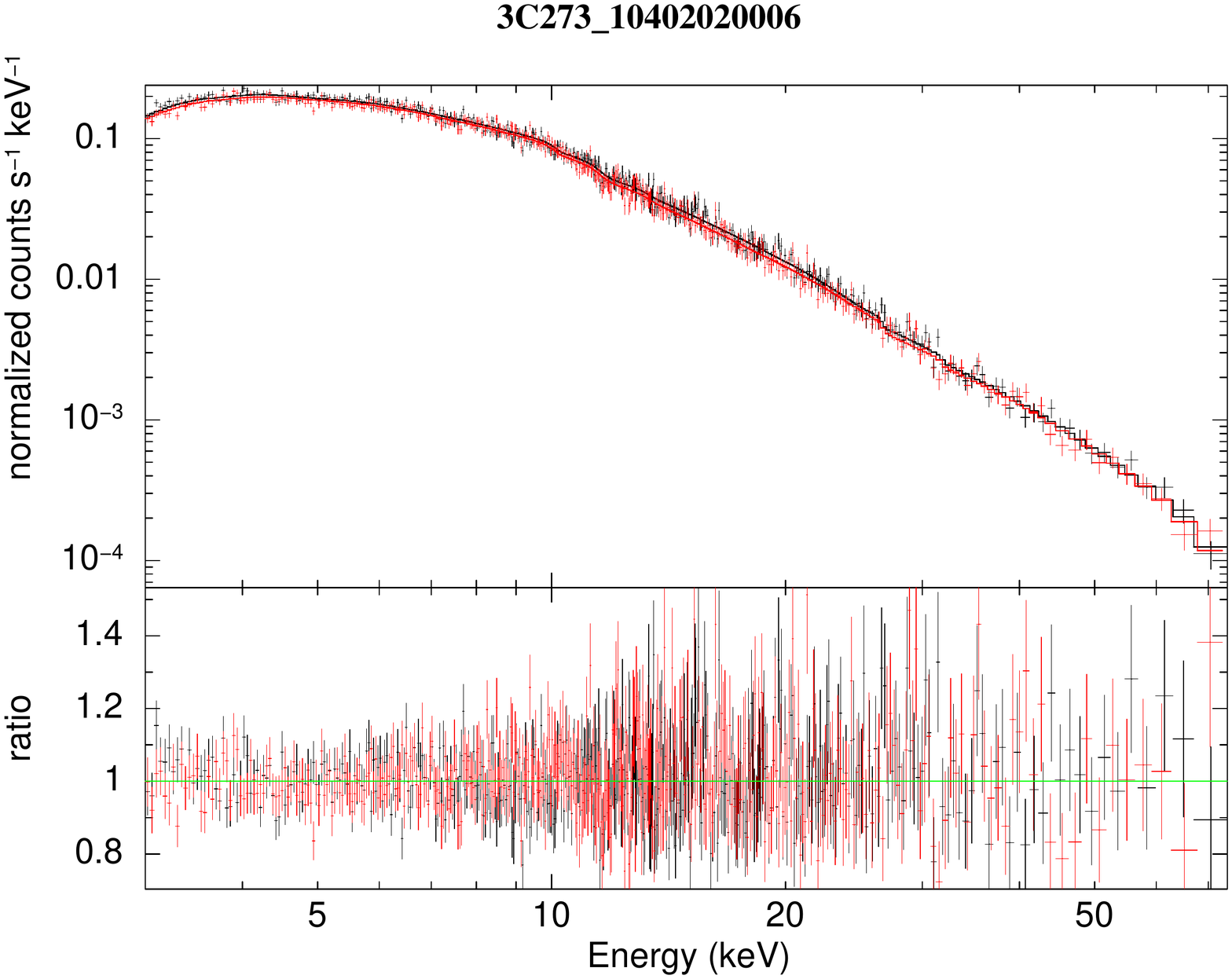}}\\
\subfloat{\includegraphics[width=0.33\textwidth]{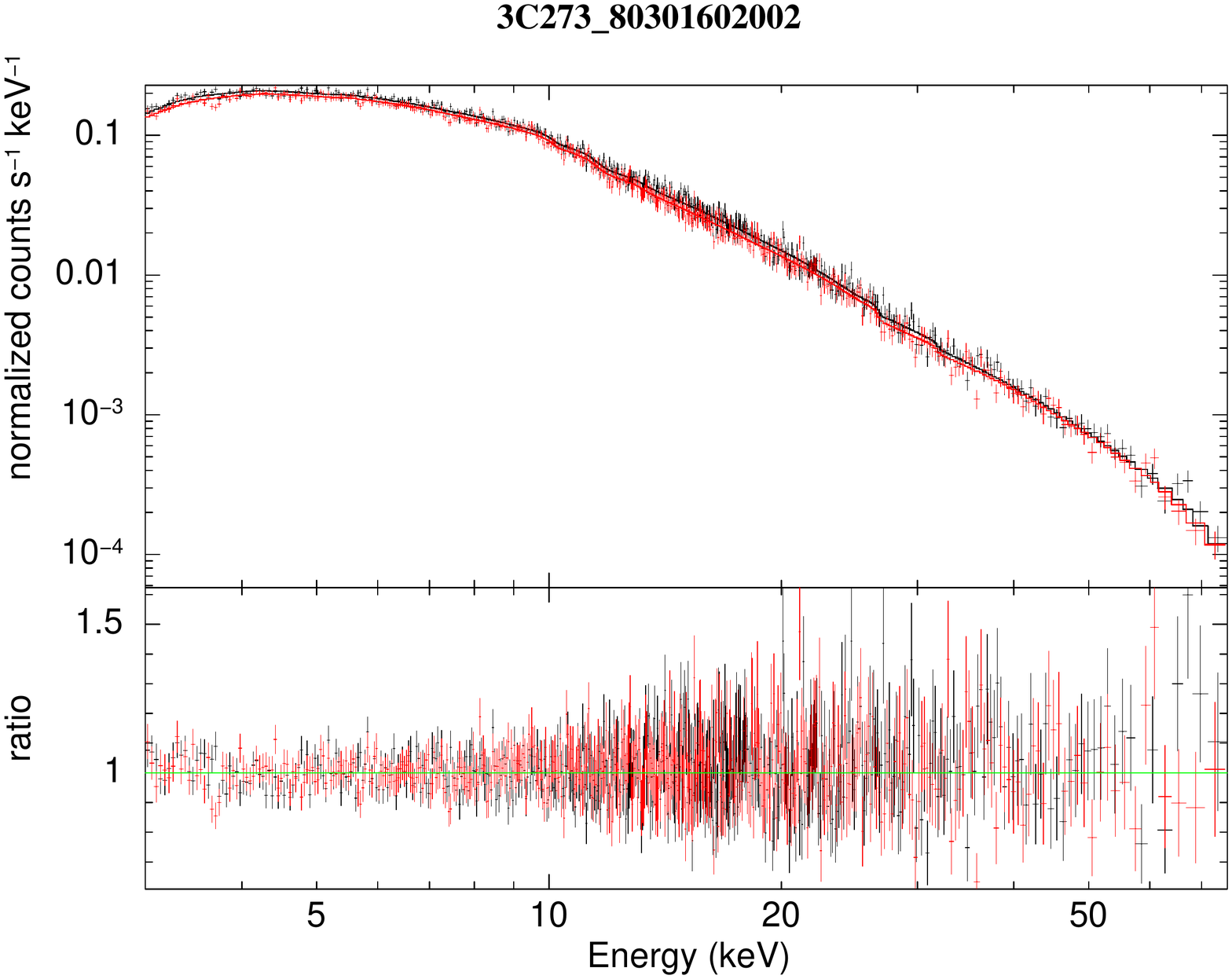}}
\subfloat{\includegraphics[width=0.33\textwidth]{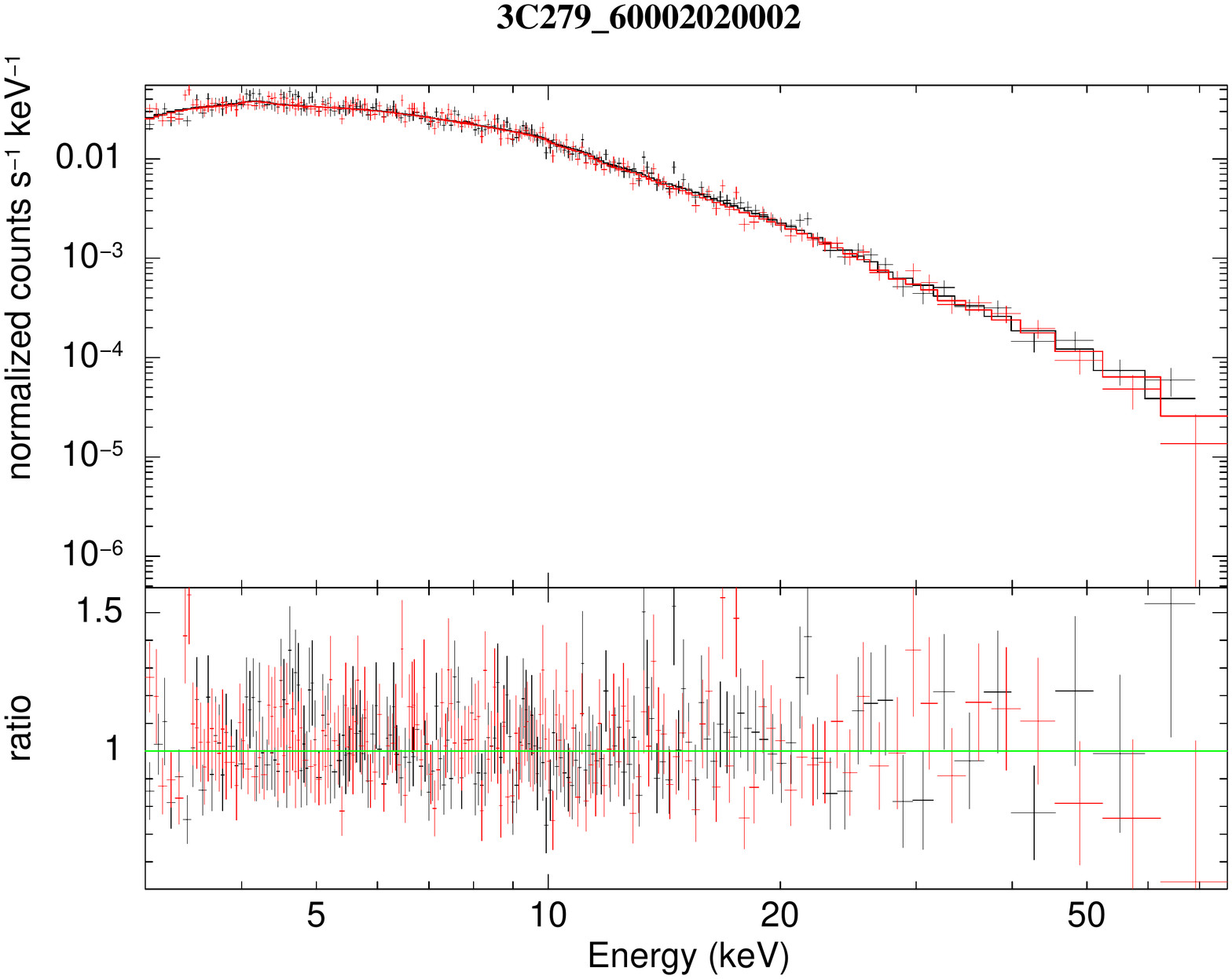}}
\subfloat{\includegraphics[width=0.33\textwidth]{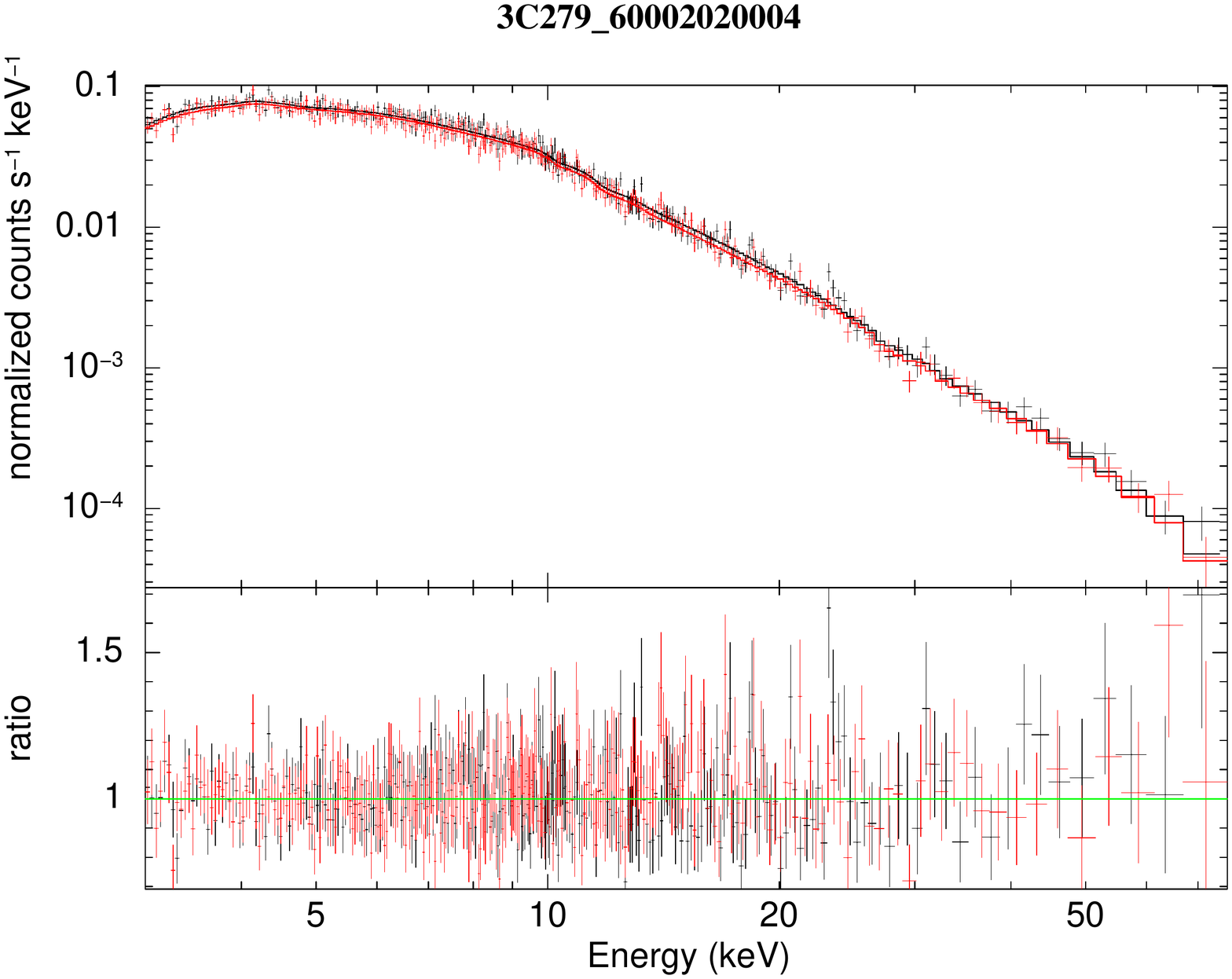}}\\
\subfloat{\includegraphics[width=0.33\textwidth]{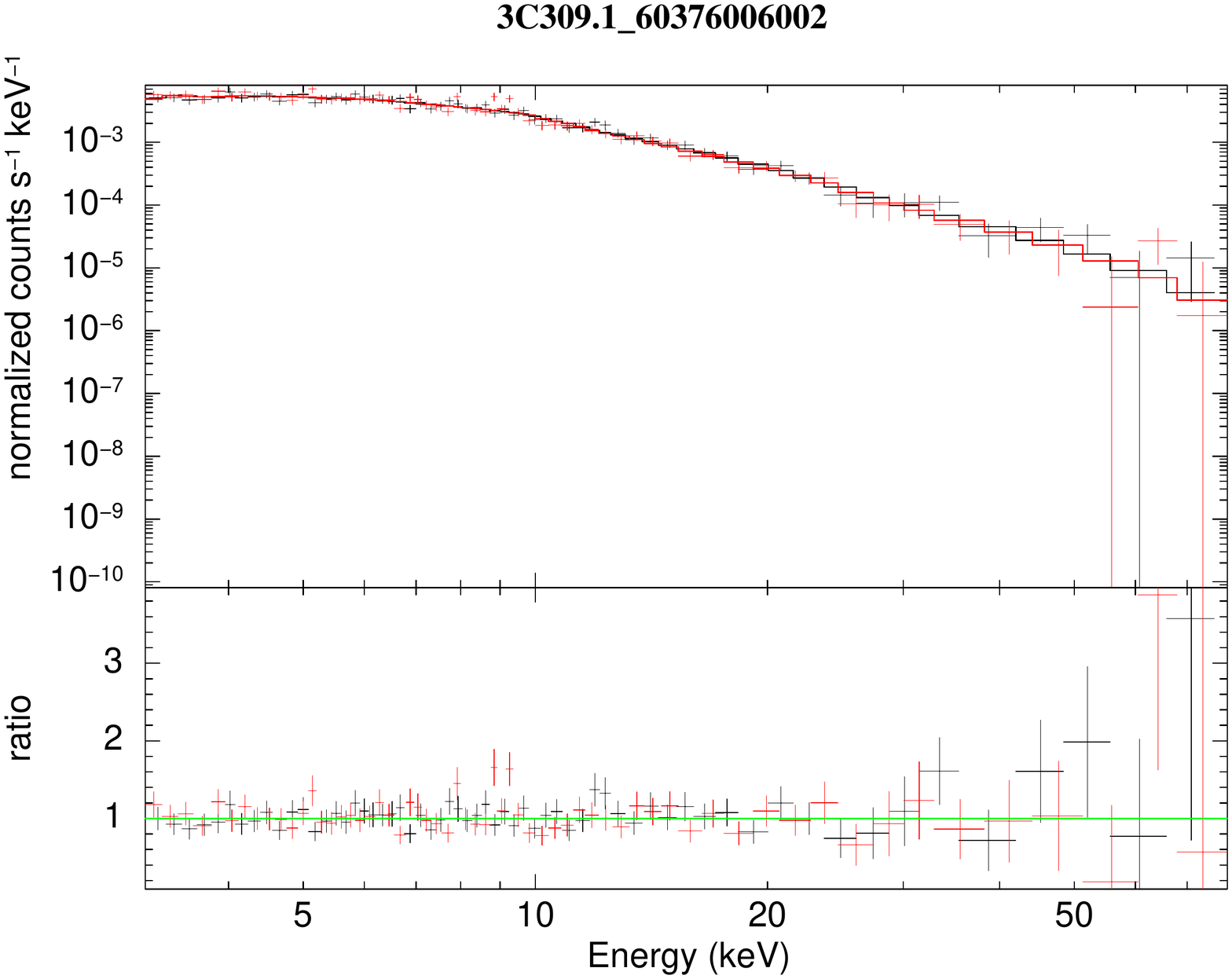}}
\subfloat{\includegraphics[width=0.33\textwidth]{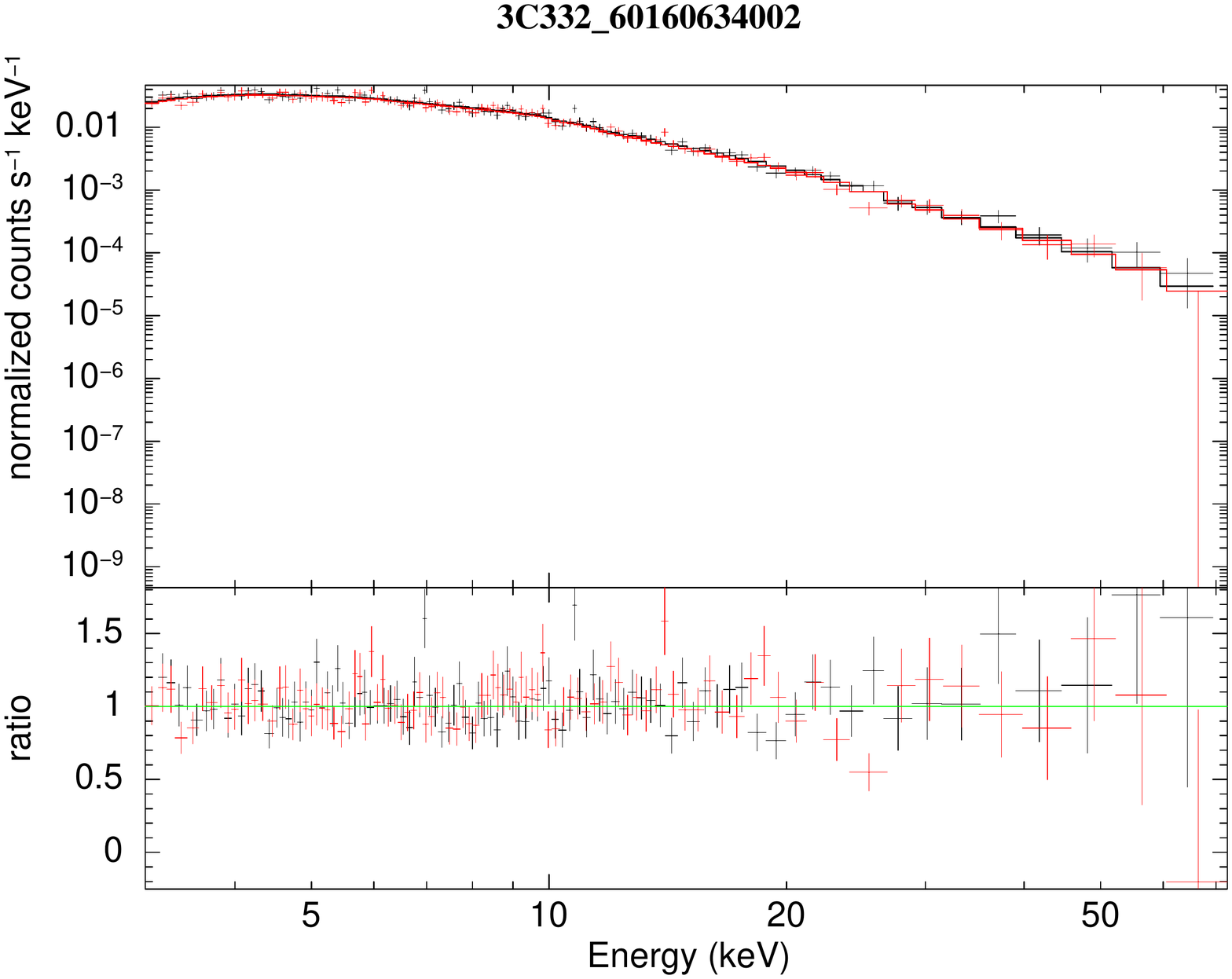}}
\subfloat{\includegraphics[width=0.33\textwidth]{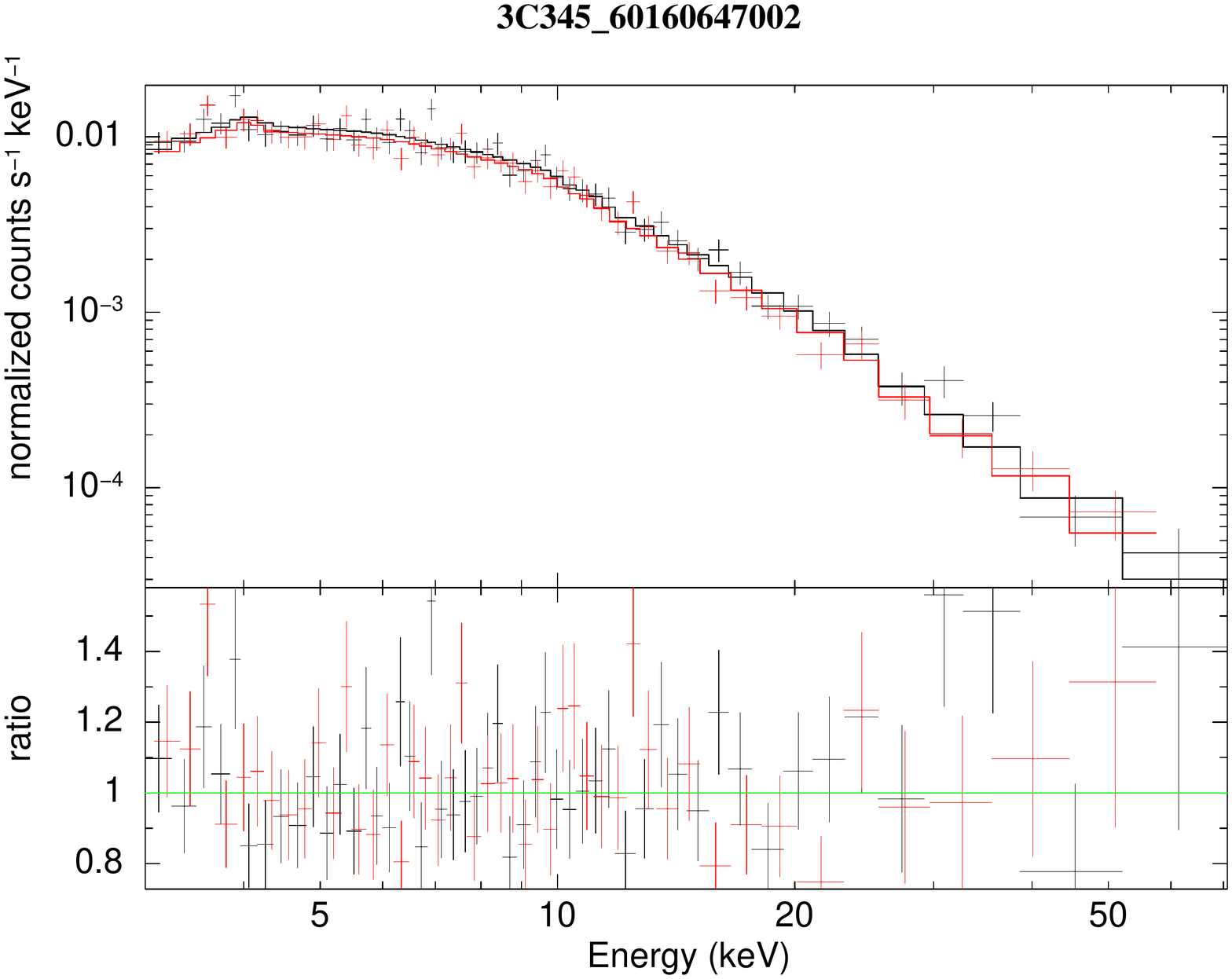}}\\
\subfloat{\includegraphics[width=0.33\textwidth]{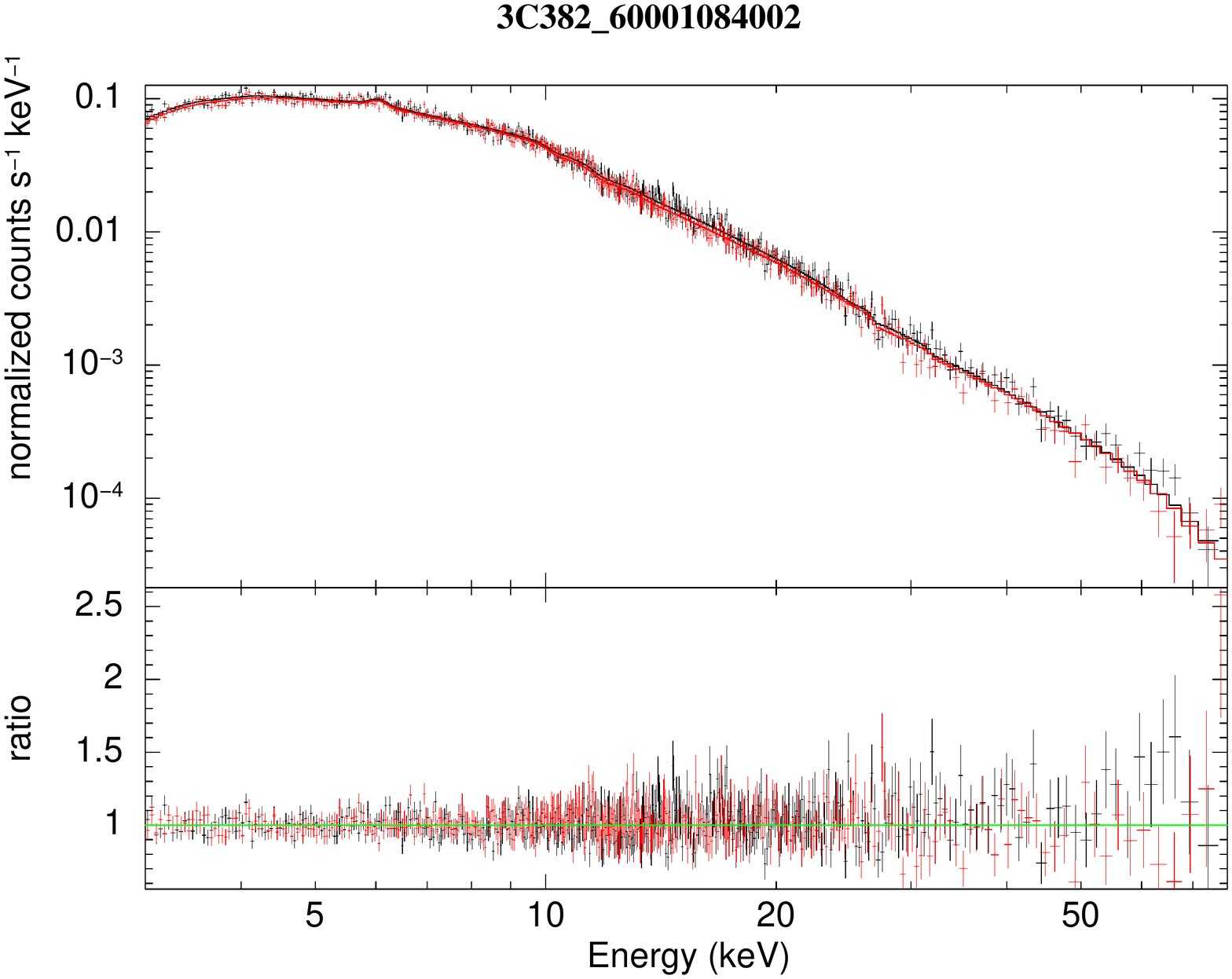}}
\subfloat{\includegraphics[width=0.33\textwidth]{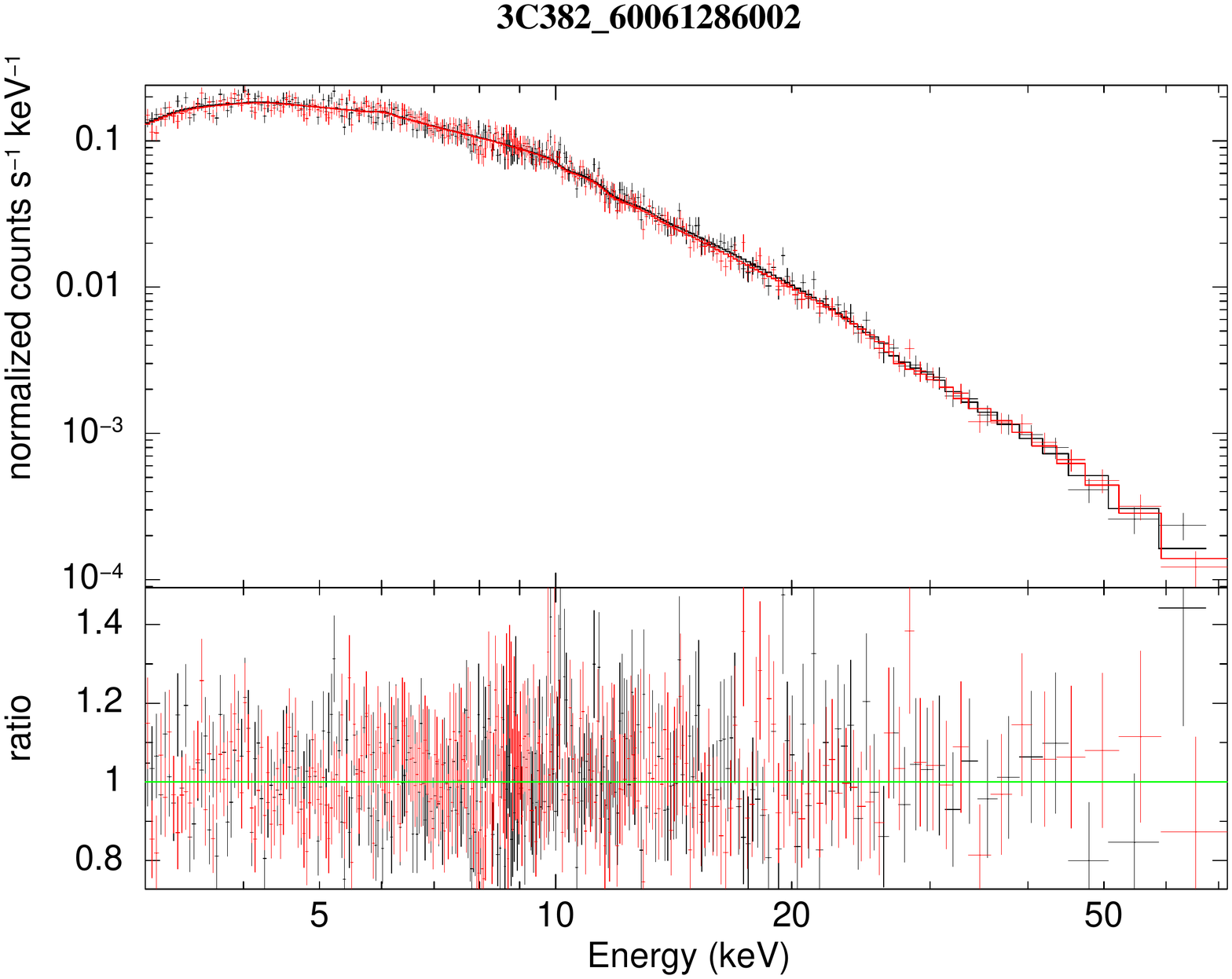}}
\subfloat{\includegraphics[width=0.33\textwidth]{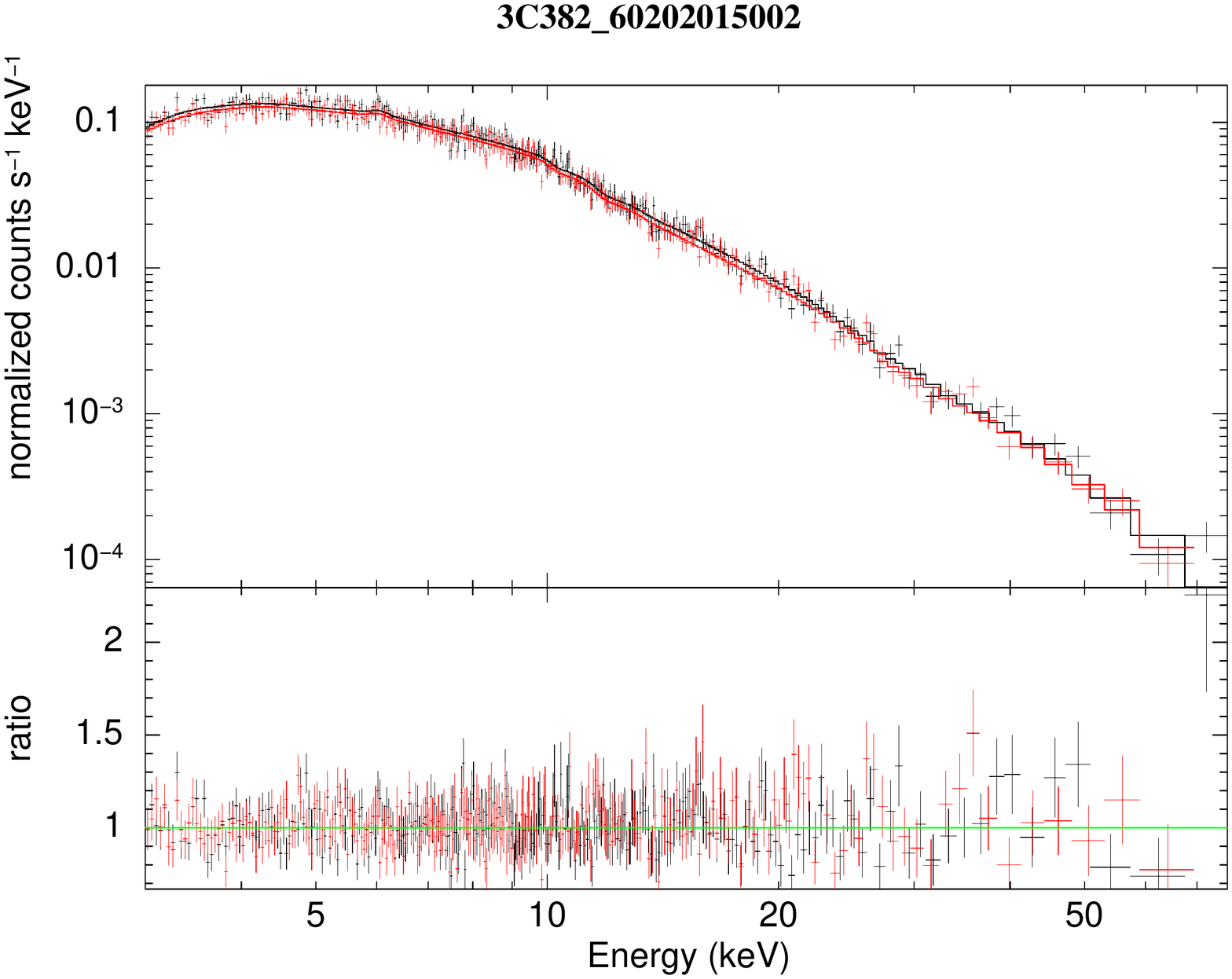}}\\
\subfloat{\includegraphics[width=0.33\textwidth]{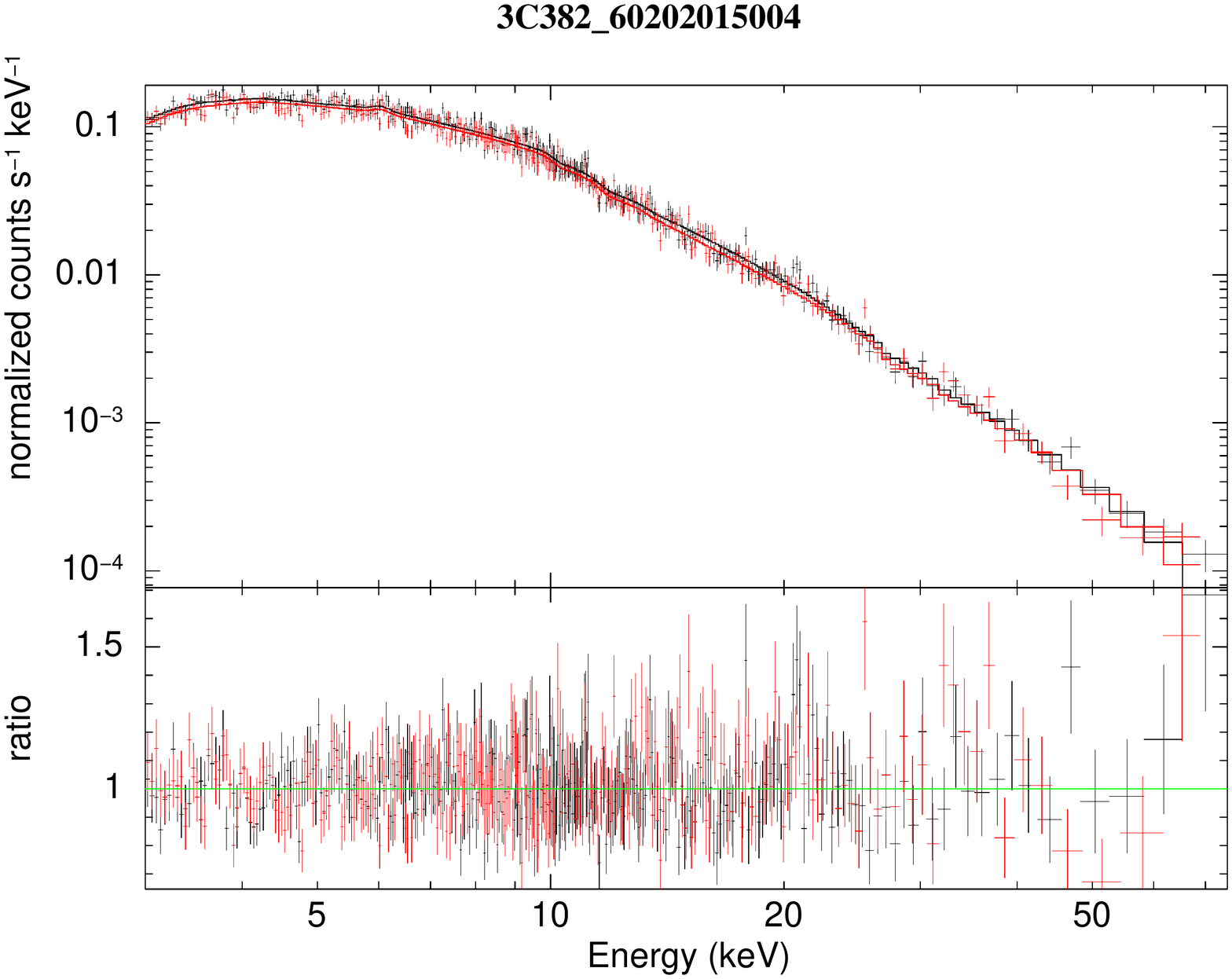}}
\subfloat{\includegraphics[width=0.33\textwidth]{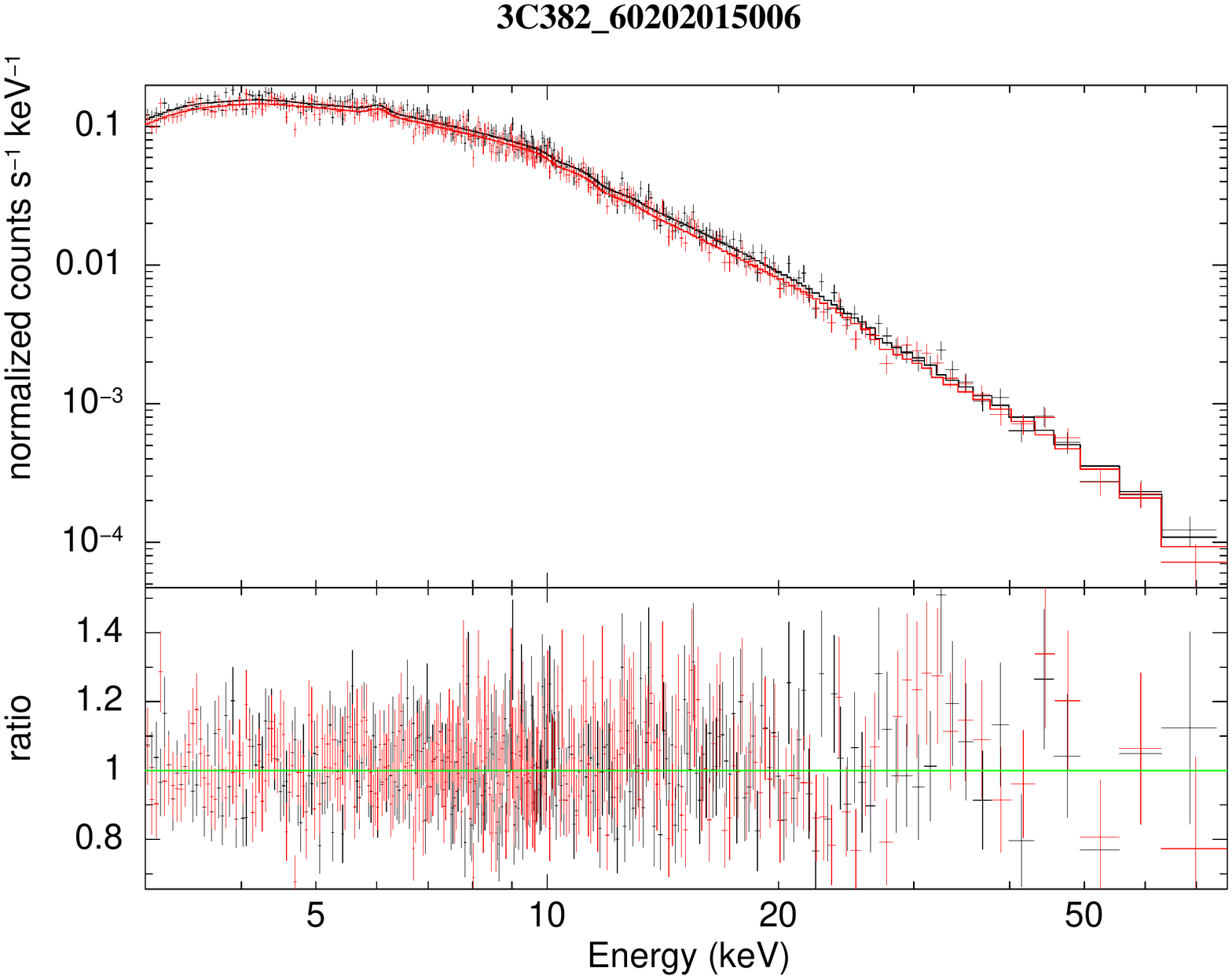}}
\subfloat{\includegraphics[width=0.33\textwidth]{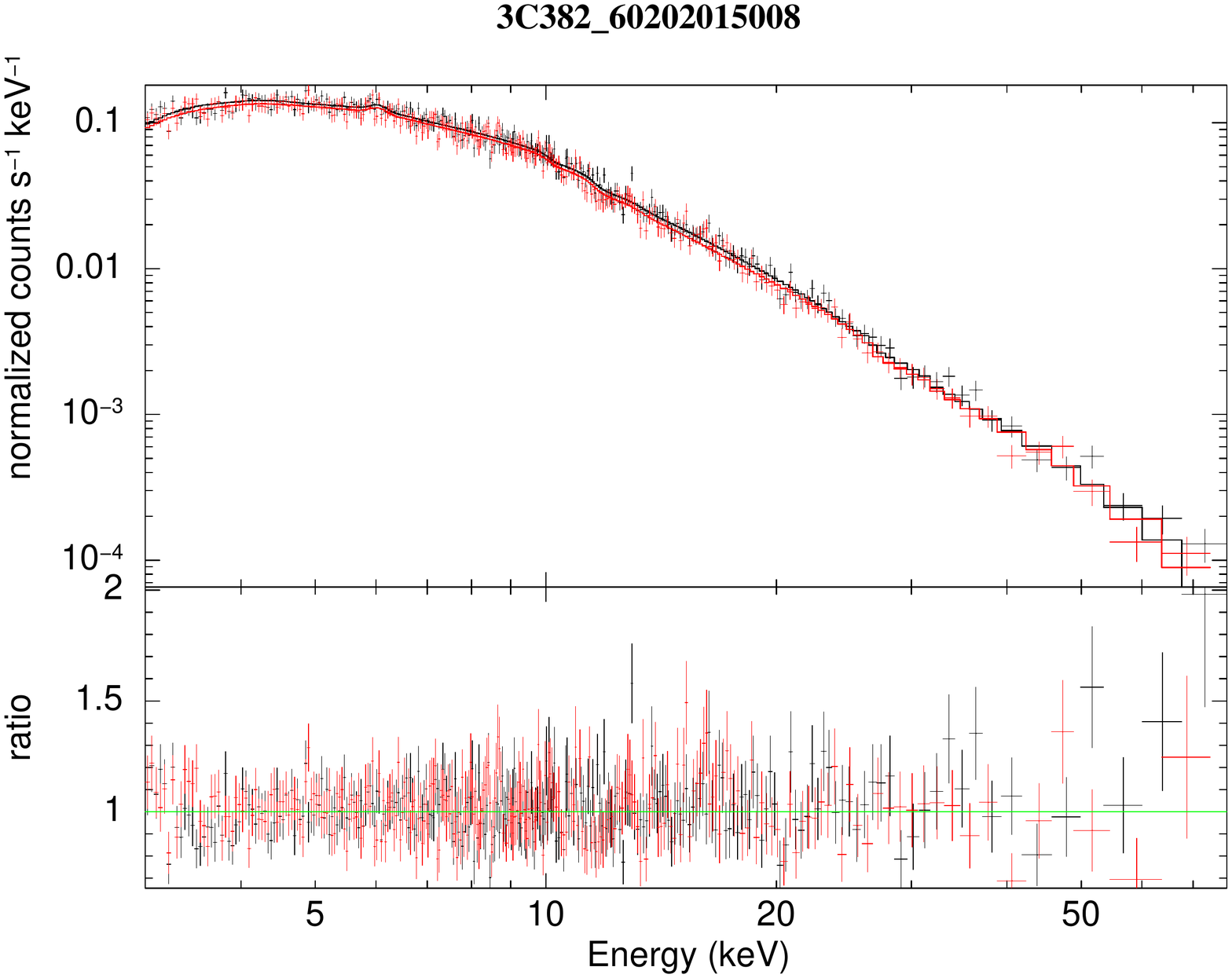}}\\
\end{figure*}

\addtocounter{figure}{-1} 
\begin{figure*}[!t]
\addtocounter{figure}{1} 
\centering
\ContinuedFloat
\subfloat{\includegraphics[width=0.33\textwidth]{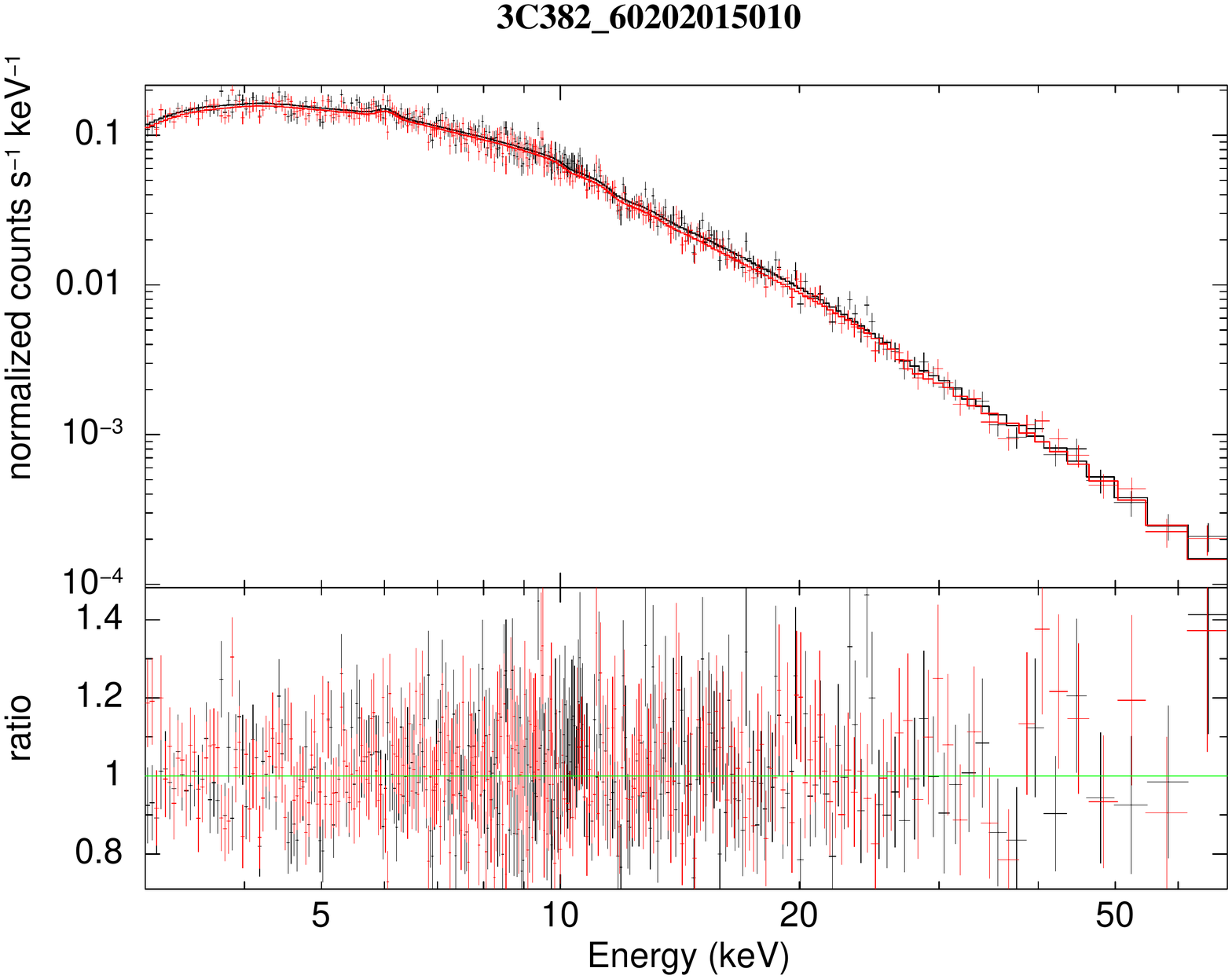}}
\subfloat{\includegraphics[width=0.33\textwidth]{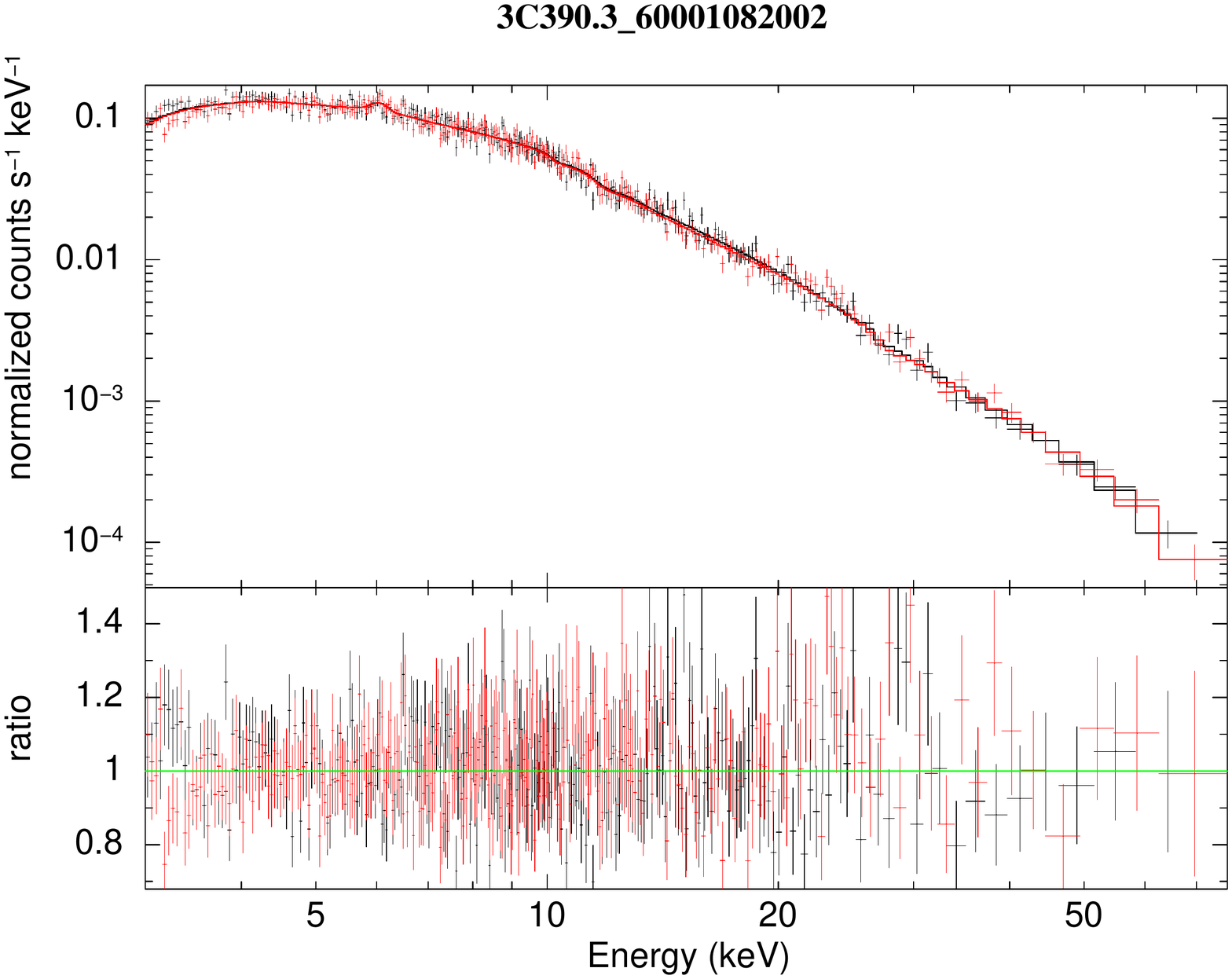}}
\subfloat{\includegraphics[width=0.33\textwidth]{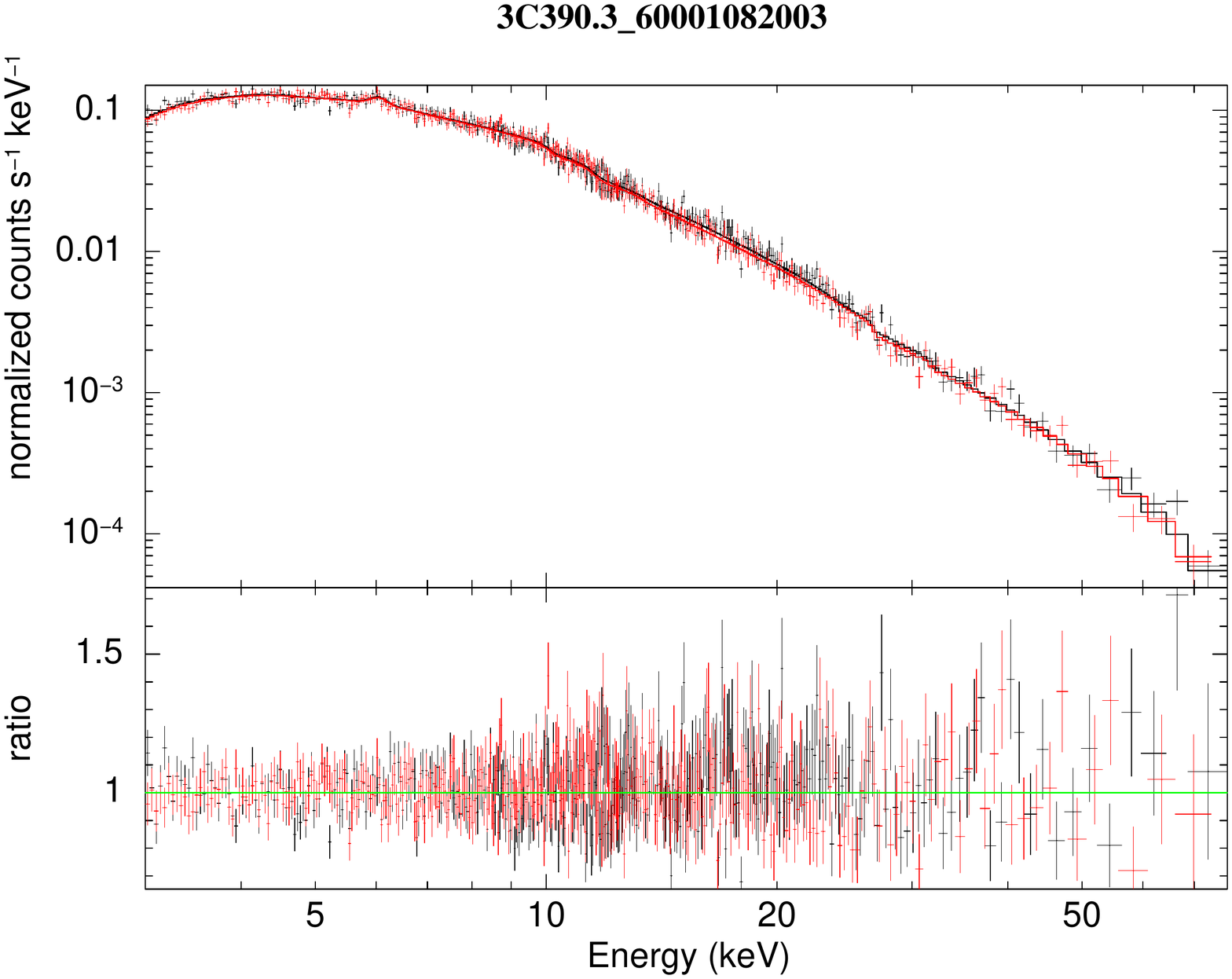}}\\
\subfloat{\includegraphics[width=0.33\textwidth]{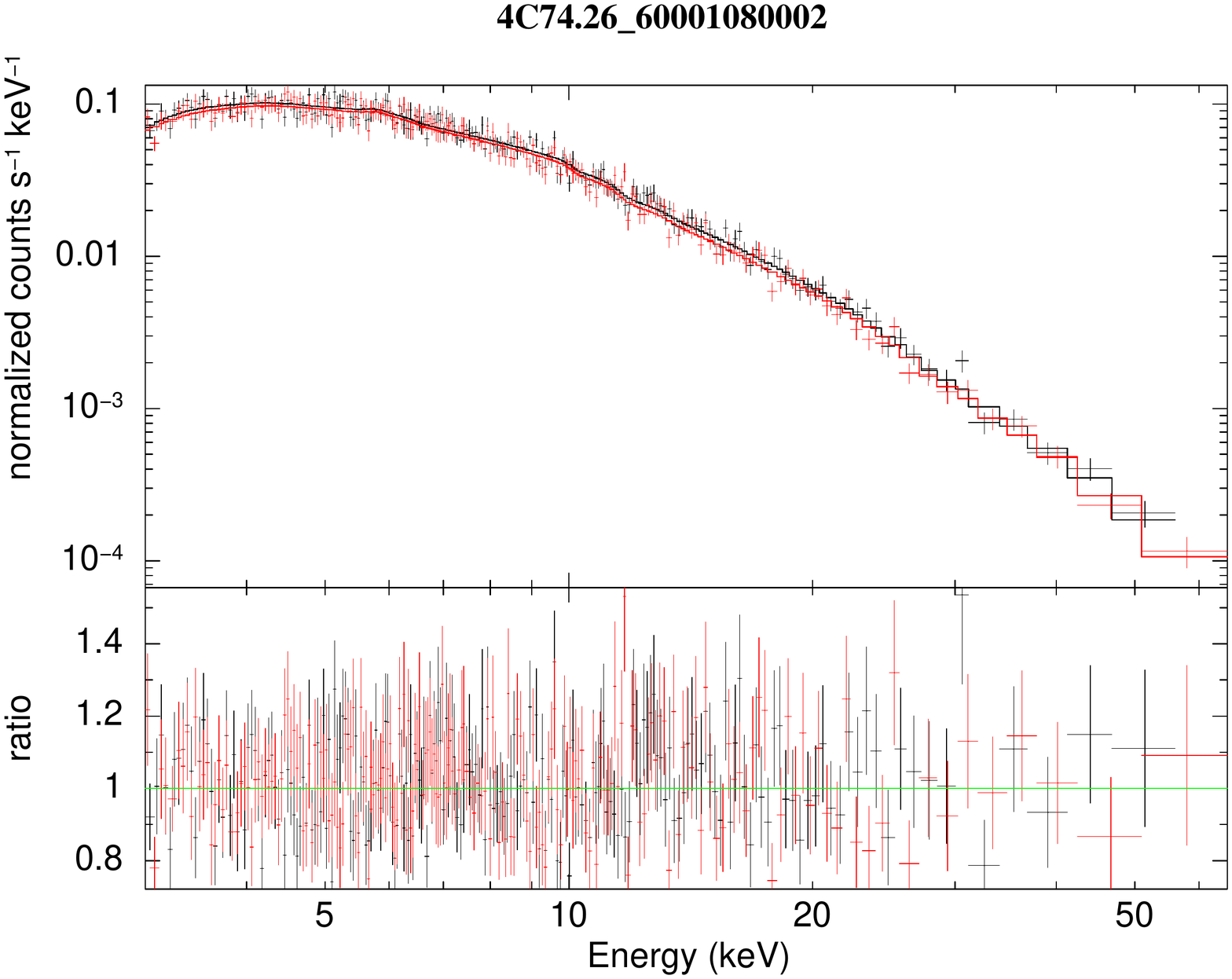}}
\subfloat{\includegraphics[width=0.33\textwidth]{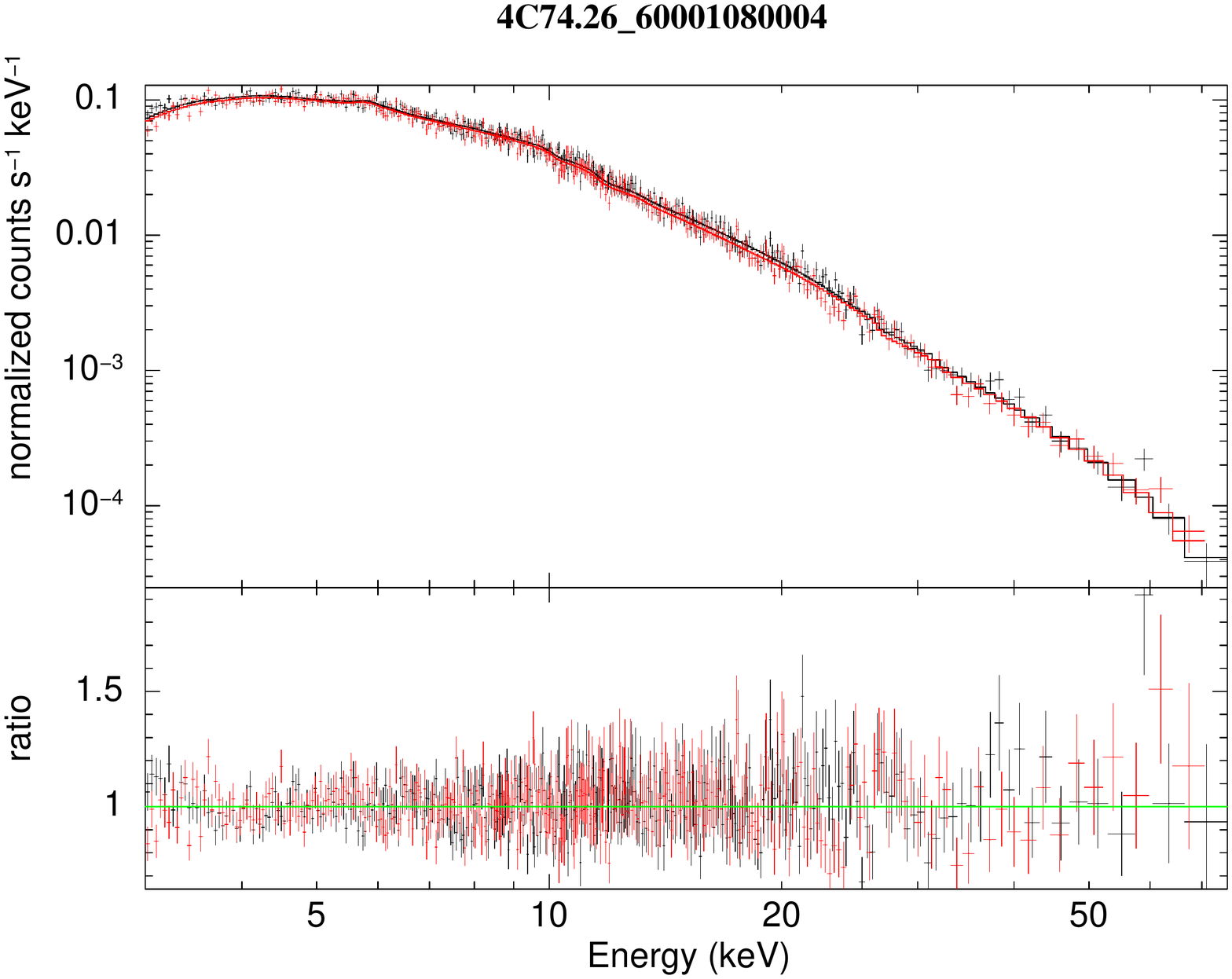}}
\subfloat{\includegraphics[width=0.33\textwidth]{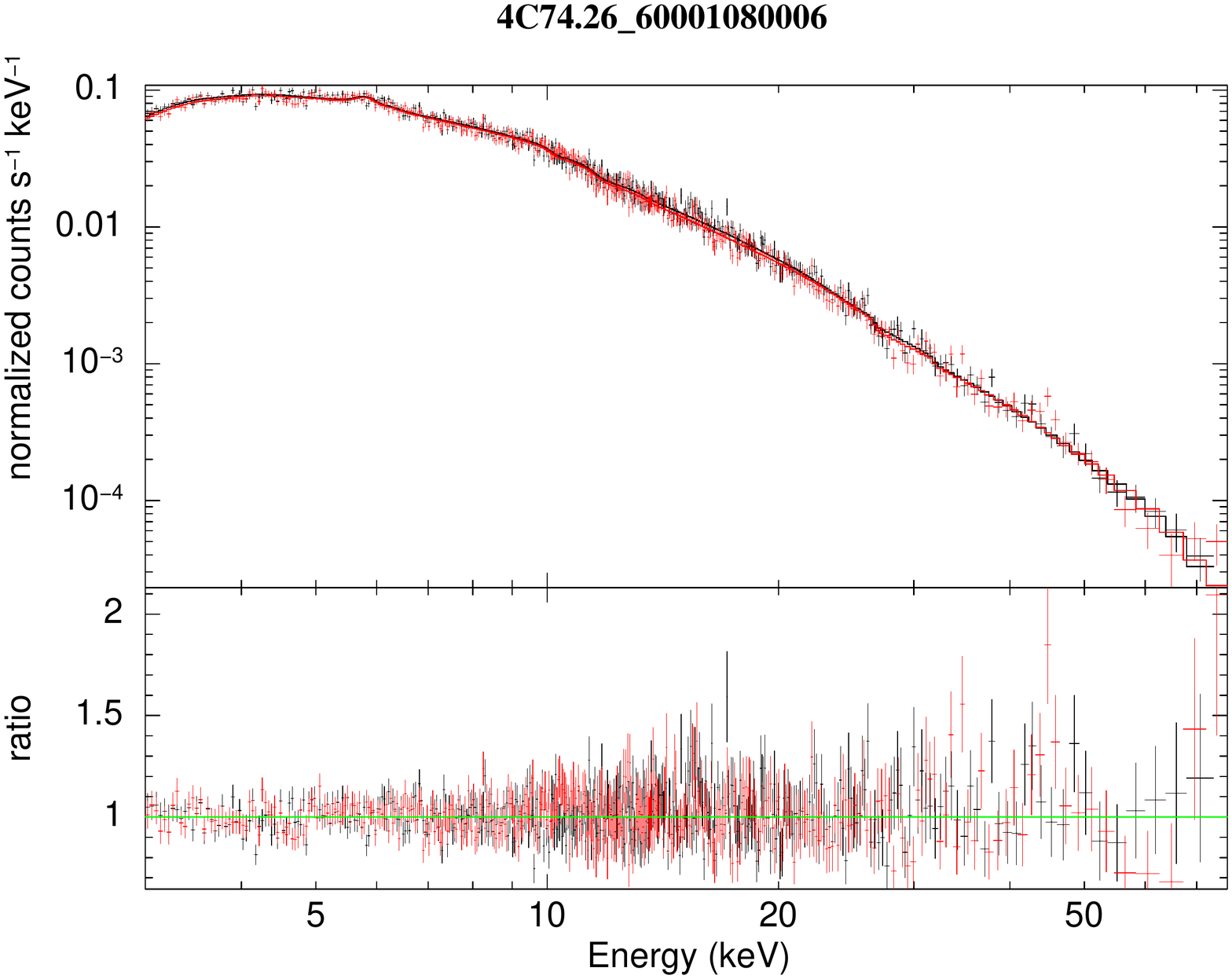}}\\
\subfloat{\includegraphics[width=0.33\textwidth]{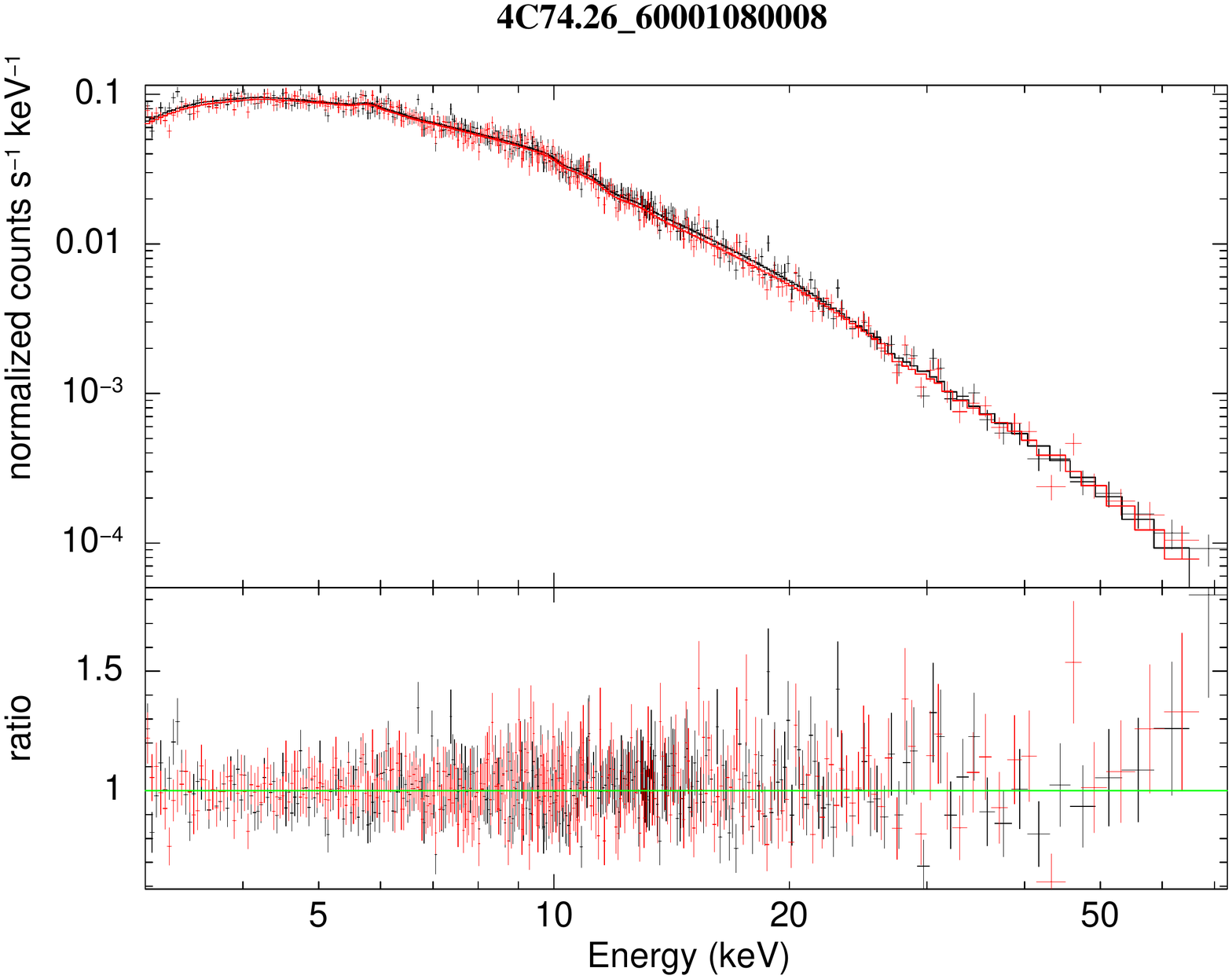}}
\subfloat{\includegraphics[width=0.33\textwidth]{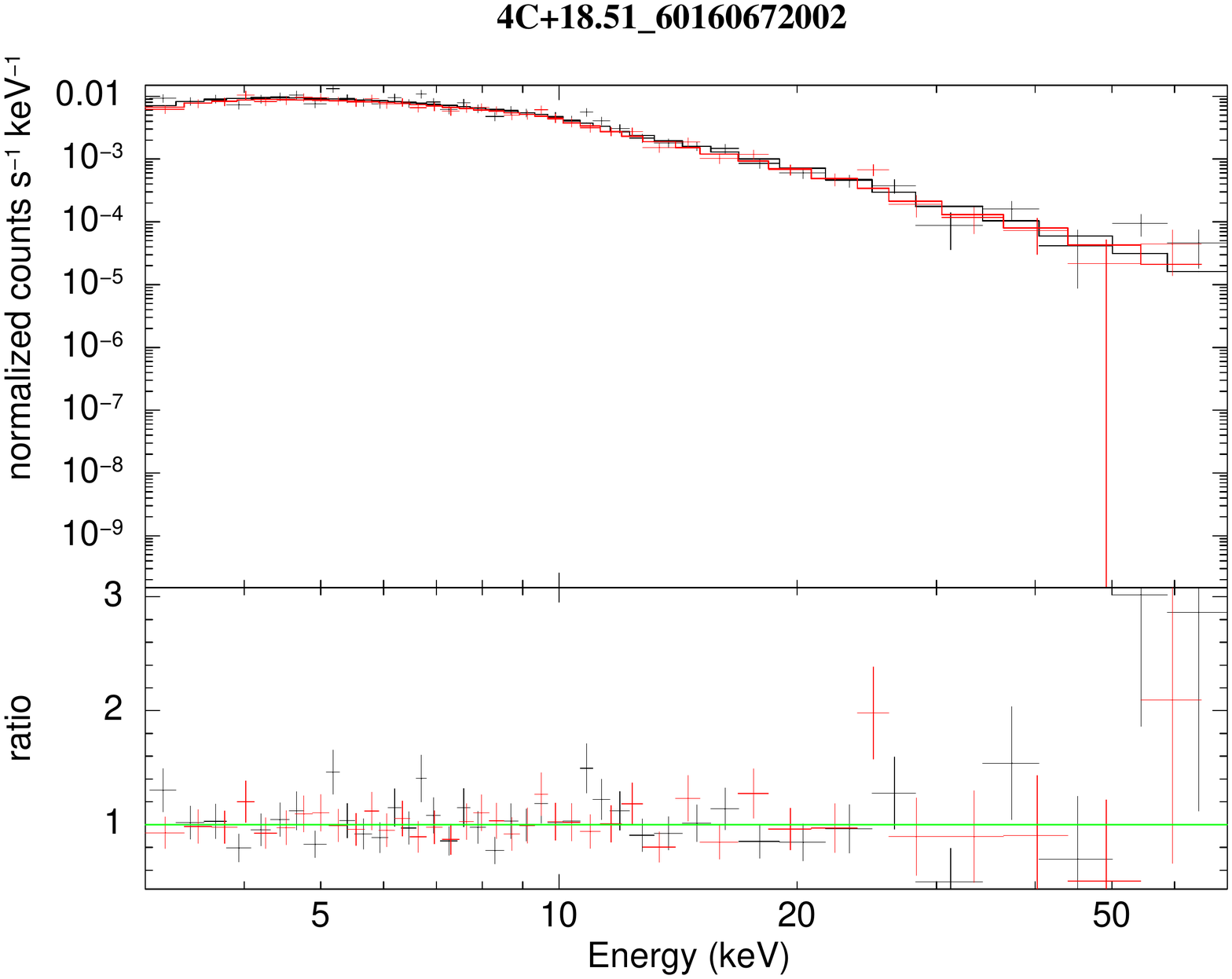}}
\subfloat{\includegraphics[width=0.33\textwidth]{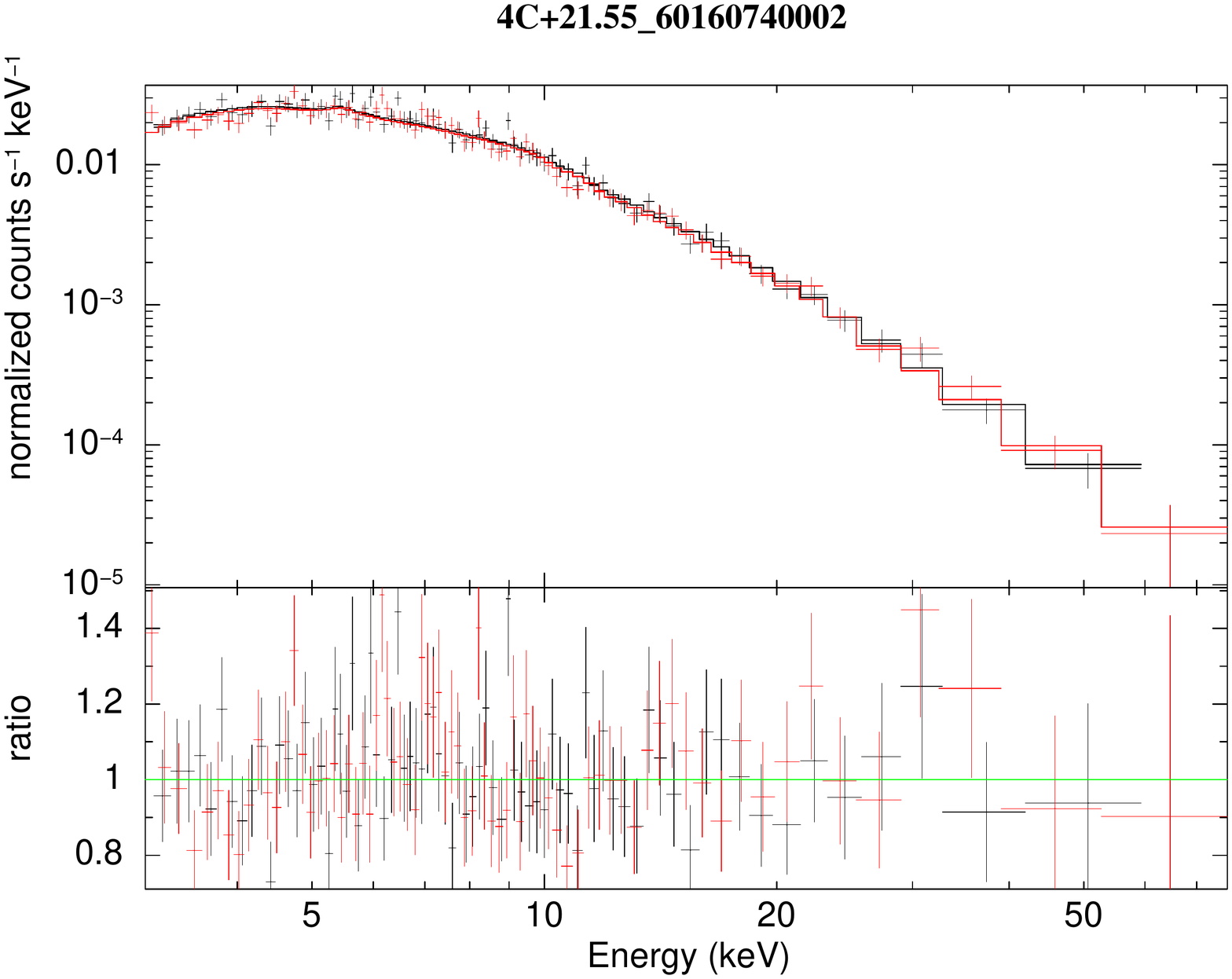}}\\
\subfloat{\includegraphics[width=0.33\textwidth]{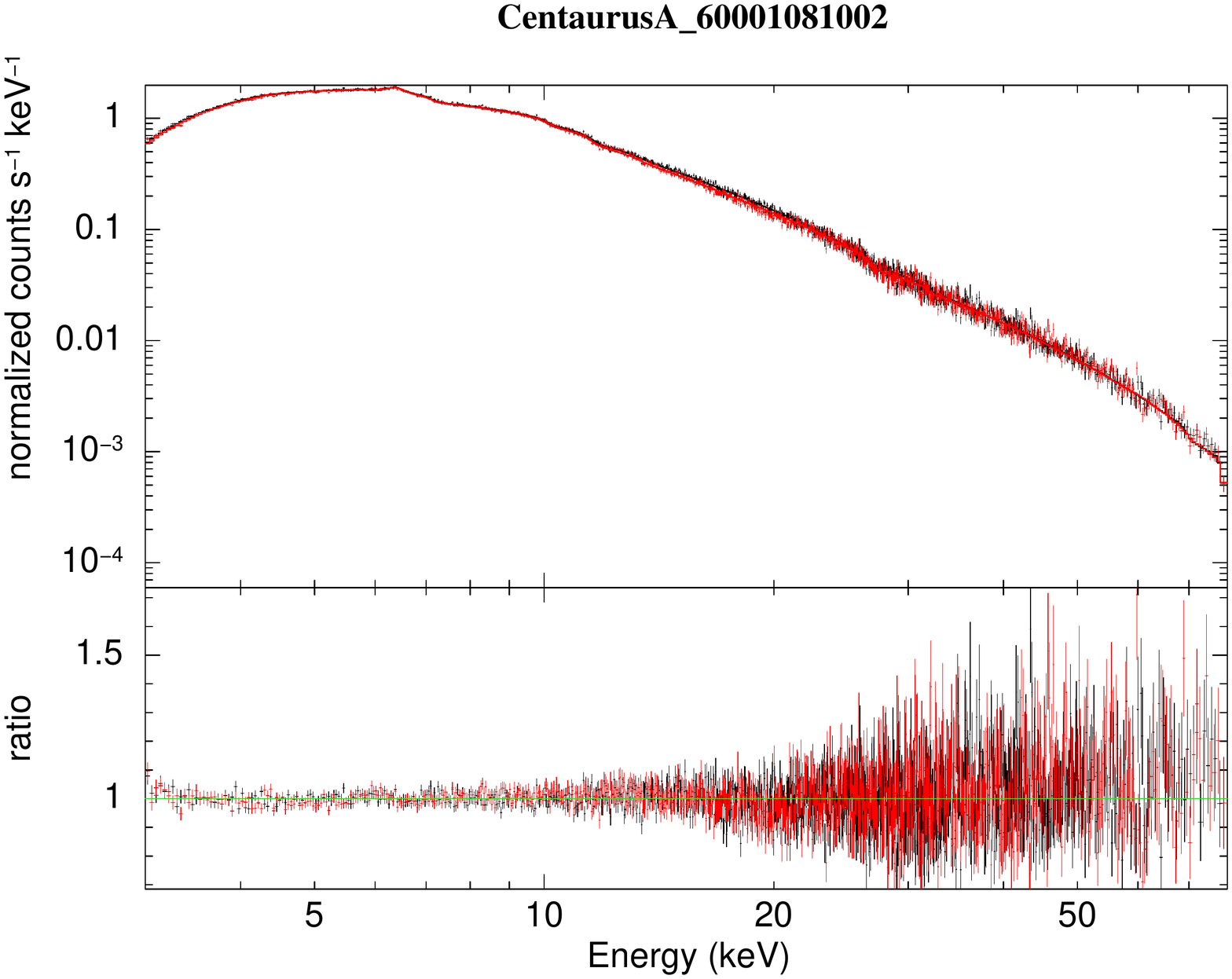}}
\subfloat{\includegraphics[width=0.33\textwidth]{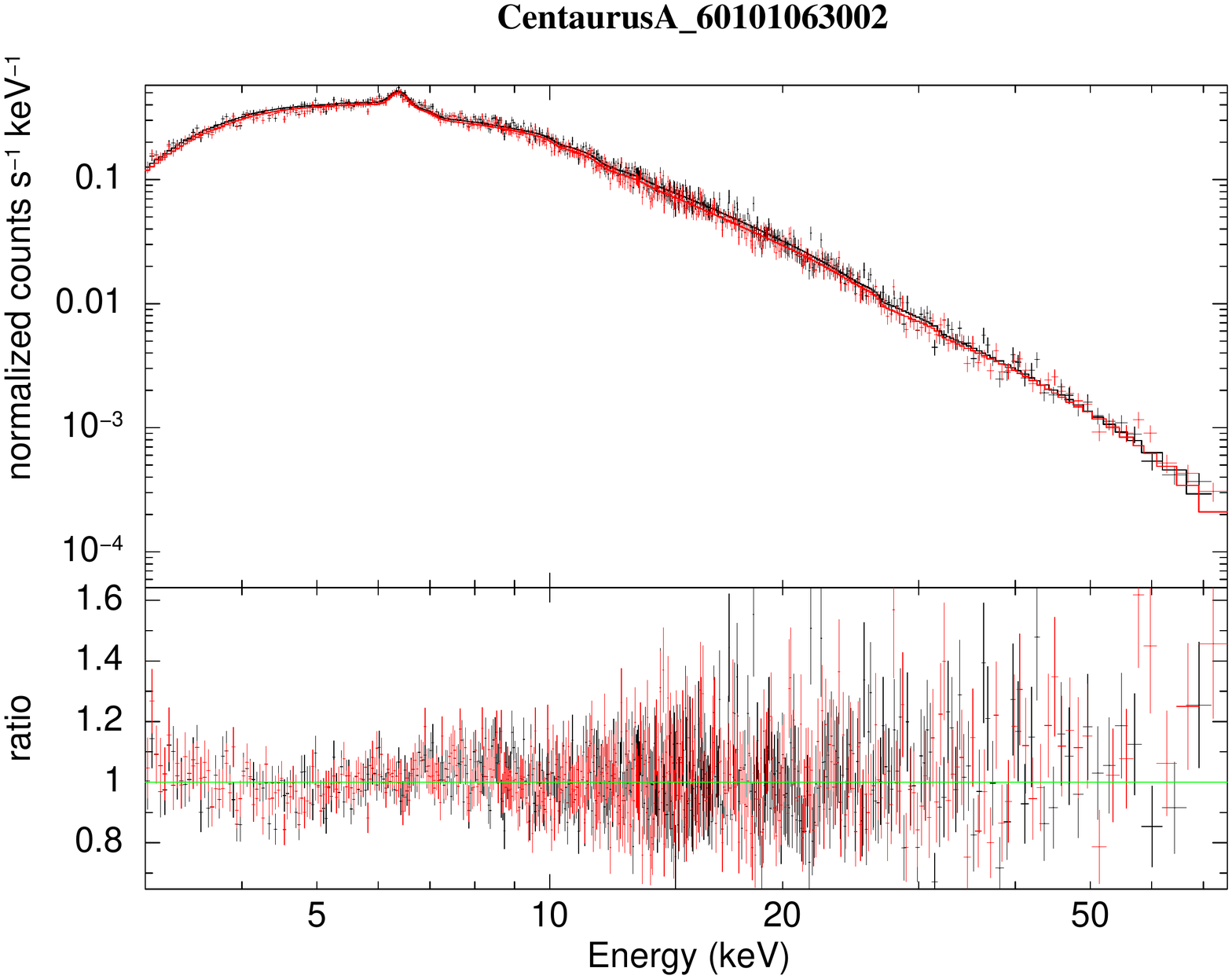}}
\subfloat{\includegraphics[width=0.33\textwidth]{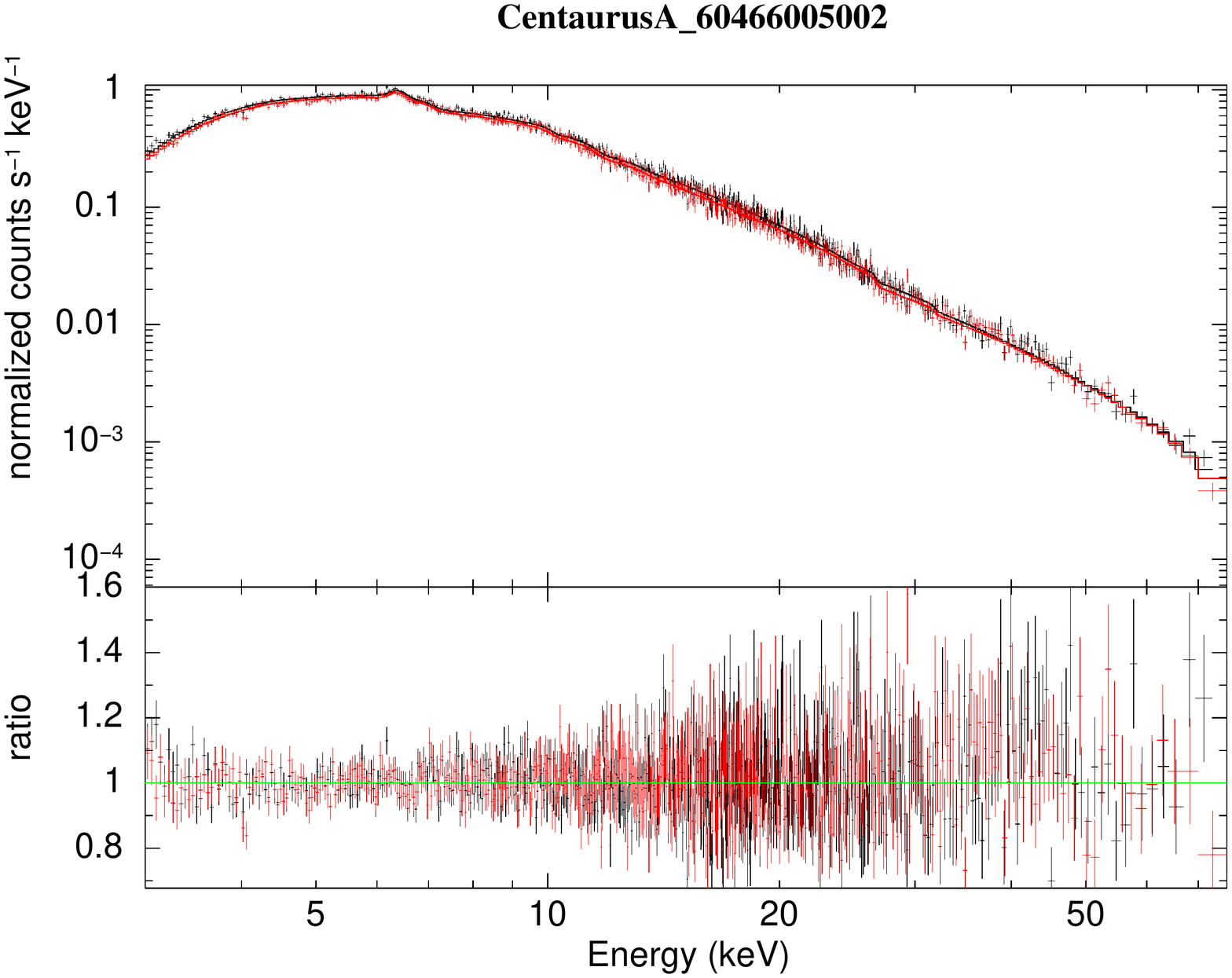}}\\
\subfloat{\includegraphics[width=0.33\textwidth]{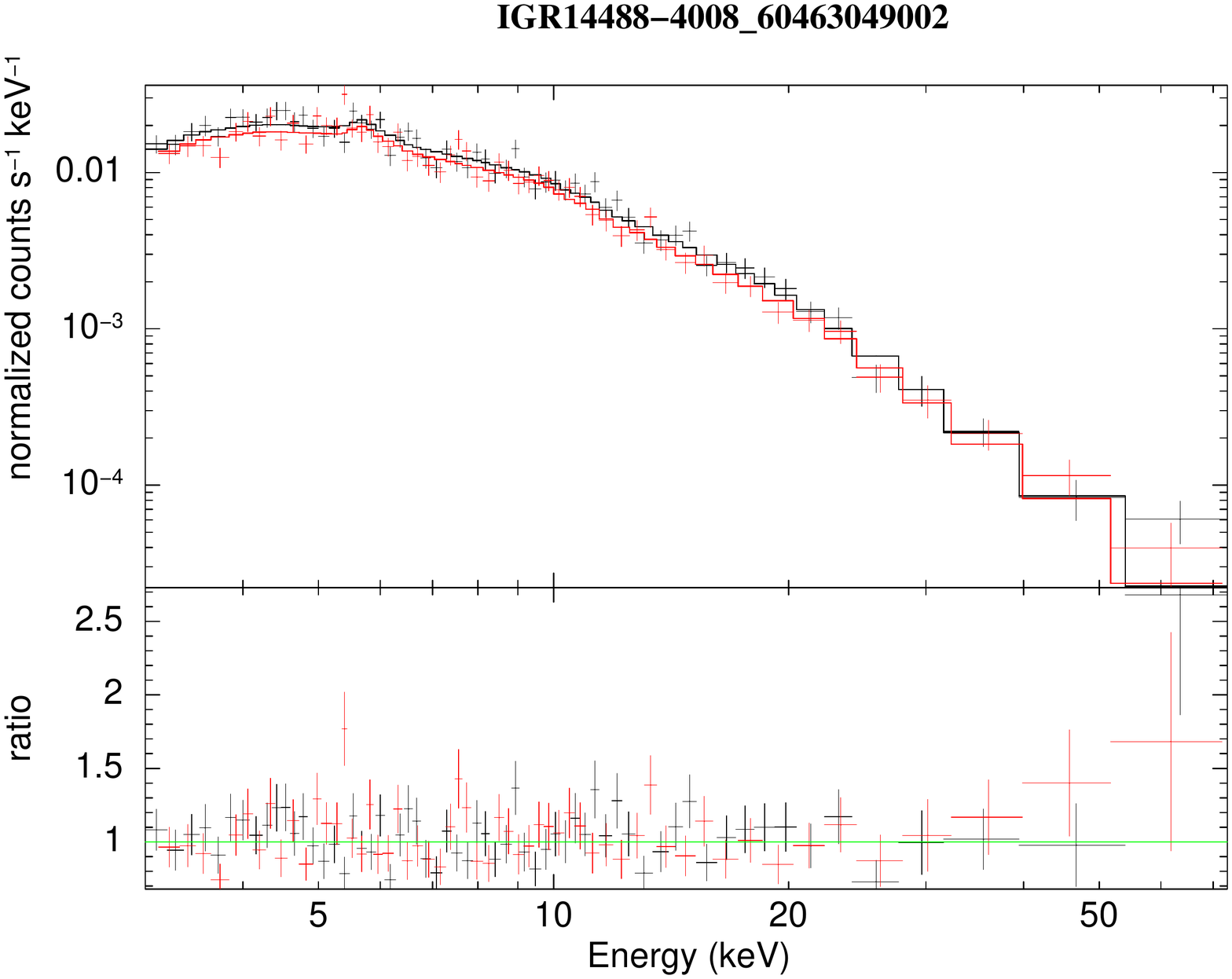}}
\subfloat{\includegraphics[width=0.33\textwidth]{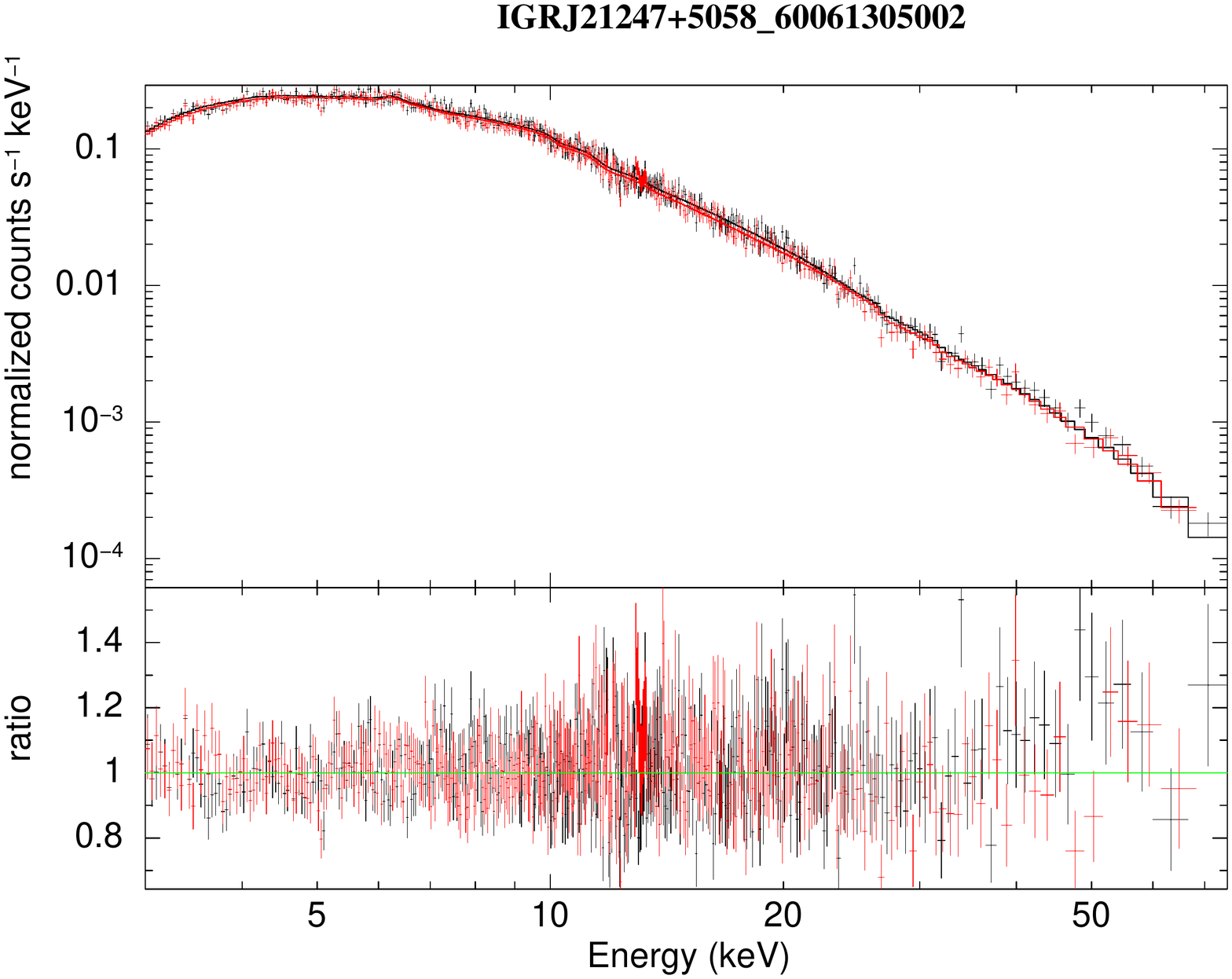}}
\subfloat{\includegraphics[width=0.33\textwidth]{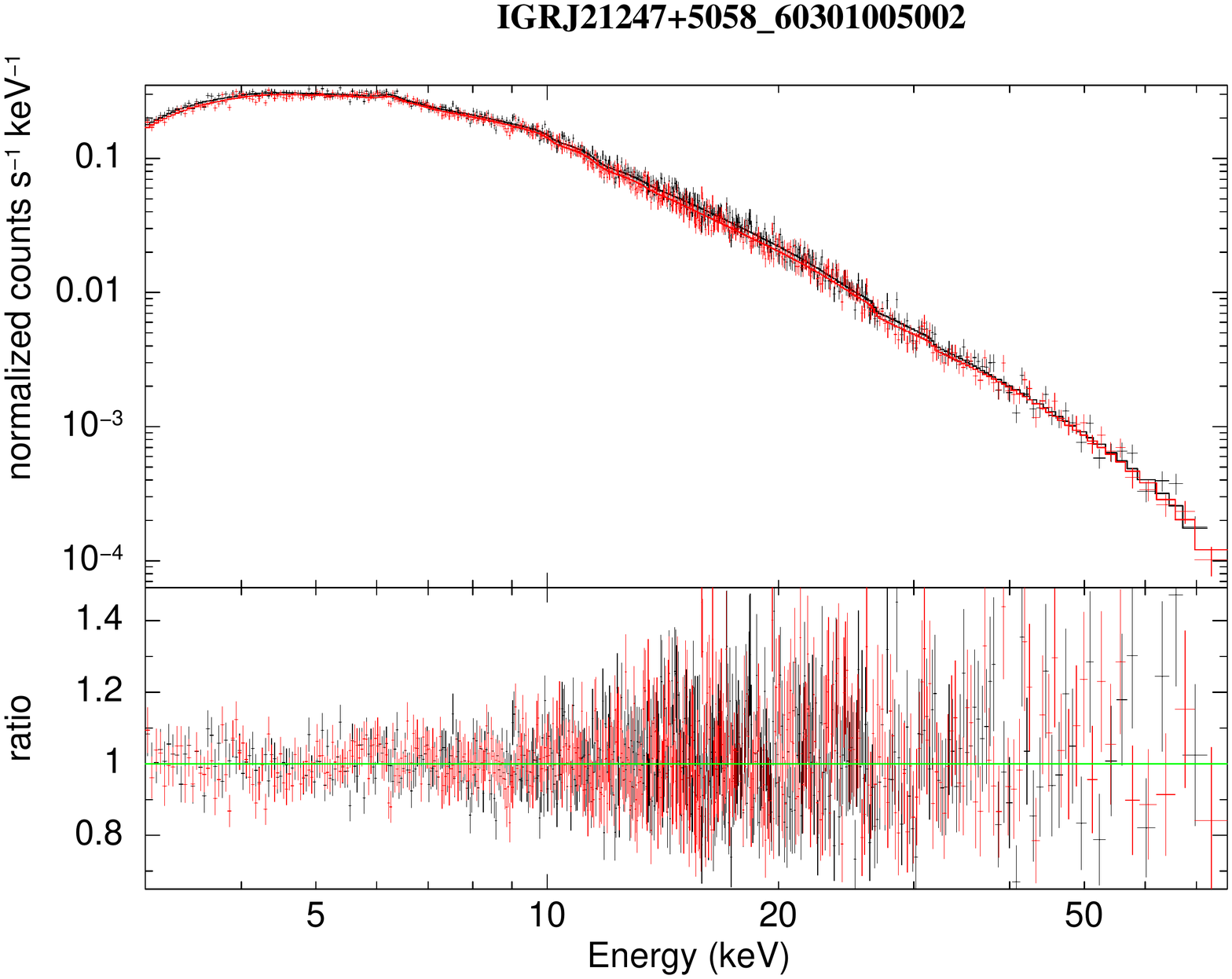}}\\
\end{figure*}

\addtocounter{figure}{-1} 
\begin{figure*}[!t]
\label{fig:spectra}
\addtocounter{figure}{1} 
\addtocounter{figure}{1} 
\addtocounter{figure}{1} 
\addtocounter{figure}{1} 
\addtocounter{figure}{1} 
\centering
\ContinuedFloat
\subfloat{\includegraphics[width=0.33\textwidth]{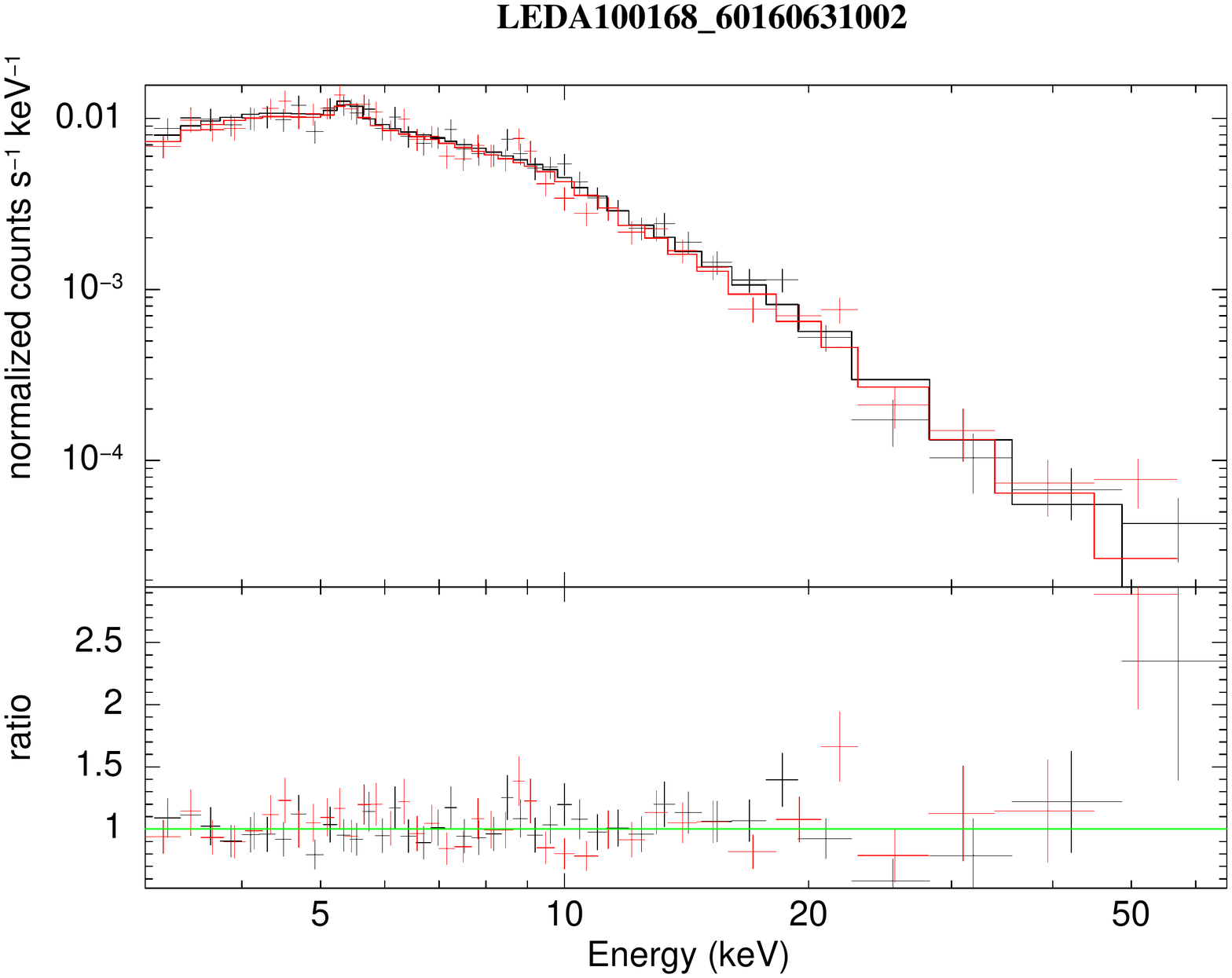}}
\subfloat{\includegraphics[width=0.33\textwidth]{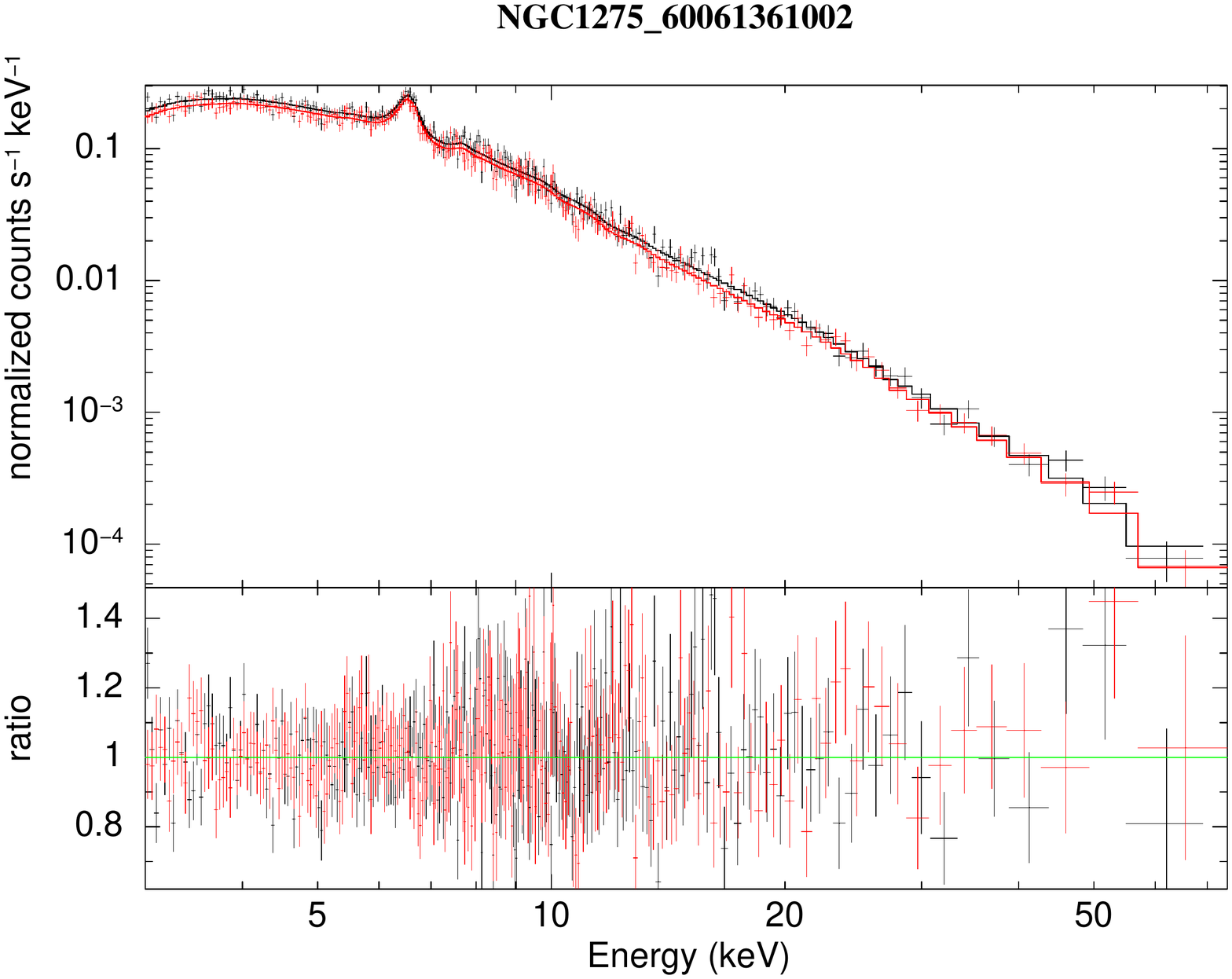}}
\subfloat{\includegraphics[width=0.33\textwidth]{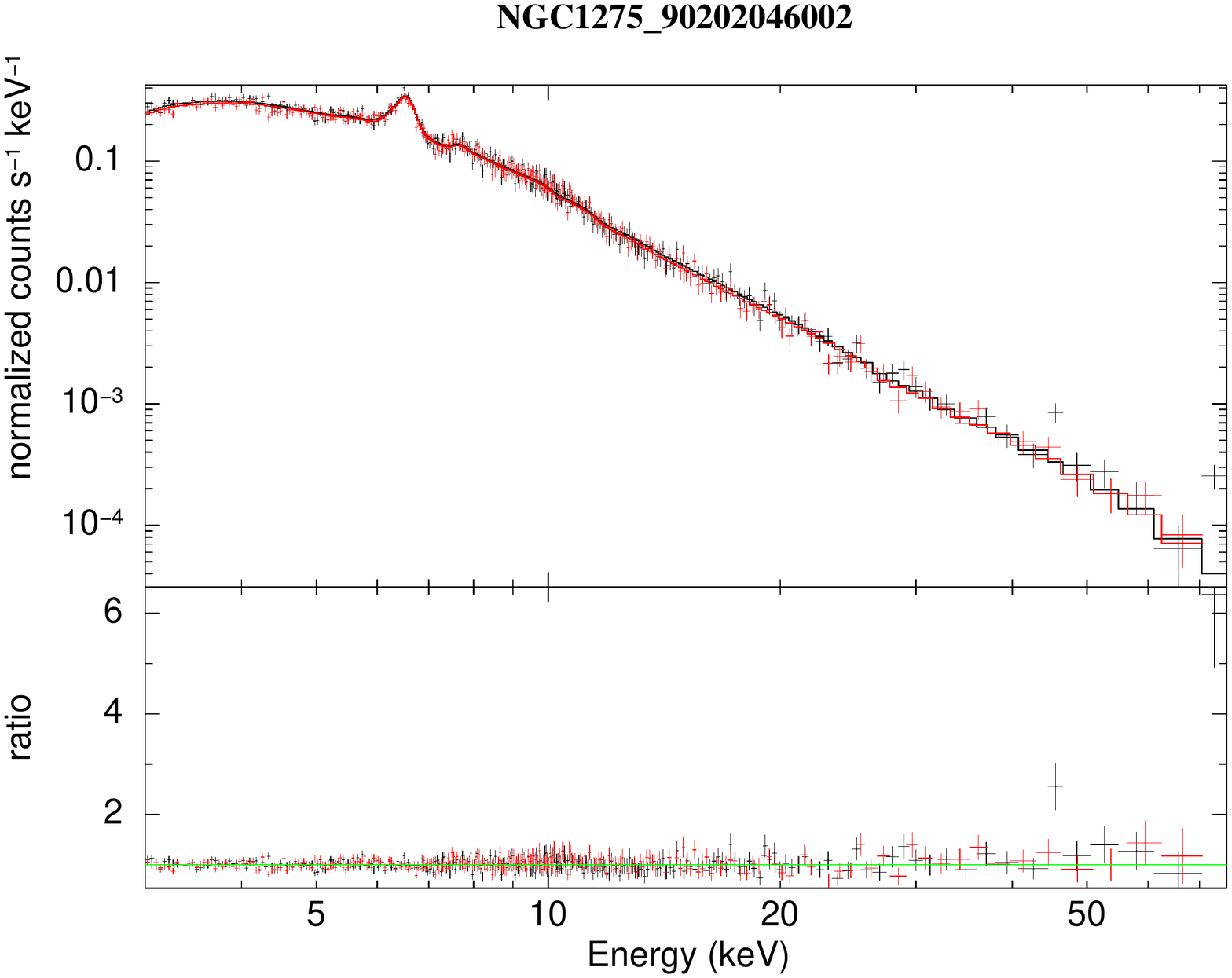}}\\
\subfloat{\includegraphics[width=0.33\textwidth]{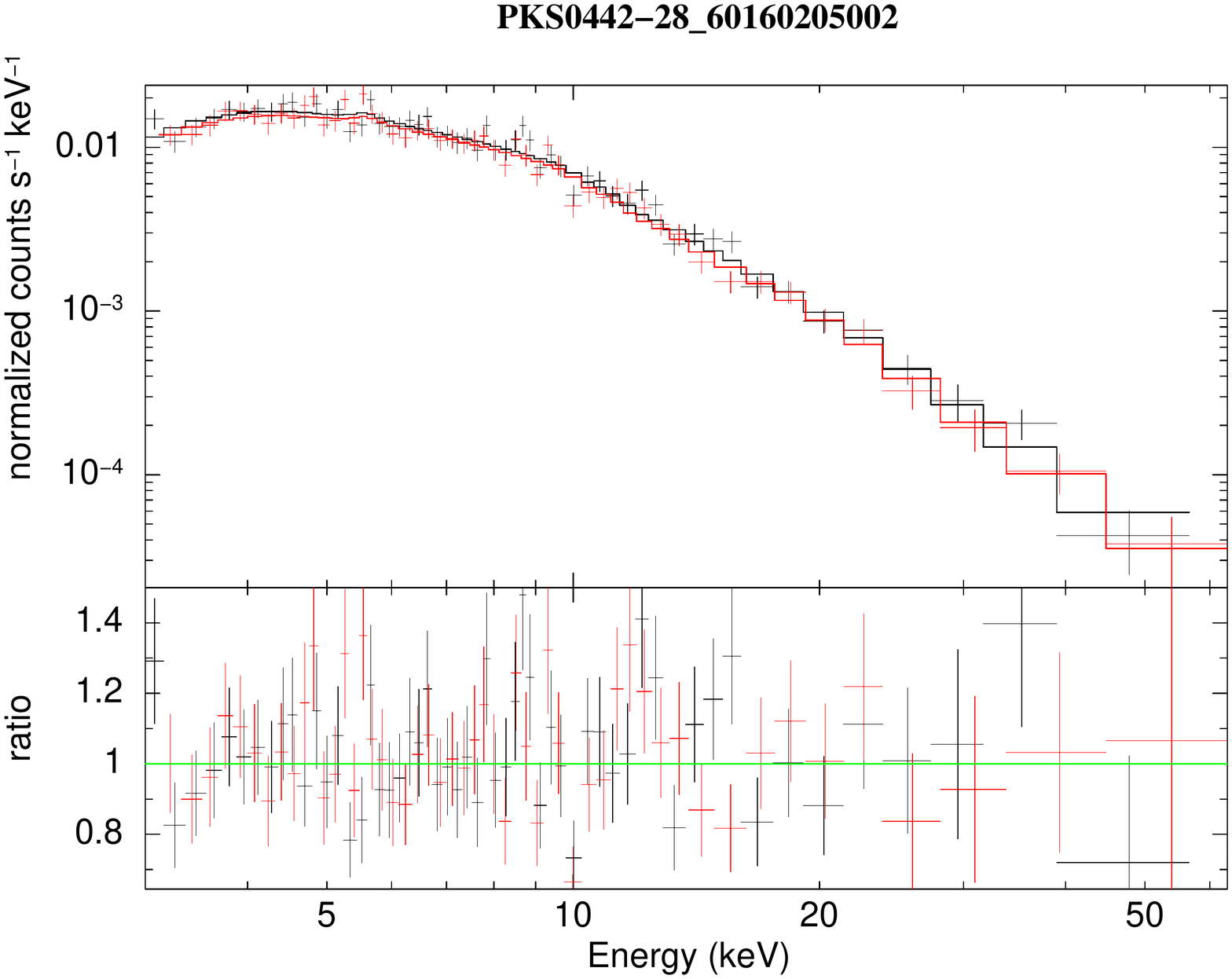}}
\subfloat{\includegraphics[width=0.33\textwidth]{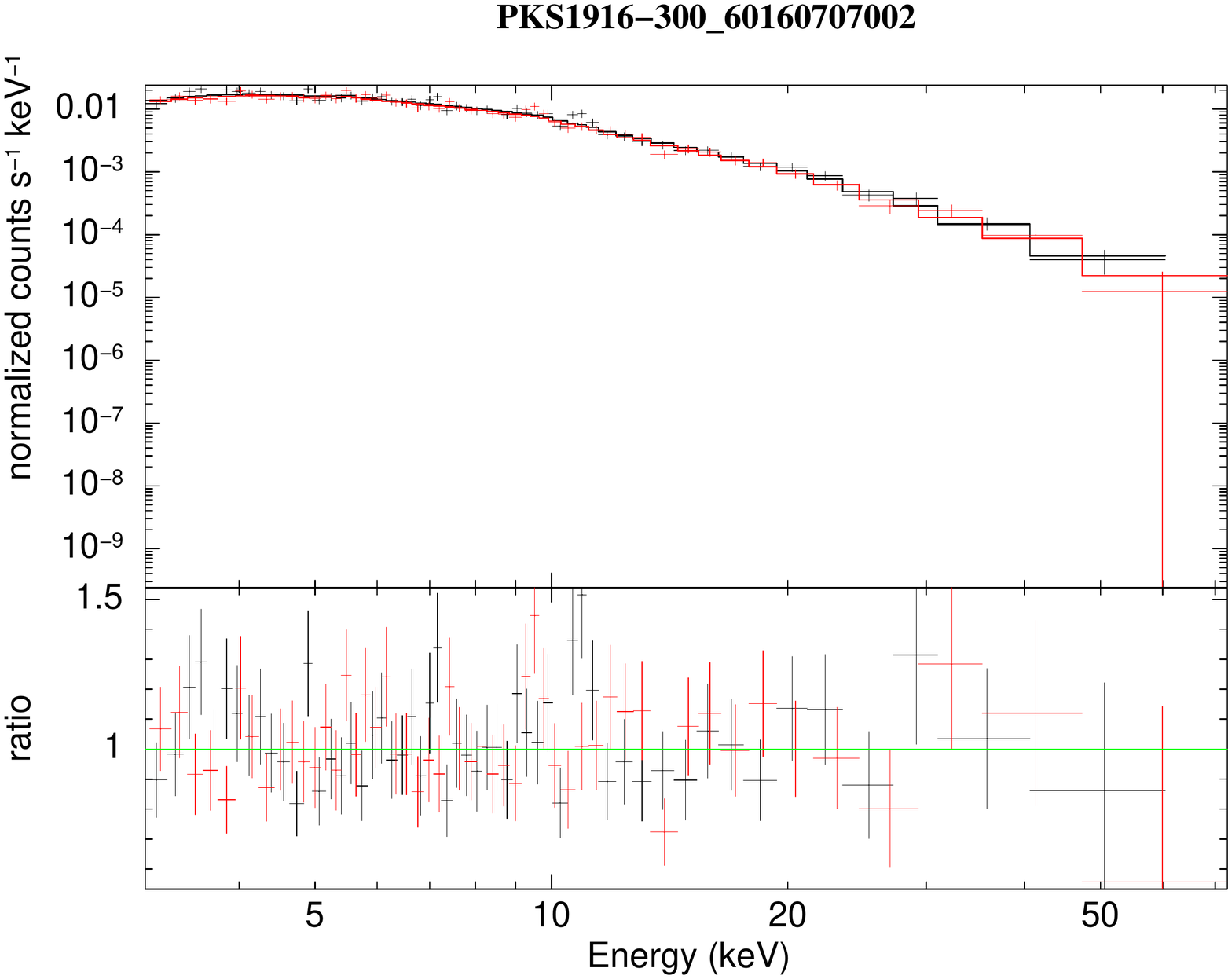}}
\subfloat{\includegraphics[width=0.33\textwidth]{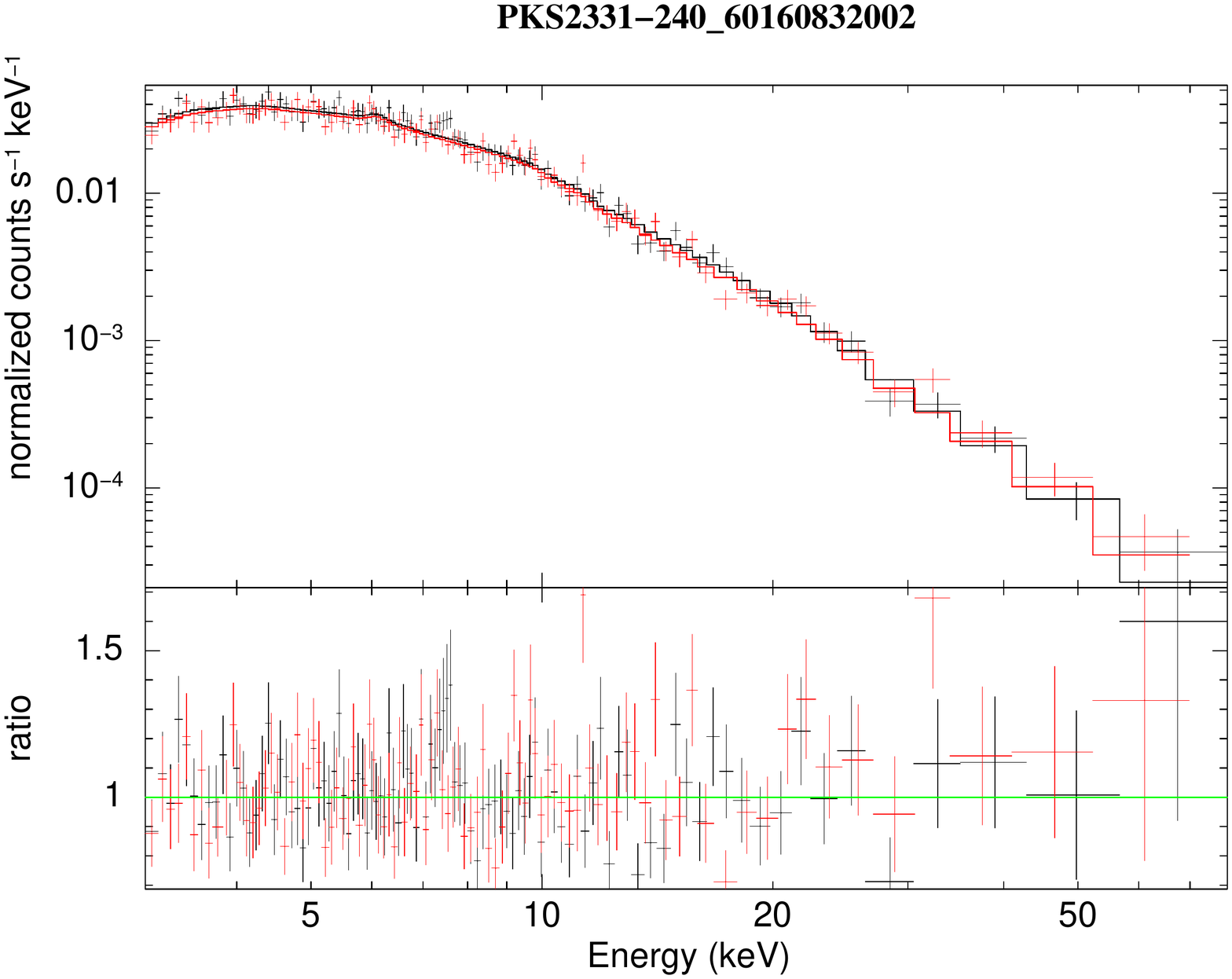}}\\
\subfloat{\includegraphics[width=0.33\textwidth]{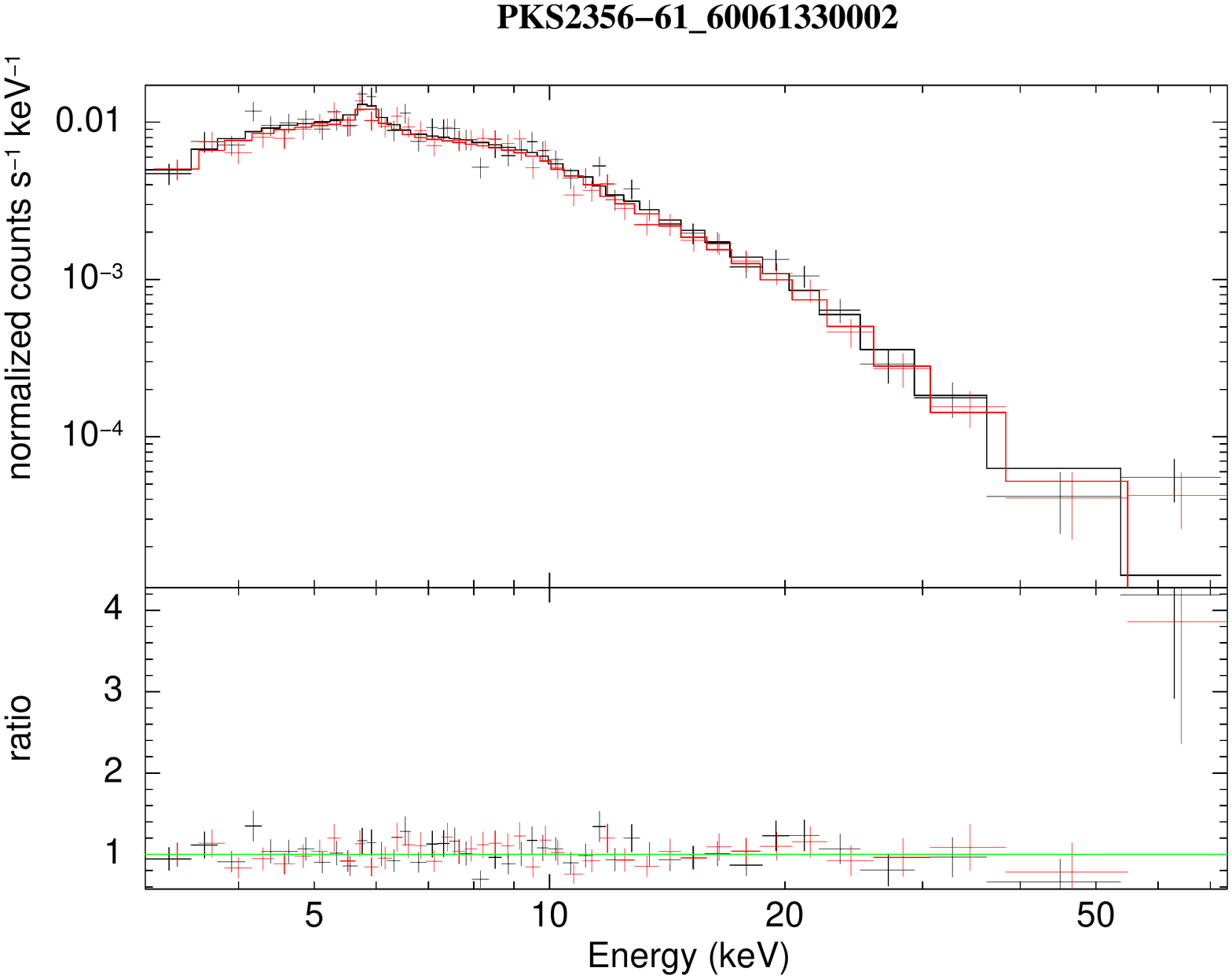}}
\subfloat{\includegraphics[width=0.33\textwidth]{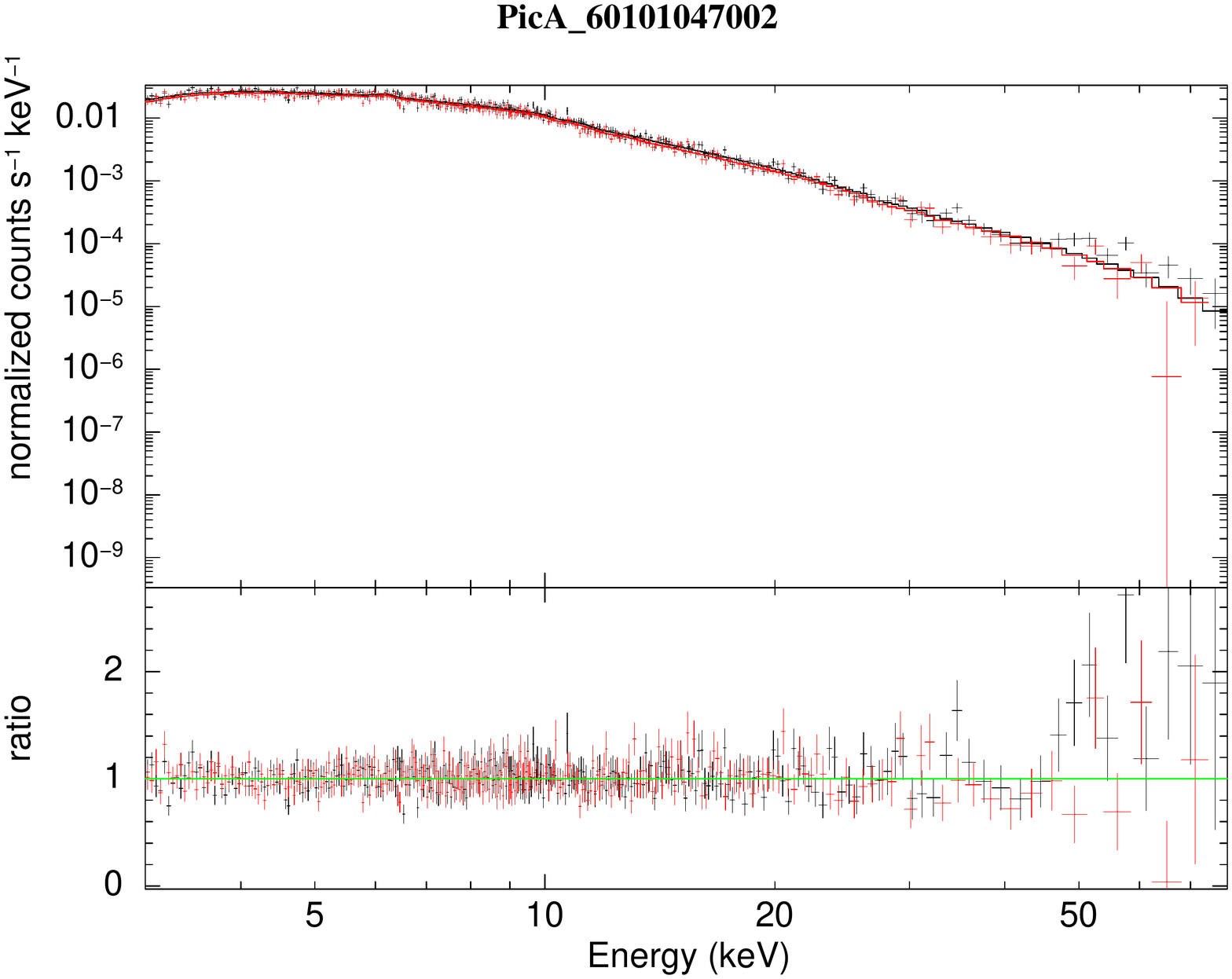}}
\subfloat{\includegraphics[width=0.33\textwidth]{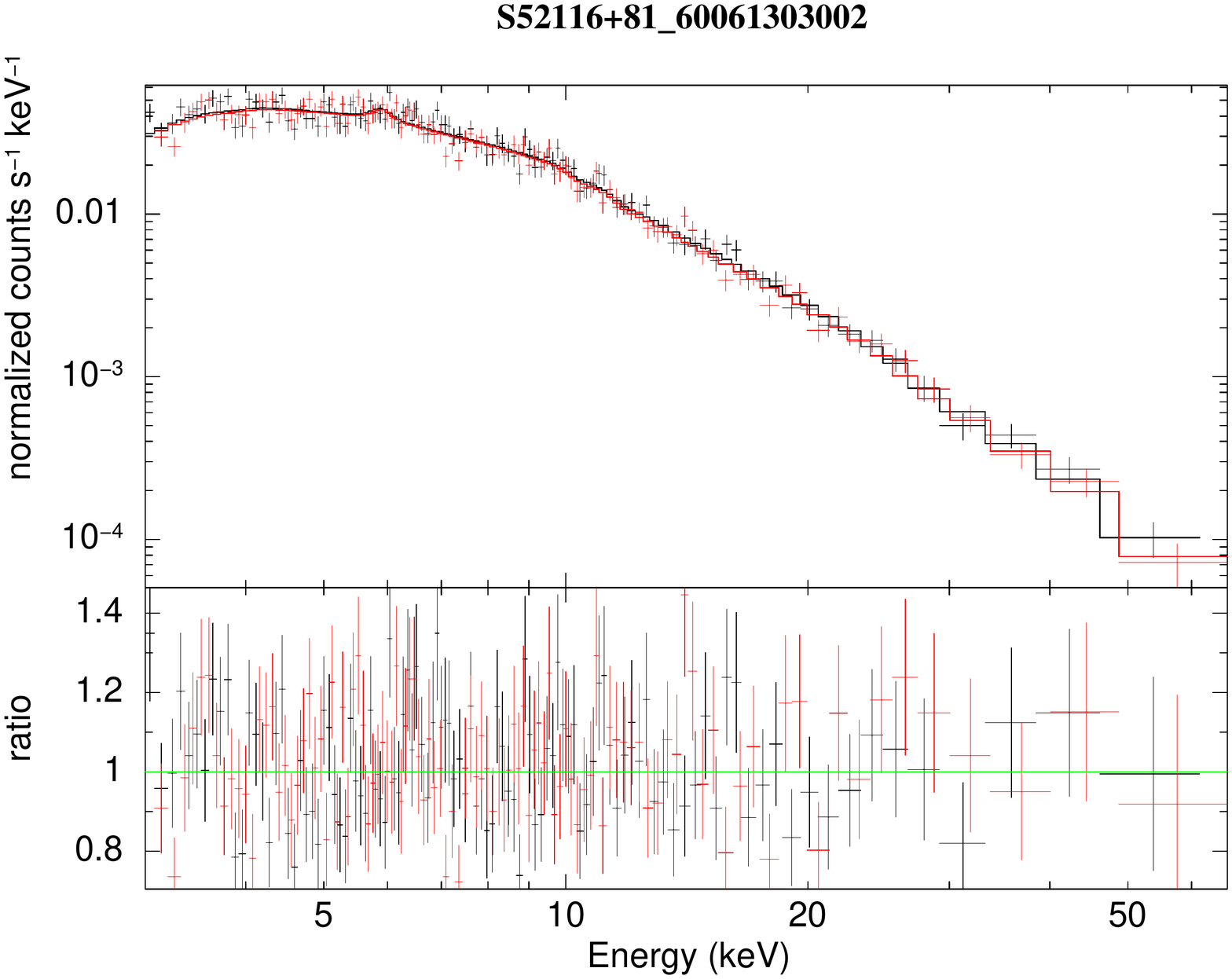}}\\
\caption{
NuSTAR spectra, best-fit models and the residual data-to-model ratios of our sample. 
Spectra from both FPMA (black) and FPMB (red) modules are given. 
Source name and NuSTAR ObsID are presented on top of each spectrum. The best-fit spectral parameters are given in Tab. \ref{tab:Results}.
 }
\label{fig:spectra}
\end{figure*}
}
\section{C: $\Gamma$ -- $E_{cut}$  contours}
\label{app:C}
We plot the $\Gamma$ vs $E_{cut}$  contours of 13 sources with $E_{cut}$ detection. For sources with multiple observations, the best constrained one is plotted. 
\begin{figure*}[!t]
\addtocounter{figure}{1} 
\centering
\ContinuedFloat
\subfloat{\includegraphics[width=0.32\textwidth]{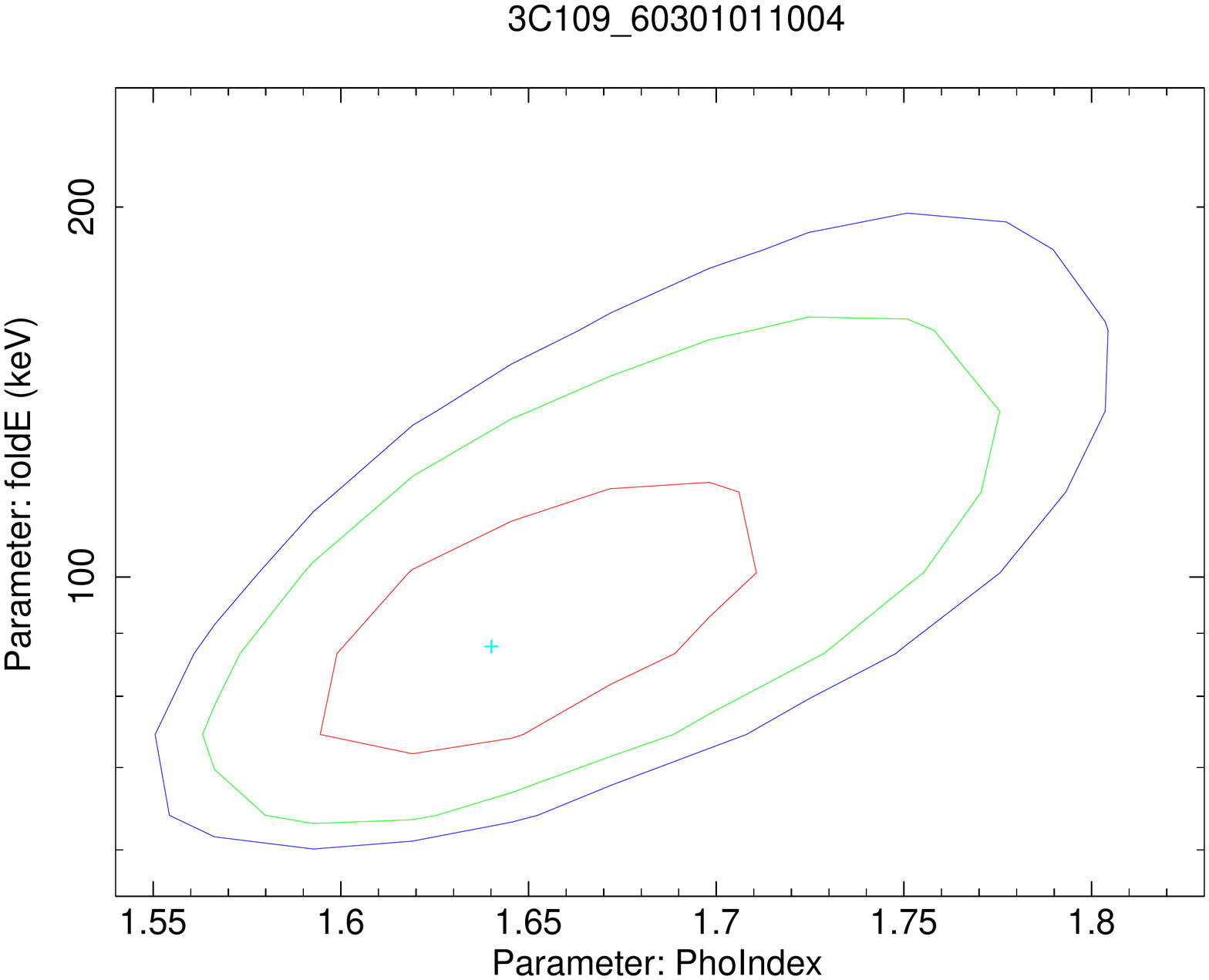}}
\subfloat{\includegraphics[width=0.32\textwidth]{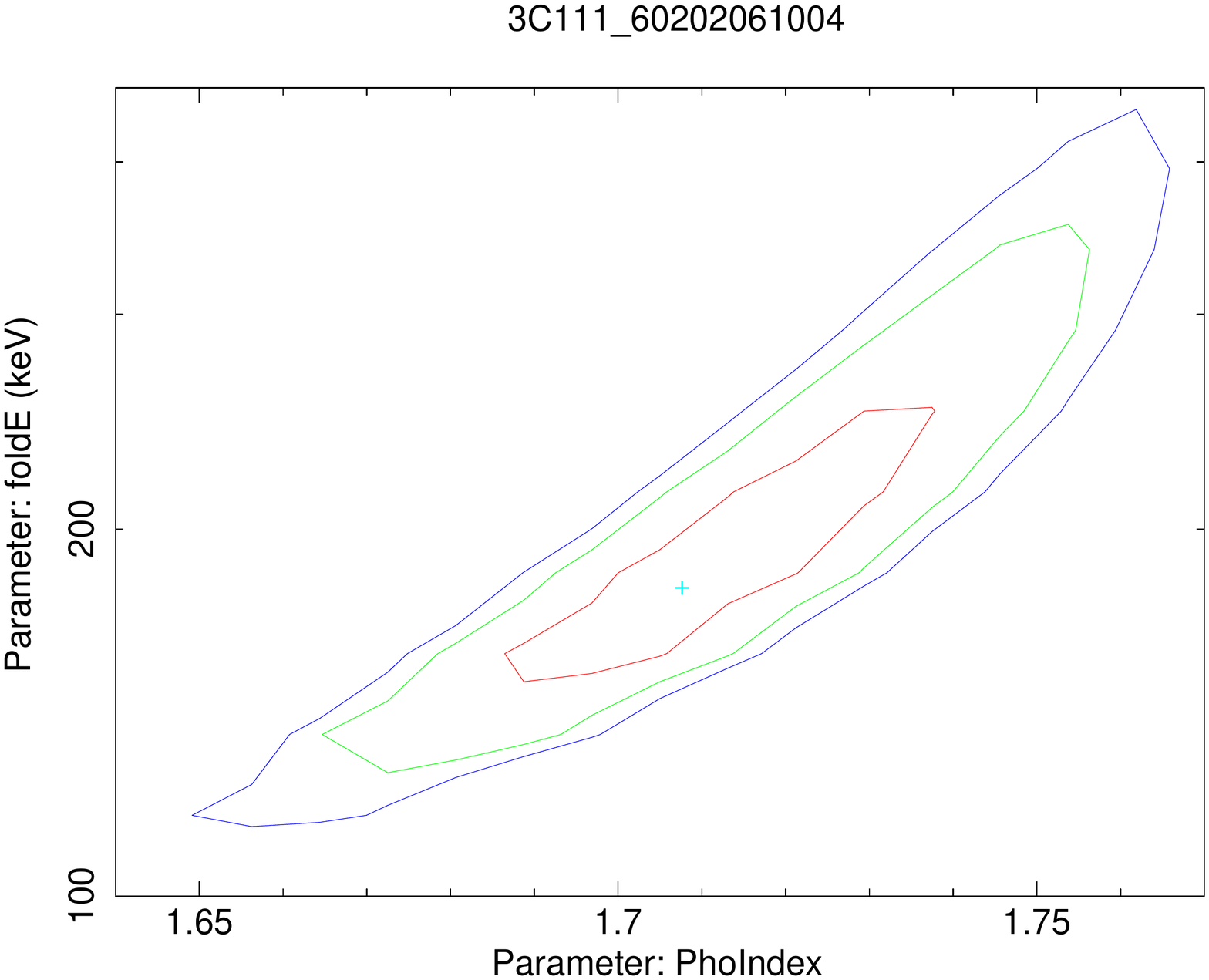}}
\subfloat{\includegraphics[width=0.32\textwidth]{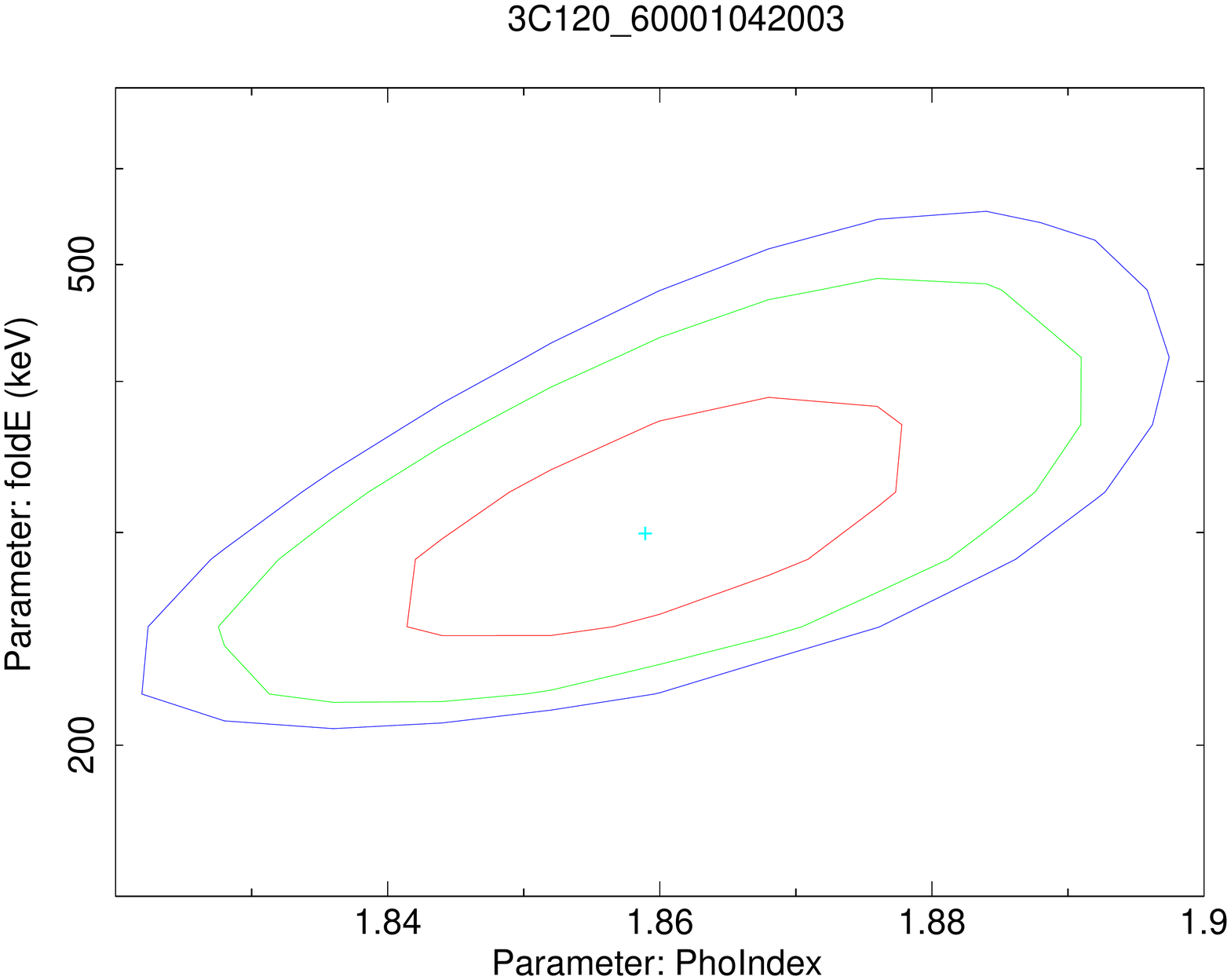}}\\
\subfloat{\includegraphics[width=0.32\textwidth]{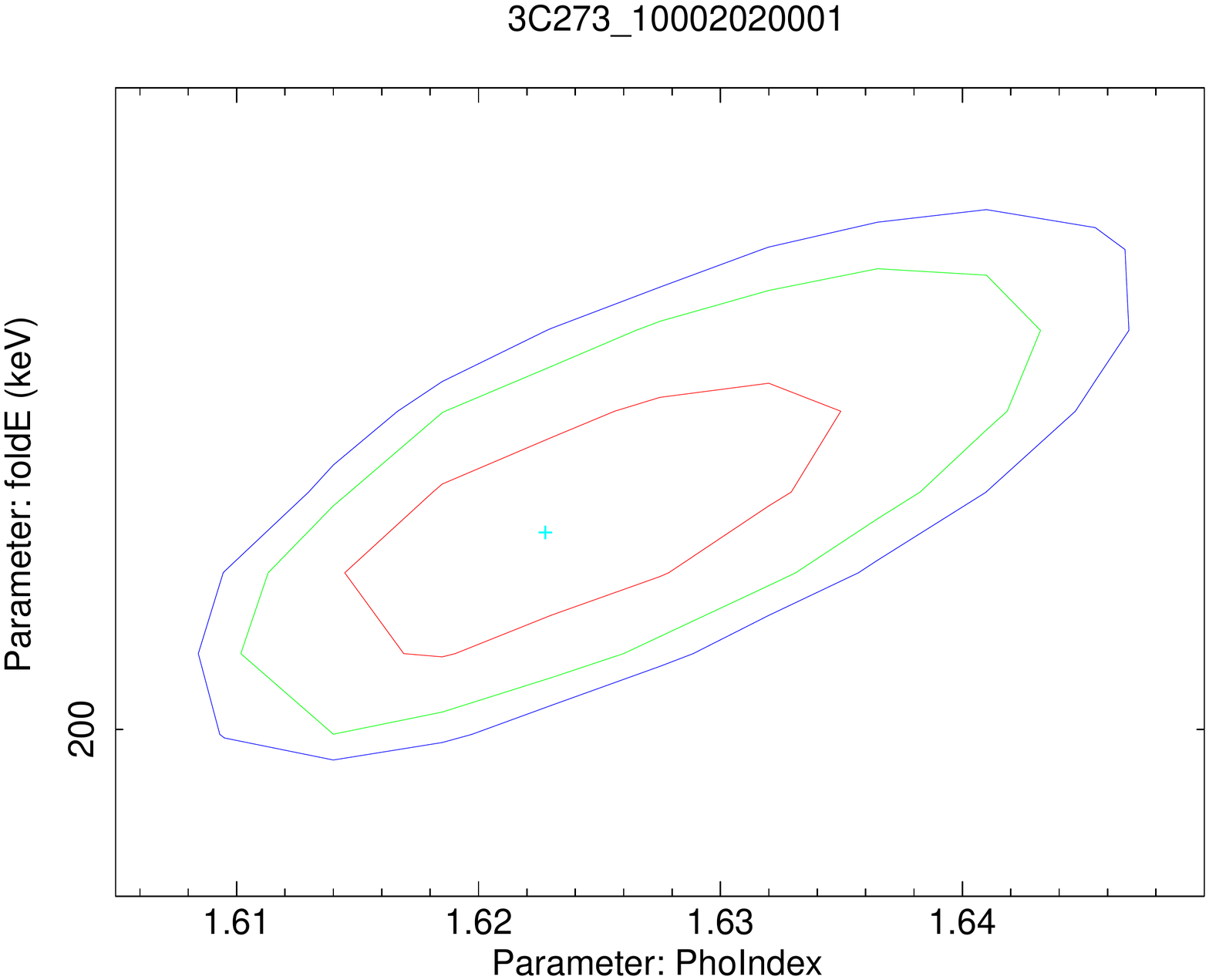}}
\subfloat{\includegraphics[width=0.32\textwidth]{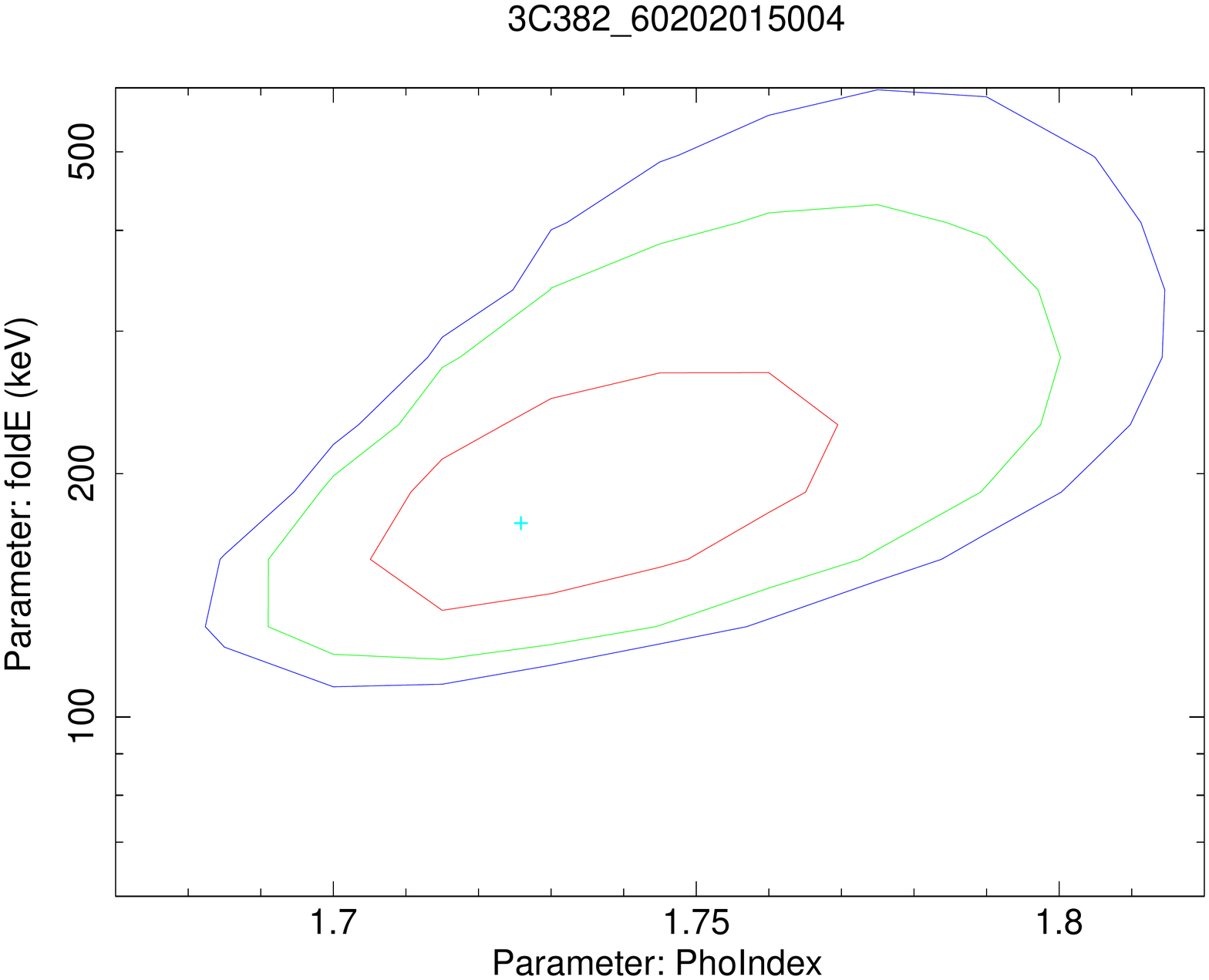}}
\subfloat{\includegraphics[width=0.32\textwidth]{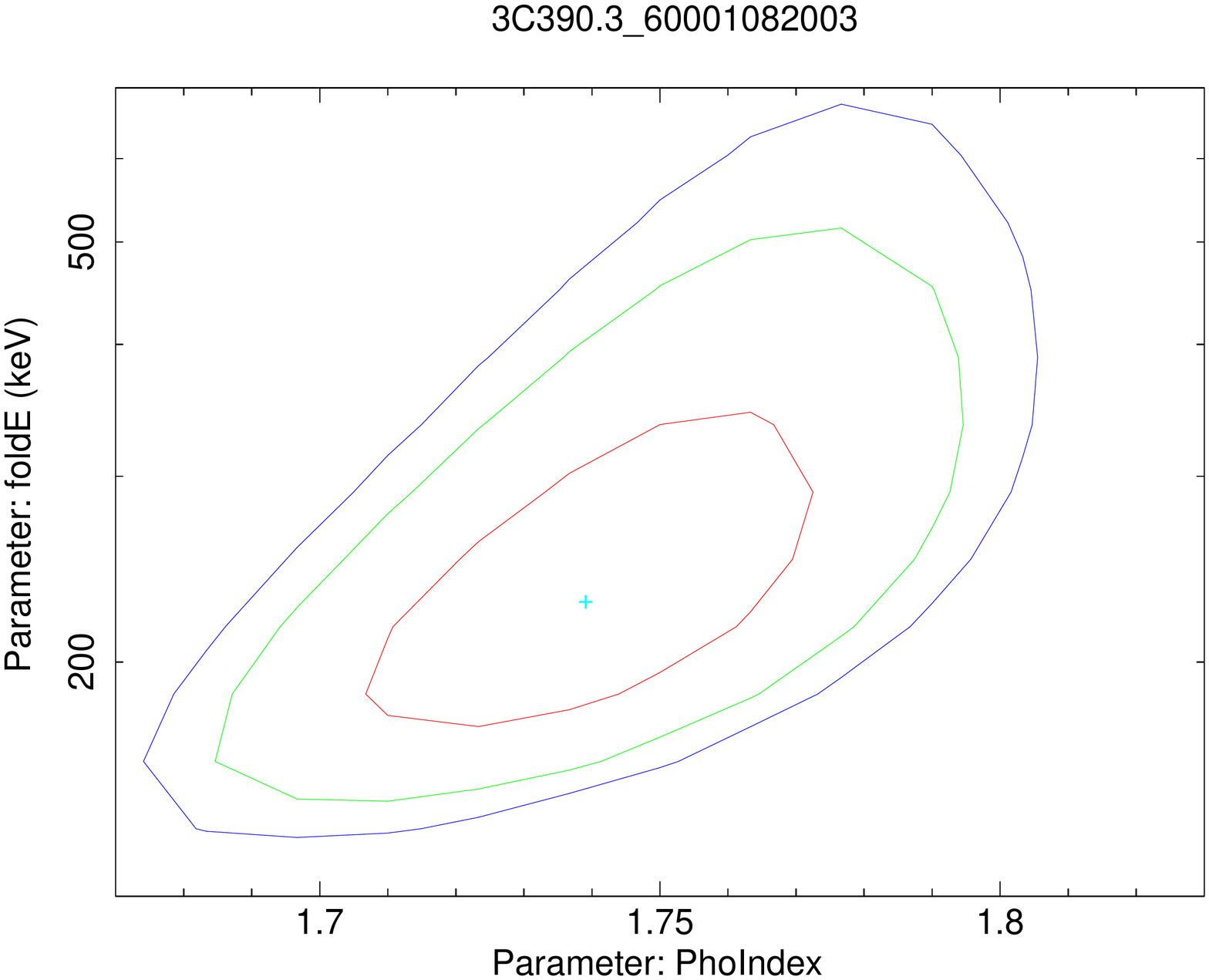}}\\
\subfloat{\includegraphics[width=0.32\textwidth]{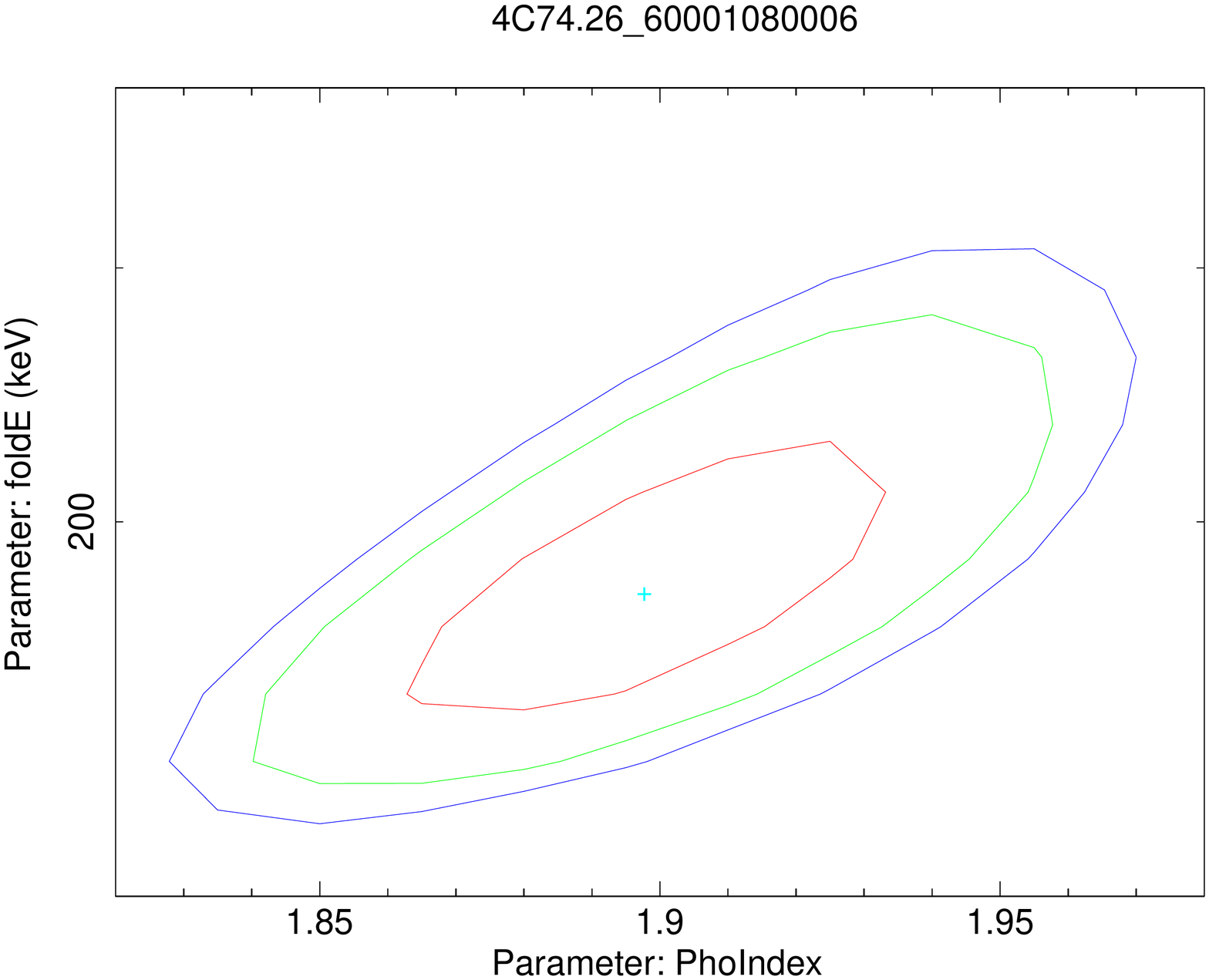}}
\subfloat{\includegraphics[width=0.32\textwidth]{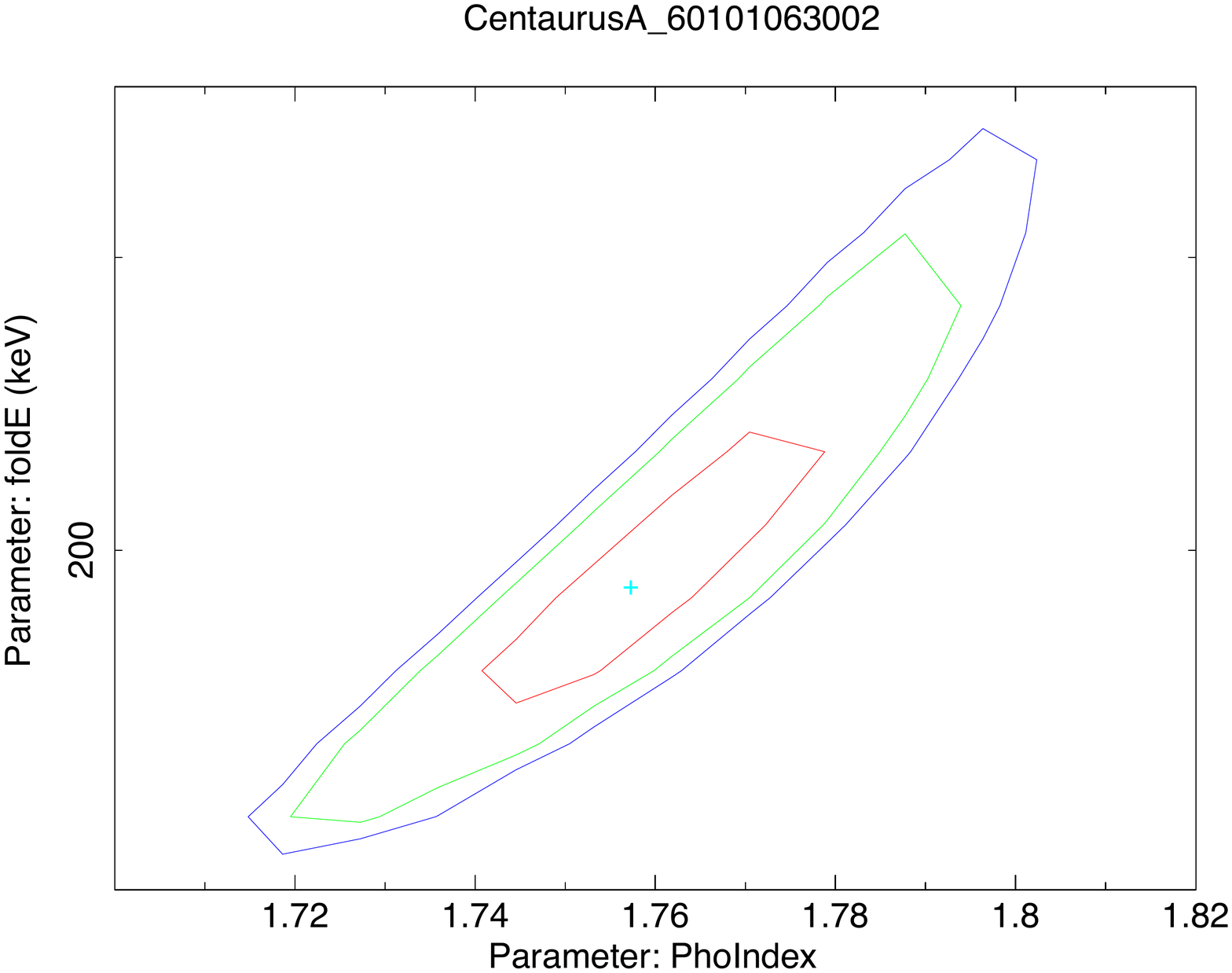}}
\subfloat{\includegraphics[width=0.32\textwidth]{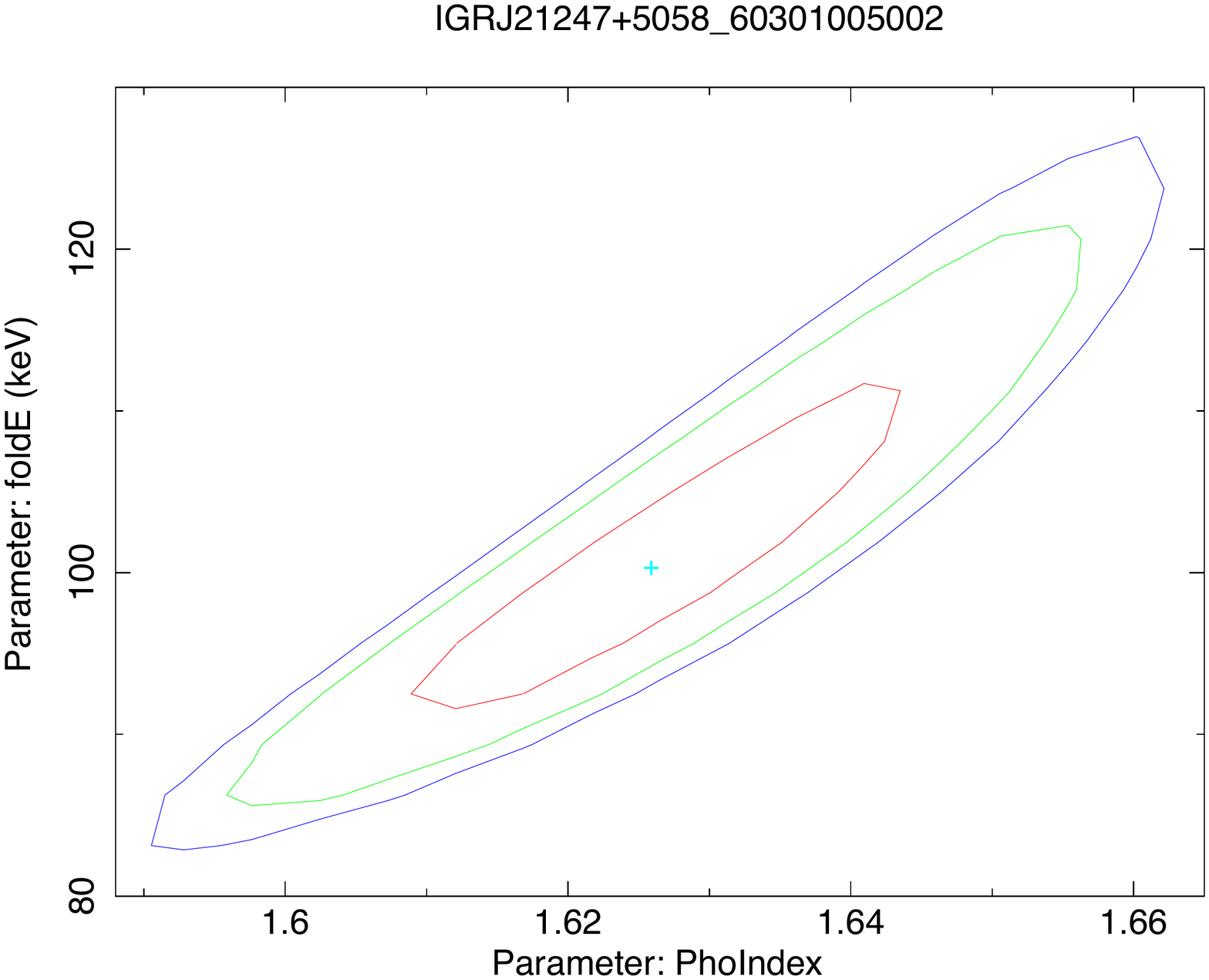}}\\
\subfloat{\includegraphics[width=0.32\textwidth]{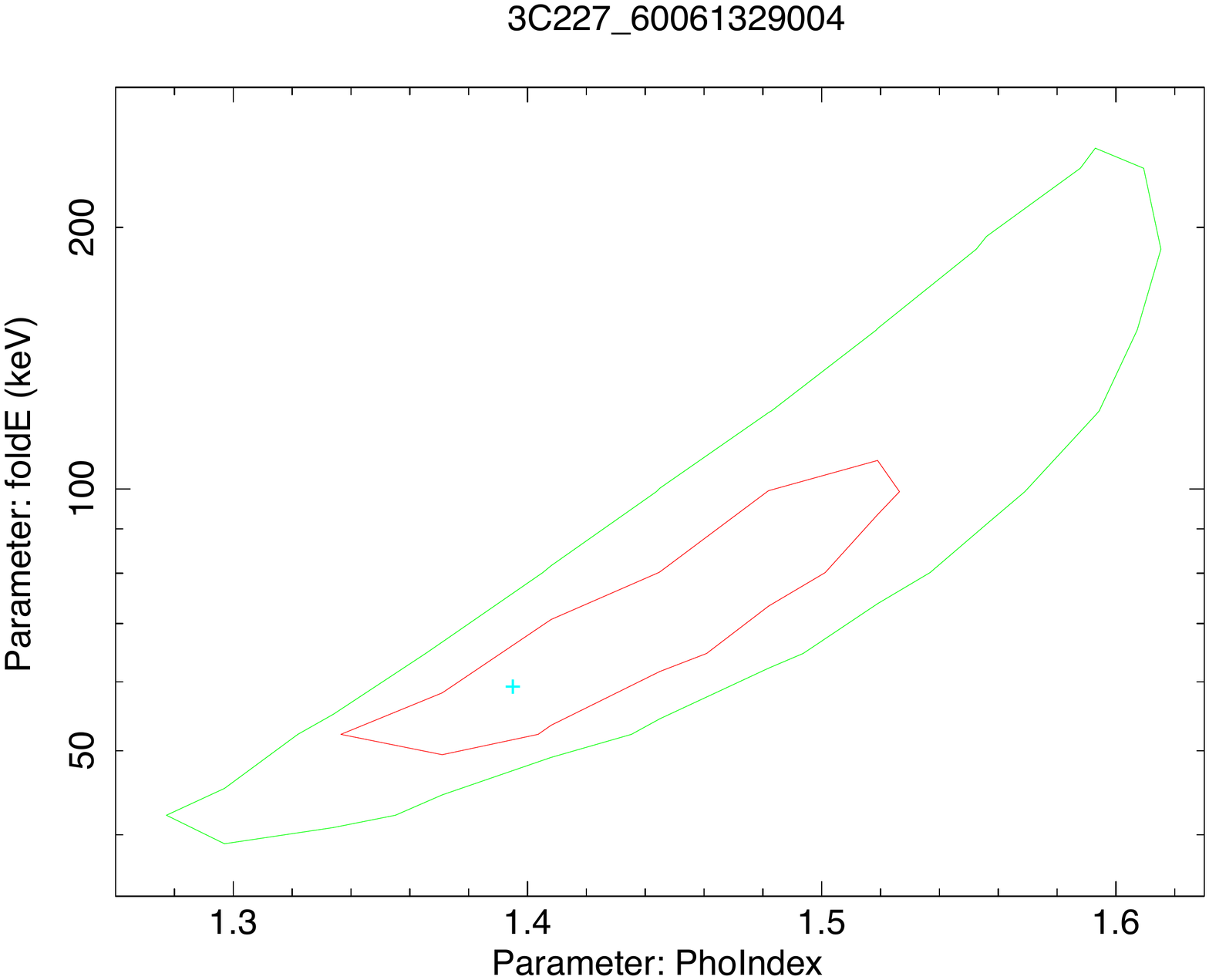}}
\subfloat{\includegraphics[width=0.32\textwidth]{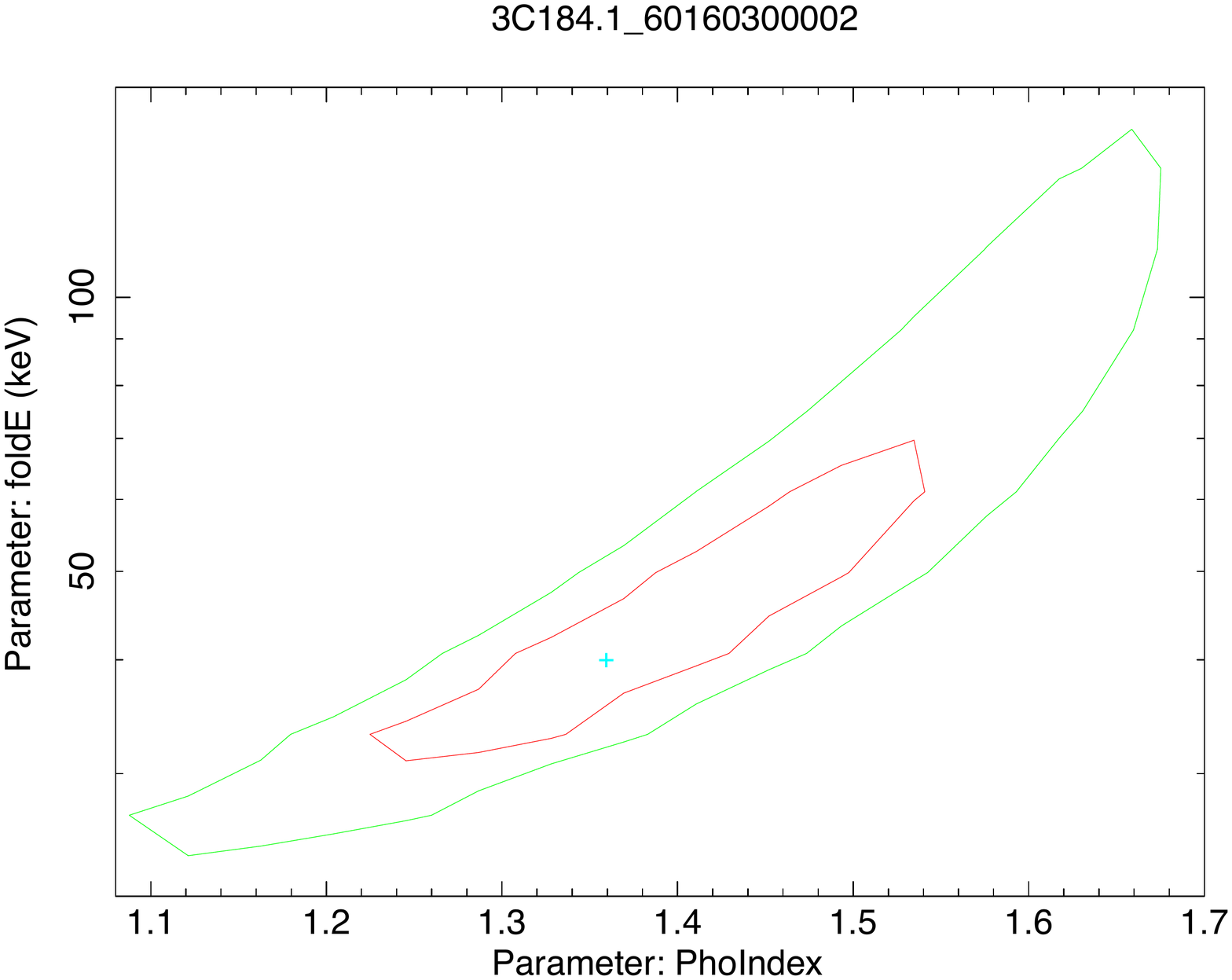}}
\subfloat{\includegraphics[width=0.32\textwidth]{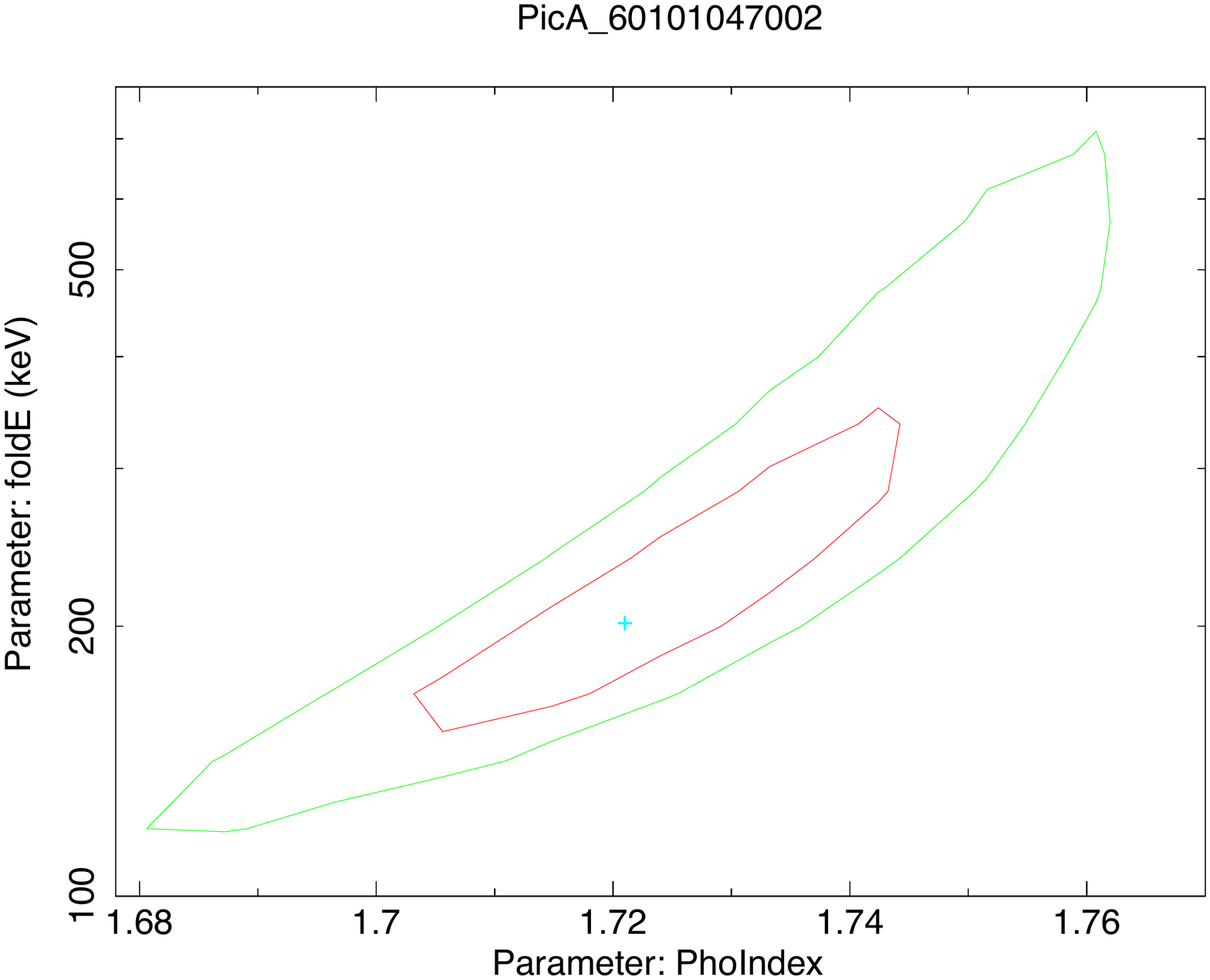}}\\
\subfloat{\includegraphics[width=0.32\textwidth]{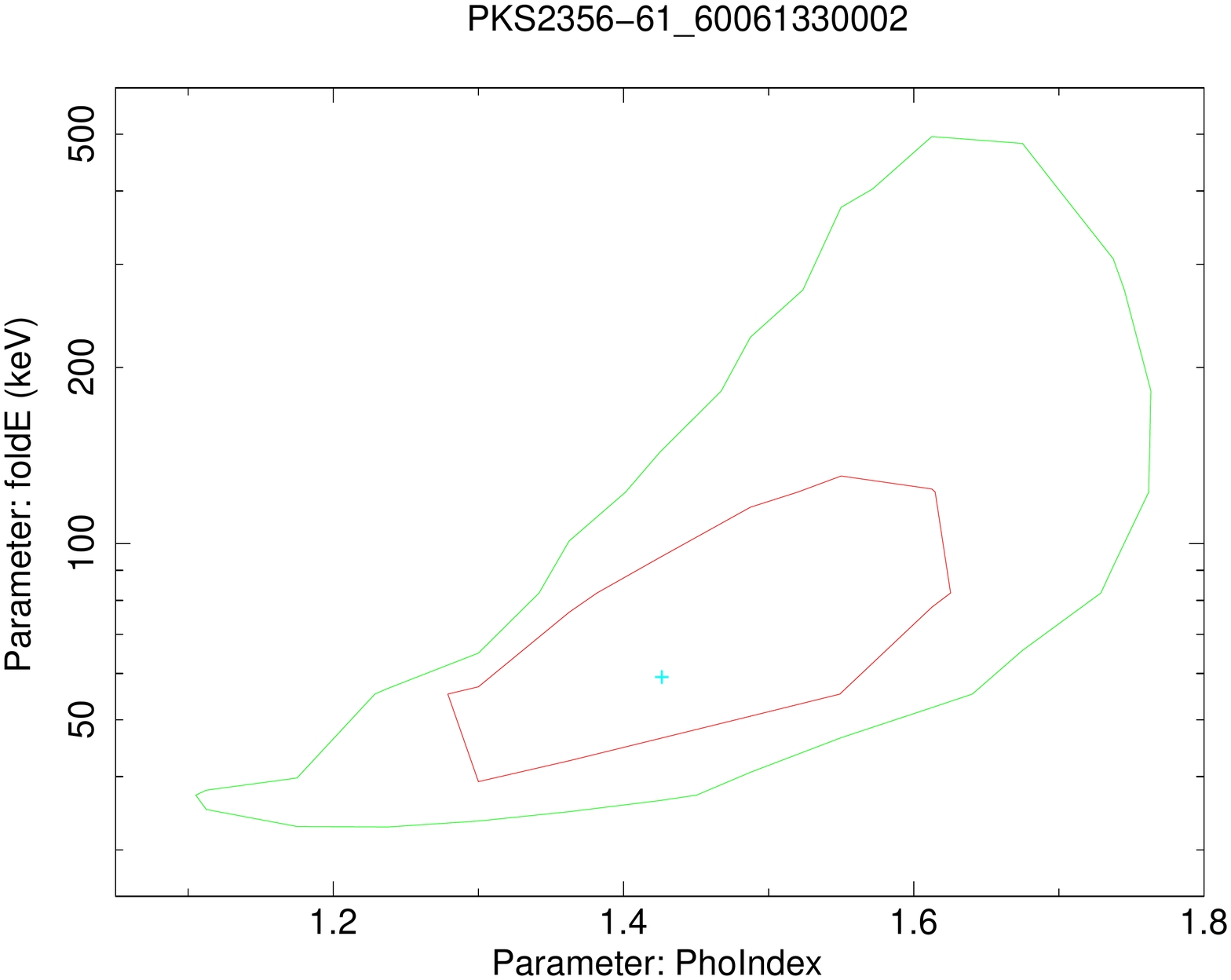}}\\

\label{fig:contour}
\caption{
 The contour plots of $\Gamma$ v.s. $E_{cut}$ of sources with $E_{cut}$ detected. For sources with multiple observations, the best-constrained one is plotted. The red, green and blue present 1$\sigma$, 90\% and 2$\sigma$ (when available) confidence level respectively.
 }
\end{figure*}

\nocite{Tange2011a}
\end{appendix}
\end{document}